\documentclass[12pt]{article}
\usepackage[utf8]{inputenc}

\usepackage{amsmath}
\usepackage{amssymb}
\usepackage{amsthm}
\usepackage{graphicx}
\usepackage{subcaption}
\usepackage{url}
\usepackage[margin=1in]{geometry}
\usepackage{multirow}
\usepackage{color}

\begin{document}
\renewcommand{\baselinestretch}{1.1}
\newcommand{\rb}{\rangle}
\newcommand{\mm}{m}
\newcommand{\lb}{\langle}
\newcommand{\ga}{{v}}
\newcommand{\la}{\lambda}
\def\cistar{\kern.2em\mbox{$\odot\kern-.67em\star\kern .4em$}}
\newcommand{\gaL}{{\gammaa \in \Gamma}}
\newcommand{\wS}{\widetilde S}
\newcommand{\om}{\omega}
\newcommand{\op}{{\rm op}}
\newcommand{\Real} {{\rm Real}}
\newcommand{\Ima} {{\rm Ima}}
\newcommand{\wtilde} {\widetilde}
\newcommand{\R} {{\mathbb R}}
\newcommand{\M} {{\cal M}}
\newcommand{\E} {{\mathbb E}}
\newcommand{\N} {{\mathbb N}}
\newcommand {\h} {{\mathfrak h}}
\newcommand{\Z} {{\mathbb Z}}
\newcommand{\C} {{\mathbb C}}
\newcommand{\La}{{E}}
\newcommand{\LaG}{{E_G}}
\newcommand{\LAG}{E_G}
\newcommand{\Ld} {{\bf L^2}}
\newcommand{\LDT} {{\bf L^2}({[0,T]}^{d})}
\newcommand{\LD} {{\bf L^2}({\mathbb R}^{d})}
\newcommand{\Ltwo} { { {\bf L}^2(\R^2)}}
\newcommand{\Lu} {{\bf L^1}}
\newcommand{\Ga} {{V}}
\newcommand{\Gaw} {{\Gamma}}
\newcommand{\1}{\mathbbm{1}}
\newcommand{\UPhi}{{\cal R}}
\newcommand{\Uphi}{{\cal R}}
\newcommand{\Ns} {{T}}
\newcommand{\W} {{\cal W}}
\newcommand{\HH} {{\cal H}}
\newcommand{\NN} {{|\La|}}
\newcommand{\Tr} {{\rm Tr}}
\newcommand{\F} {{\cal F}}
\newcommand{\ns} {{t}}
\newcommand{\Q} { 2L}
\newcommand{\B} {\Lambda }
\newcommand{\Cov} {{\mbox{\rm Cov}}}
\newcommand{\Var} {{\mbox{Var}}}
\newcommand{\Si}{{\widehat{K}}}
\newcommand{\SiM}{{\widehat{M}}}
\newcommand{\SiC}{{\widehat{C}}}
\newcommand{\SiU}{{\widehat{U}}} 
\newcommand{\SiH}{{\widehat{H}}} 
\newcommand{\OmN}{\widehat{\Omega}_N}
\newcommand{\what}{\widehat}
\newcommand{\wH}{\widehat H}
\newcommand{\flip}{ \overset{\circ}}
\newcommand{\bunderline}[1]{\underline{#1\mkern-4mu}\mkern4mu }
\newcommand{\jmax}{{J}}
\newcommand{\Voi}{{\mathcal{N}}}

% Keywords command
\providecommand{\keywords}[1]
{
  \small	
  \textbf{\textit{Keywords---}} #1
}

\newcommand*{\figuretitle}[1]{%
    {\centering%   <--------  will only affect the title because of the grouping (by the
    \textbf{#1}%              braces before \centering and behind \medskip). If you remove
    \par\medskip}%            these braces the whole body of a {figure} env will be centered.
}
\newtheorem{theorem}{Theorem}[section]
\newtheorem{proposition}{Proposition}[section]
\newtheorem{corollary}{Corollary}[section]
\newtheorem{lemma}[theorem]{Lemma}
\newtheorem{definition}{Definition}[section]

\title{\vspace{-0.7in} 
Maximum Entropy Models from Phase Harmonic Covariances
}

\author{Sixin Zhang$^{1,4}$, St\'ephane Mallat$^{1,2,3}$\thanks{This work is supported by the ERC InvariantClass 320959. This work was supported by the PRAIRIE 3IA Institute of the French ANR-19-P3IA-0001 program.}\\
\it $^1$ ENS, PSL University, Paris, France \\
\it $^2$ Coll\`ege de France, Paris, France\\
\it $^3$ Flatiron Institute, New York, USA\\
\it $^4$ Center for Data Science, Peking University, Beijing, China\\
}

\maketitle

\abstract{The covariance of a stationary process $X$ is diagonalized by a Fourier transform. It does not take into account the complex Fourier phase and
defines Gaussian maximum entropy models.
We introduce a general family of phase harmonic covariance moments, which rely
on complex phases to capture non-Gaussian properties.
They are defined as the covariance of $\what \HH(L X)$,
where $L$ is a complex linear operator and $\what \HH$ is a non-linear phase harmonic operator which multiplies the phase of each complex coefficient
by integers.
The operator $\what \HH$ can also be calculated from
rectifiers, which relates $\what \HH(L X)$ to neural network coefficients.
If $L$ is a Fourier transform then the covariance is a sparse matrix whose
non-zero off-diagonal coefficients capture dependencies between frequencies. These coefficients have
similarities with high order moment, but  
smaller statistical variabilities because $\what \HH$ is Lipschitz. If $L$ is a complex wavelet transform then off-diagonal coefficients
reveal dependencies across scales, which specify the geometry of local
coherent structures. We introduce maximum entropy models 
conditioned by these wavelet phase harmonic covariances.
The precision of these models is numerically evaluated
to synthesize images of turbulent flows and other stationary processes.
}


\section{Introduction}

Many phenomena in physics, finance, signal processing and image 
analysis can be modeled as realizations of a non-Gaussian stationary
process $X$. We often observe a single
realization of dimension $d$ from which the model must be estimated.
Models of stochastic processes may be defined as maximum entropy
distributions conditioned by a predefined family of moments
\cite{jaynes1957information,Cover2006}.
To estimate accurately these moments from a single or few realizations,
the number of moments should not be too large.
The Fourier transform diagonalizes the covariance of stationary processes but
a maximum entropy model conditioned by these second order moments is Gaussian.
The main difficulty is
to specify moments which characterize important classes of
non-Gaussian stationary distributions.

Second order moments do not depend upon
the complex phase of Fourier coefficients. 
We shall see that this complex
phase plays a crucial role to specify non
Gaussian properties of probability distributions, which can be
captured with appropriate covariance moments. We define a 
phase harmonic covariances as
the covariance of $\widehat \HH ( L X)$, where $L$ is a complex linear
operator and $\widehat \HH$ is a non-linear phase harmonic operator 
introduced in \cite{mzr-18}. The operator $\widehat \HH$
multiplies by integers the phase of each complex
coefficient of $L X$. The paper studies general properties
of phase harmonic covariances when $L$ is a Fourier or a wavelet transform,
to approximate non-Gaussian distributions through maximum entropy models.

If $L$ is a Fourier transform,
Section \ref{secpower} proves
that the covariance of $\widehat \HH ( L X)$ is a sparse
matrix whose non-zero off-diagonal coefficients reveal
non-linear dependencies between different frequencies.
Similarly to high order moments,
the covariance of
$\widehat \HH(L X)$ captures this dependence across frequencies
by cancelling relative phase fluctuations.
A high order moment of a complex random variable
$z = |z| \, e^{i \varphi(z)} \in \C$ computes an exponentiation
$z^k = |z|^k\, e^{i k \varphi(z)}$, which amplifies the variability
of large $|z|$
if $k > 1$. A phase harmonic operator $\widehat \HH$ performs a similar
transformation of the phase by calculating
$|z| \,e^{i k \varphi(z)}$. It also
produces non-zero off-diagonal covariance
coefficients without
amplifying $|z|$. It is a Lipschitz operator
as opposed to high order exponentiations, which reduces the variance
of empirical statistical estimators.

The error of a maximum entropy model can be
measured with a Kullback-Leibler divergence relatively to the true
probability distribution of $X$.
Section \ref{maxentsec} explains that this error
can be reduced by improving the sparsity of
$\what \HH(L X)$. For large classes of random processes $X$,
sparsity is improved by defining $L$ as a complex wavelet transform as opposed
to a Fourier transform.
High order moments of wavelet coefficients
have been used to characterize non-Gaussian
multifractal properties of random 
processes and turbulent flows \cite{leader4286563,farge1999non}.
We replace these high order moments by the covariance matrix of
$\widehat \HH(L X)$, to capture non-linear dependencies across scales,
without suffering from high statistical variabilities. Dependencies across
scales reveal the existence of coherent signal
structures as opposed to Gaussian fluctuations.
Section \ref{s:wavecov} studies the covariance
of this wavelet phase harmonic operator.

Wavelet phase harmonics are related to 
the pioneer work of Grossmann and Morlet \cite{1989wtfm}, who showed
qualitatively that complex wavelet transforms have a phase whose
variations across scales provide important information on
the geometry of transient structures. In image processing,
Portilla and Simoncelli \cite{portilla2000parametric} brought
a new perspective by showing that
one can synthesize non-Gaussian image textures from 
the covariance of
the modulus of wavelet coefficients and by using their
phase at different scales.
We shall see that Portilla and Simoncelli
covariances coefficients \cite{portilla2000parametric} are low order
phase harmonic covariances $\what \HH(L X)$, computed with
a wavelet transform $L$.
Other remarkable image texture syntheses have been obtained
in \cite{Ustyuzhaninov2017a}, by computing the covariance of 
output coefficients of an one-layer convolutional neural network
with a rectifier.
Section \ref{phaharmcov} proves that a non-linear phase harmonic operator
$\widehat \HH$ can be calculated with a rectifier followed by a Fourier
transform along a phase variable. These neural network covariances are
therefore equivalent to phase harmonic covariances, 
but for a convolutional operator $L$ which is not a wavelet transform.
The analysis of phase harmonic operators and the properties of covariance
matrices gives a general mathematical framework to understand these 
algorithms. It guides the choice of the linear transform $L$,
of the diagonal and off-diagonal covariance moments,
to model specific classes of random processes.

Section \ref{s:maxent} reviews the properties of
maximum entropy models conditioned by empirical estimators of 
covariance moments. Sampling a maximum entropy distribution
requires to use expensive 
algorithms that iterate over Gibbs samplers \cite{slowmcmc}, which
is not feasible over large size images. 
We thus rather
use microcanonical models studied in \cite{bruna2018multiscale}.
Samples of microcanonical models are estimated by
transporting an initial maximum entropy
measure with a gradient descent which adjusts its moments.
Section \ref{approxnsec} defines low-dimensional 
foveal models conditioned by a small subset of
wavelet phase harmonics covariances. These moments
capture local dependencies within and across scales. 
We evaluate the numerical precision of foveal wavelet
phase harmonic models,
to approximate two-dimensional non-Gaussian processes
including images of turbulent flows and multifractals. 
All calculations can be reproduced by a Python software
available at \url{https://github.com/kymatio/phaseharmonics}.

{\bf Notations:} 
We write $z^*$ the complex conjugate of $z \in \C$, $Y^*$ is the complex transpose of a matrix $Y$  and ${\cal A}^*$ is the 
adjoint of an operator ${\cal A}$. The covariance of two random variables $A$ and $B$ 
is written $\Cov(A,B) = \E(A\,B^*) - \E(A)\,\E(B)^*$. 
An inner product 
is written $\lb x , y \rb = \int x(u)\, y(u)\, du$ in $\LD$
and $\lb x , y \rb = \sum_u x(u)\, y(u)$ in $\mathbb R^d$. The cardinal of a set $S$ is $|S|$. The norm of $x$ in $\LD$ or $\mathbb R^d$ is written $\| x \| = \lb x, x \rb^{1/2}$.

\section{Fourier Phase Harmonic Representation}
\label{secpower}

Section \ref{maxentsec}
introduces general properties of maximum entropy models
conditioned by covariance coefficients of a non-linear representation
$\UPhi(X)$ over a graph. 
Section \ref{psdofnsodfs} then shows that the Fourier coefficients
of a stationary process are uncorrelated because of phase fluctuations. 
It motivates the use of higher order
moments to define a representation which
can cancel random phase fluctuations and capture dependence
across frequencies.
Section \ref{waveharmosec} 
reviews the properties of phase harmonic operators $\widehat \HH$ introduced
in \cite{mzr-18}.
If $L$ is a Fourier transform,  Section \ref{sndfoihs8as} shows
that $\UPhi(X) = \widehat \HH( L X)$ is a bi-Lipschitz representation,
whose covariance
captures non-Gaussian properties through dependencies across frequencies,
similarly to high order moments. 

\subsection{Maximum Entropy Covariance Graph Models}
\label{maxentsec}

We introduce maximum entropy models of a stationary random vector $X$, 
conditioned by covariance 
coefficients of a representation $\UPhi(X)$ which is also a random
vector:
\[
K_\UPhi =  \Cov( \Uphi (X) ,  \Uphi (X) ) =
\E \Big( (\UPhi (X) - M_\UPhi )( \UPhi(X) - M_\UPhi )^* \Big)
\]
with $M_\UPhi = \E(\UPhi (X))$.
We suppose that $X(u)$  is defined for
$u$ in a cube $\Lambda_d \subset \Z^r$
with $d$ grid points,
for example in $[1,d^{1/r}]^r$. If $r = 2$ then each realization is an
image of $d$ pixels. This section reviews general properties
of maximum entropy models defined on a graph, which take advantage of
known symmetries of the distribution of $X$.

\paragraph{Maximum entropy on a graph with symmetries}
Let us write $\UPhi(X) = \{\UPhi_v (X) \}_{v \in V}$, where $\UPhi_v (X) \in \C$ and
$V$ is a finite set of vertices in a graph that we shall define.
The representation $\UPhi(X)$ is a complex-valued vector in $\C^{|V|}$.

We introduce a maximum entropy model 
conditioned by the covariance between vertices
\[
K_\UPhi(\ga, \ga') = \Cov( \Uphi_v (X) ,  \Uphi_{v'} (X) ),
\]
where $v'$ is in a neighborhood $\Voi_v \subset V$ of $v$.
We suppose that $v \in \Voi_v$.
If $v' \in \Voi_{v}$ then $v \in \Voi_{v'}$ so
it defines a reflexive and symmetric undirected graph $(V,E)$, 
where the set of edges relates all neighbors
$E = \{(v,v') : v \in V\,,\,v' \in \Voi_v \}$.
The edge weights are the covariances $K_\UPhi(\ga, \ga')$ over $E$.
In statistical physics,
$(\UPhi_v (x) - M_\UPhi(v) )( \UPhi_{v'}(x) - M_\UPhi(v') )^*$ 
is called the interaction potential of $(v,v') \in E$, 
where the mean $M_\UPhi$ is considered here 
as a predefined constant vector.

Let $p$ be the probability density of $X$.
We can reduce the constraints of the maximum entropy model if 
we also know a finite
group $G$ of symmetries of $p$. We consider 
linear unitary symmetries $g$ from $\R^d$ to $\R^d$, which satisfy 
$p(g.x) = p(x)$ for all $x \in \R^d$.
The stationarity of $X$ in $\Lambda_d$ means that 
$p$ is invariant to periodic translations, that
we write $g.x(u) = x(u-g)$, so $G$ includes all translations.
If $p$ is invariant to rotations 
then $G$ also includes rotations. 
Since $p$ is invariant to the action of $G$,
its covariance is also invariant to the action of any $g \in G$
\[
\Cov ( \UPhi_{v}(g.X) , \UPhi_{v'}(g.X)) = 
\Cov ( \UPhi_{v}(X) , \UPhi_{v'}(X)).
\]
Let $|G|$ be the total number of symmetries. 
These $|G|$ conditions will be included in the maximum entropy model.

The entropy of a density $\tilde p$ on $\R^d$ is
\[
H(\tilde p) = - \int \tilde p(x)\, \log \tilde p(x)\,dx.
\]
A maximum entropy macrocanonical model $\widetilde X$ conditioned 
by the covariance $K_\UPhi$ over $\La$ and by the symmetry group $G$
has a probability density
$\tilde p$ which maximizes $H(\tilde p)$ and satisfies covariance
moment conditions for all $(\ga,\ga',g) \in \La \times G$
\begin{equation}
\label{macrosndsf0}
\int (\UPhi_\ga(g.x) - M_\UPhi(\ga))\, (\UPhi_{\ga'} (g.x) - M_\UPhi (\ga'))^*\, \tilde p(x)\, dx 
= K_\UPhi (\ga,\ga') .
\end{equation}
If there exist Lagrange multipliers $\beta_{\ga,\ga'} \in \C$ such that
\begin{equation}
\label{macrosndsf}
\tilde p(x) = Z^{-1} \, \exp\Big(- \frac 1 {|G|}
 \sum_{(\ga,\ga',g) \in \La \times G}  \beta_{\ga,\ga'}
\,(\UPhi_\ga(g.x) - M_\UPhi(\ga))\, (\UPhi_{\ga'}(g.x) - M_\UPhi(\ga'))^* \Big)
\end{equation}
is a probability distribution which satisfies each equality constraint \eqref{macrosndsf0},
then it is the unique solution to the maximum-entropy problem \cite{Cover2006}.
The sum in the exponential is the Gibbs energy, and $Z$ is the partition function. Since $K_\UPhi (\ga,\ga') = K^*_\UPhi (\ga',\ga)$
we have $\beta_{\ga,\ga'} = \beta_{\ga',\ga}^*$.
We verify that for all $g \in G$, $\tilde p(g.x) = \tilde p(x)$ so
$G$ is also a group of symmetries of $\tilde p$. If $G$ includes all translations then $\tilde p$ is stationary.

In the particular case where all 
$\UPhi_\ga (x)$ are linear operators then the Gibbs
energy is bilinear and $\tilde p(x)$ 
is thus a Gaussian distribution.
Appendix \ref{appendixmaxent} shows that 
the Lagrange multipliers can then 
be computed efficiently \cite{Cover2006}.
If some of the $\UPhi_\ga(x)$ are non-linear then
computing the Lagrange multipliers is computationally very expensive
when the dimension $d$ of $X$ and the total
number $|\La|$ of moments is large. 

\paragraph{Covariance estimation from symmetries}
The covariance estimation from a single realization $\bar x$ of $X$
is calculated with an average over the symmetries of $G$.
The orbit of the action of $G$ on $\bar x$
is the set of $\{g.\bar x \}_{g \in G}$. 
The empirical estimation of $M_{\UPhi} (\ga) = \E(R_\ga (X))$ is
computed as an empirical average over this orbit
\begin{equation}
\label{phiFouriensfmean6}
\widetilde M_{\UPhi}  = 
\frac {1} {|G|} \sum_{g \in G} \UPhi(g. \bar x) .
\end{equation}
Similarly, $K_\UPhi$ is estimated
with an average on the same orbit
\begin{equation}
\label{phiFouriensf6}
\widetilde K_{\UPhi \bar x} = 
\frac {1} {|G|} \sum_{g \in G} 
\Big( \UPhi (g. \bar x) - \widetilde M_{\UPhi}  \Big)\,
\Big( \UPhi (g. \bar x) - \widetilde M_{\UPhi} \Big)^*\,.
\end{equation}
Summing over all 
transformations of $\bar x$ by $g \in G$ 
is called ``data augmentation'' in machine learning. 
The larger $|G|$ the more accurate the estimation. The maximum entropy model is estimated
from a single realization $\bar x$ 
by replacing the mean $M_\UPhi$ and covariance $K_{\UPhi}$ 
in (\ref{macrosndsf0}) by their 
estimation $\widetilde M_\UPhi$ and $\widetilde K_{\UPhi \bar x}$.

The existence of symmetries also reduces the number of covariances that
must to be estimated.
We define $\LaG$ as a minimum set of edges $(v,v')$ such that
for any $(v,v') \in E$ there exists $(v_1,v_1',g) \in E_G \times G$
such that $\UPhi_{v} (x) = \Uphi_{v_1} (g.x)$ and 
$\UPhi_{v'} (x) = \Uphi_{v'_1} (g.x)$.
Since $p$ is invariant to the action of $G$,
its covariance $K_\UPhi$ is also invariant:
\[
K_\UPhi(v,v') = \Cov ( \UPhi_{v_1}(g.X) , \UPhi_{v'_1}(g.X)) = 
K_\UPhi(v_1,v'_1) .
\]
It is therefore sufficient to estimate covariance coefficients indexed 
by $E_G$ to specify all covariances
indexed by $E$. The set $\LaG$ is interpreted as a set of 
sufficient statistics.
Given a realization $\bar x$ of dimension $d$, to control the overall
covariance estimation errors we must insure that $|E_G| \ll d$.

\paragraph{Bi-Lipschitz continuity}
The estimation of covariance coefficients may have a large variance
in the presence of rare outliers. These outliers 
induce a large variability
in the empirical sum (\ref{phiFouriensf6}) depending upon the
realization $\bar x$ of $X$. To avoid amplifying these outliers, we
impose that $\UPhi$ is bi-Lipschitz, so that the 
variations of $\UPhi(X)$ are of the same order as the variations of $X$.
It also insures that $\UPhi$ is an invertible operator. 

The representation $\UPhi$ is bi-Lipschitz if there exists
$A_\UPhi > 0$ and $B_\Uphi$ such that for all $(x,x') \in \R^{2d}$
\begin{equation}
\label{biLips}
A_\UPhi\, \| x - x' \|^2 \leq \| \UPhi(x) - \UPhi(x') \|^2 \leq B_\UPhi \, \|  x - x' \|^2 . 
\end{equation}

For any random vector $Z$, we write $\sigma^2(Z) = \E(\|Z - \E(Z) \|^2)$,
which is the trace of its covariance. 
The following proposition proves that
the variance of $\UPhi (X)$ and $X$ 
have the same order of magnitude.

\begin{proposition}
	\label{bioLihscionfth}
	If $\UPhi$ is bi-Lipschitz then
	\begin{equation}
	\label{traninequali}
	A_\UPhi\, \sigma^2(X) \leq \sigma^2(\UPhi( X))\leq  B_\UPhi \,\sigma^2(X) .
	\end{equation}
\end{proposition}

{\it Proof:}
If $X'$ and $X$ are two independent random vectors having same 
probability distribution then the
bi-Lipschitz
bounds (\ref{biLips}) of $\UPhi$ imply that 
\begin{equation}
\label{eandf08s}
A_\UPhi \,
\E(\| X - X' \|^2) \leq
\E(\|\UPhi( X) - \UPhi( X') \|^2) \leq 
B_\HH\,\E(\| X - X' \|^2) .
\end{equation}
If $Y$ and $Y'$
are two i.i.d random variables then we verify that
\[
\E(|Y - Y'|^2) = 2\, 
\E(|Y - \E(Y)|^2) .
\]
where the first expected value is taken relatively to the joint
distribution of $Y$ and $Y'$. It results that
$ \E(\| X - \E(X) \|^2) = 2^{-1}\,
\E(\| X - X' \|^2)$ and
$ \E(\|\UPhi( X) - \E(\UPhi(X)) \|^2) = 2^{-1}\,
\E(\|\UPhi( X) - \UPhi(X')) \|^2)$.
Inserting these equalities in (\ref{eandf08s}) proves (\ref{traninequali}).
$\Box$

\paragraph{Maximum entropy reduction with sparsity}
Zhu, Wu and Mumford \cite{zhuminimax} have proposed to 
optimize maximum entropy parameterized models by minimizing the resulting
maximum entropy. Indeed, 
if $X$ has a probability density $p$ then model error can be evaluated with a Kullback-Leibler divergence
\[
D_{KL} (p||\tilde p) = \int p(x)\, \log \frac{p(x)} {\tilde p(x)}\, d x .
\]
We verify
that $\int p(x)\, \log {\tilde p(x)} \, d x = 
\int \tilde p(x)\, \log {\tilde p(x)} \, d x$ 
by inserting (\ref{macrosndsf}) and by using
the equalities (\ref{macrosndsf0}).
Since $H(p) = \int p(x) \,\log p(x)\, dx$,
it results that
the Kullback-Leibler divergence is equal to the excess of entropy of
the maximum entropy model:
\begin{equation}
\label{KLdivers}
D_{KL} (p||\tilde p) = H(\tilde p) - H(p) \geq 0 .
\end{equation}
We thus reduce the model error by reducing 
the maximum entropy $H(\tilde p)$ to obtain a lower entropy model.
The minimum $H(\tilde p) = H(p)$ is reached if and only if
$\tilde p = p$ almost-surely. 

In our case, the model depends upon the choice of $\UPhi$ and of the
edges $E \subset V^2$ of the covariance graph. Optimizing $\UPhi$ by
calculating $H(\tilde p)$ is computationally too expensive. However,
we explain below that we can minimize an upper bound of $H(\tilde p)$ by
finding a representation such that $\UPhi(X)$ is as sparse as possible,
which gives a 
partial control on the maximum entropy.
Let $\wtilde X$ be
a maximum entropy model of density $\tilde p$. If $\UPhi$ has
a stable inverse, which we guarantee with the bi-Lipschitz condition,
then an upper bound of $H(\tilde p)$ can be computed from the 
entropy of each marginal $\UPhi_v (\wtilde X)$. Such an entropy is
small if $\UPhi_v (\wtilde X)$ is sparse, because it then
has a narrow probability density centered at $0$. 
To impose this sparsity we must find
a representation $\UPhi$ such that $\UPhi_v (X)$ is sparse.
This sparsity must also be 
captured by diagonal covariance coefficients $K_\UPhi (v,v)$ so that
the maximum entropy model $\wtilde X$, conditioned by these moments retains this sparsity. This is the case for Fourier and wavelet
phase harmonic representations studied in Sections \ref{sndfoihs8as} and 
\ref{phaharmcov}. For non-Gaussian processes, 
it amounts to represent ``coherent structures''
with as few non-zero coefficients as possible. 

Increasing the number $|\La|$ of moment conditions
can further reduce $H(\tilde p)$ but it also increases the statistical error when estimating these moments from a single realization of $X$. The choice of $E$ is thus a trade-off between the model error (bias) and the estimation error (variance).

\subsection{Fourier Phase and High Order Moments}
\label{psdofnsodfs}

The covariance of a stationary random vector 
is diagonalized by the discrete Fourier transform, because of random
phase fluctuations.
This suggests defining a covariance representation in a Fourier basis.
For non-Gaussian processes, the 
dependence of Fourier coefficients is partly
captured by high order moments which cancel random
phase fluctuations.

For $\UPhi = Id$, since $X(u)$ for $u \in \Lambda_d$ is stationary, 
$K_{\UPhi} (u,u') =  \Cov ( X(u) , X(u') )$
only depends on $u-u'$.
We write $|u|$ the norm of $u \in \Lambda_d$ and $u'.u$ the inner product.
A low-dimensional maximum entropy model is constructed on $V = \Lambda_d$
by restricting covariances to neighborhoods of fixed radius:
$\Voi_v = \{v' : |v-v'| \leq c \}$. The radius $c$ is chosen to be large
enough to capture long range correlations.  
Since $X$ is stationary, the symmetry group includes translations and
we can thus define $E_G$ by setting $v = 0$.
The maximum entropy model then defines a Gauss-Markov random vector \cite{Cover2006} specified by covariances in a small neighborhood. This model
does not capture non-Gaussian properties. If $\Lambda_d$ is an r-dimensional
grid, then the model size is $|\LaG| = O(c^r)$. It
may be large if the process has long-range spatial correlations.

To better understand how to capture non-Gaussian properties, 
we study these
covariance coefficients in a Fourier basis. 
We write $\what x = \F_u x$ the discrete Fourier transform of $x$
over $\Lambda_d$:
\begin{equation}
\label{dionsdfsdf}
\widehat x(\om) = \sum_{u\in \Lambda_d} x(u)\, e^{-i  \om. u }~~\mbox{for}~~
\om = 2 \pi d^{-1/r} m ~~\mbox{with}~~ m \in \Lambda_d.
\end{equation}
The Fourier representation $\UPhi = \F_u$ is indexed by 
$\om \in V = 2 \pi d^{- 1/r} \Lambda_d$. 
The covariance for $(\om,\om') \in V^2$ is
\[
K_\UPhi (\om,\om') = \Cov(\what X(\om) , \what X(\om')) .
\]

If $\om \neq 0$ then $\E(\what X(\om)) = 0$. If $\om \neq \om'$ then
$K_\UPhi (\om,\om') = 0$ because of random phase 
fluctuations. Indeed, translating $X(u)$ by any $\tau \in G_q$ multiplies
$\what X(\om)$ by $e^{-i \tau. \om }$. Since $X$ is stationary, this 
translation is a symmetry which
does not modify $\Cov(\what X(\om), \what X(\om'))$. It 
results that
\begin{equation}
\label{high-ordercov00}
\Cov(\what X(\om), \what X(\om'))
 = e^{i  (\om - \om'). \tau}\, 
\Cov(\what X(\om), \what X(\om')) .
\end{equation}
Since this is true for any $\tau \in \Lambda_d$ it implies that
\begin{equation}
\label{high-ordercov0}
\Cov(\what X(\om) , \what X(\om')) = 0~~\mbox{if}~~\om \neq \om'.
\end{equation}
Since the discrete Fourier transform is periodic beyond
$[0,2\pi]^r$, any equality or non-equality between frequencies
must be understood modulo $2 \pi$ along the $r$ directions.
If $X$ is Gaussian then $\what X$ is also Gaussian so this non-correlation 
implies that $\what X(\om)$ and $\what X(\om')$ are independent. 
However, if $X$ is non-Gaussian then 
$\what X(\om)$ and $\what X(\om')$ are typically not independent.

\paragraph{Phase cancellation with higher order moments}
To capture the dependencies of Fourier coefficients across frequencies, 
one can use high order moments \cite{cramerstat}. 
For any $k \in \Z$,
similarly to (\ref{high-ordercov00}), translating $X$ by 
$\tau \in \Lambda_d$ yields
\begin{equation}
\label{high-ordercov001}
\Cov(\what X(\om)^k, \what X(\om')^{k'})
 = e^{i  (k \om - k' \om'). \tau}\, 
\Cov(\what X(\om)^k, \what X(\om')^{k'}) .
\end{equation}
A high-order Fourier representation 
$\UPhi_v (X) = \what X(\om)^k$ is indexed
by $v = (\om,k) \in V$ with $0 \leq k \leq k_{\max}$.
It results from  (\ref{high-ordercov001}) that
\begin{equation}
\label{high-ordercov}
K_\UPhi(v,v') = \Cov(\what X(\om)^k, \what X(\om')^{k'}) = 0~~\mbox{if}~~k \om \neq k' \om'~.
\end{equation}
If $k \om = k' \om'$ and $X$ is not Gaussian then
this covariance is typically non-zero because the
phase variations of $\what X(\om')^{k'*}$
cancel the phase variations of $\what X(\om)^k$. This is also the key
idea behind the use of bi-spectrum moments \cite{rao2012introduction}.
By adjusting $(k,k')$ these moments
provide some dependency information between $\what X(\om)$ and
$\what X(\om')$ for $\om \neq \om'$.

For example, let $X(u) = x(u-S)$ be a random shift vector, where $x(u)$ is
a fixed signal supported in $\Lambda_d$ and $S$ is a random periodic 
shift which is uniformly distributed in $\Lambda_d$. It is a
stationary process whose Fourier coefficients have a random phase:
$\what X(\om) = \what x(\om)\, e^{- i S \om}$. In this case,
if $k \om = k' \om'$ then
\[
\Cov(\what X(\om)^k, \what X(\om')^{k'}) = \what x(\om)^k\, \what x(\om')^{*k} .
\]
It is non-zero at frequencies where $\what x$ does not vanish.
This shows that covariances of the high order Fourier exponents
$\UPhi(X)$ can capture the dependence
of Fourier coefficients at different frequencies. However, this
high order Fourier representation is not Lipschitz. Indeed, when 
$k > 1$, the exponent $k$ amplifies the variability
of each random variables $\what X(\om)$, so
estimators of covariance coefficients have a large variance.

\subsection{Phase Harmonic Operator}
\label{waveharmosec}

We review the properties of phase harmonic operators $\widehat \HH$
introduced in \cite{mzr-18}, which
replace high order exponents by an exponent on the
phase only. These operators are bi-Lipschitz and can be computed
by applying a rectifier followed by a Fourier transform on a phase
variable. If $L$ is a Fourier tranform,
next section shows that $\widehat \HH( L X)$ can have
non-zero covariance coefficients across frequencies because
of relative phase cancellations.

If $z = |z| e^{i \varphi(z)} \in \C$ then
$z^k= |z|^k\, e^{i k \varphi(z)}$.
If $k > 1$ then $|z|^k$
amplifies large values of $|z|$ and hence
the variance of a random variable $z$. 
A phase harmonic computes a power 
$k \in \Z$ of the phase only:
\[
[z]^k = |z|\, e^{i k \varphi(z)}~.
\]
It preserves the modulus: $|[z]^k| = |z|$.
A phase harmonic operator computes all such phase harmonics
\begin{equation}
\label{Harmondsdef00}
\what \HH (z) = \{ \what h(k)\, [z]^{k} \}_{k \in \Z}~,
\end{equation}
with finite energy weights $\|\what h \|^2 = \sum_{k \in \Z} |\what h(k)|^2 < \infty$. The harmonic weights $\what h(k)$ amplify
or eliminate different phase harmonics.

\paragraph{Phase windows and phase Fourier transform}
We show that $\what \HH$ is the Fourier transform along a phase
variable of a non-linear phase windowing operator, which can be computed
with rectifiers used in neural networks.

Let us write $\what h = \F_\alpha (h)$ the Fourier transform along phases:
\begin{equation}
\label{andfphsFOurier}
\what h(k) = \frac 1 {2 \pi} \int_0^{2 \pi} h(\alpha)\, e^{-i k \alpha}\, d \alpha .
\end{equation}
Observe that
$\F_\alpha \, h(\varphi(z) + \alpha) = \{ \hat h(k)\, e^{i k \varphi(z)} \}_{k \in \Z}$.
It results that
\[
\what \HH (z) = \{ \what h(k)\, [z]^{k} \}_{k \in \Z} = \F_\alpha\, \HH (z)
\]
with
\begin{equation}
\label{Harmondsdef}
\HH(z) = \{ |z|\, h(\varphi(z) + \alpha) \}_{\alpha \in [0,2\pi]}.
\end{equation}
The operator $\HH$
translates the phase $\varphi(z)$ of $z$ by 
$\alpha \in [0,2\pi]$ and applies a $2 \pi$ periodic
phase windowing $h(\alpha)$.

A rectifier $\rho (a) = \max(a,0)$ is an important example
of non-linearity which acts as a phase windowing when applied on the
real part of $z$. Indeed
\[
\rho( {\rm Real}(z )) = |z|\, \rho(\cos\varphi(z))~,
\]
so
\begin{equation}
\label{phase-shift-filt}
\{ \rho( {\rm Real}( e^{i \alpha} z )) \}_{\alpha \in [0,2\pi]} = \HH (z)~~
\mbox{with}~~h(\alpha) = \rho(\cos \alpha) .
\end{equation}
The rectifier phase window $\rho(\cos \alpha)$ is positive and supported
in $[-\pi/2,\pi/2]$. 
The corresponding harmonic weights are computed in \cite{mzr-18} 
with the Fourier
integral (\ref{andfphsFOurier}):
\begin{equation}
\label{nsdfousdfs}
\widehat {h}(k) = \left\{
\begin{array}{ll}
\frac{(-1)^{k/2+1}} {\pi (k^2-1)} & \mbox{if $k$ is even}\\
\frac{1} 4 & \mbox{if $k= \pm 1$}\\
0 & \mbox{if $|k|>1$ is odd}
\end{array}
\right. .
\end{equation}
Their decay is slow because 
$h(\alpha)$ has discontinuous derivatives at $\pm \pi/2$.

\paragraph{Bi-Lipschitz continuity}
We mentioned that
polynomial exponents amplify the variability of random variables around
their mean. Indeed if $(z,z') \in \C^2$ then
$|z^k - z'^k|/|z-z'|$ may be arbitrarily large if $k > 1$.
On the contrary, a phase harmonic preserves the modulus which provides
a bound on such amplification.
It is proved in \cite{mzr-18} that it is Lipschitz continuous
\begin{equation}
\label{lipsdfnsdf}
\forall (z,z') \in \C^2~~,~~|[z]^k - [z']^k| \leq \max(|k|,1)\, |z - z'| .
\end{equation}
The distance $|z-z'|$ is therefore amplified by at most $|k|$.

This Lipschitz continuity 
is extended to phase windowed Fourier transforms,
which are shown to be bi-Lipschitz.
The Fourier transform preserves norms and distances.
Calculating the norm of (\ref{Harmondsdef00}) gives
\[
\|\HH(z)\| = \|\what \HH (z) \| = \|h\|^2 \, |z|^2 ,
\]
with $\|h\|^2 = \sum_k |\what h(k)|^2$. Following \cite{mzr-18},
we also derive from (\ref{lipsdfnsdf}) that
a phase windowed Fourier transform is
bi-Lipschitz over complex numbers $(z,z') \in \C^2$
\begin{equation}
\label{nsdfsdfa8sdfs8sd0}
A_{\HH} \, |z-z'|^2 \leq 
\|\what \HH (z) - \what \HH (z)' \|^2 \leq B_{\HH} \, |z - z'|^2~,
\end{equation}
with $A_{\HH} = 2\, |\what h(1)|^2$ and 
$B_{\HH} = |\widehat h(0)|^2 + \sum_{k\in \Z} k^2\, |\widehat h(k)|^2$.
The upper bound $B_\HH$ is derived from (\ref{lipsdfnsdf})
and the lower bound is obtained isolating the terms corresponding
to $k = \pm 1$.
For the rectifier filter (\ref{nsdfousdfs}), we get
$A_{\HH} = 1/8$ and $B_{\HH} = 1/4 + 1/\pi^2$.
The following theorem gives a tight
upper Lipschitz constant for the rectifier filter, 
proved in Appendix \ref{Relu}.

\begin{theorem}
The rectifier phase window $h(\alpha) = \max(\cos\alpha,0)$ 
has a Lipschitz upper-bound $B_{\HH} = 1/4$.
\label{Reluthm}
\end{theorem}

\subsection{Phase Harmonics of Fourier Coefficients}
\label{sndfoihs8as}

If $L = \F_u$ computes the Fourier transform of $X(u)$ then
the resulting phase harmonic representation is
\[
\UPhi(X) = \what \HH (\F_u X)  = 
\Big\{ \what{h}(k)\, [\what X (\om)]^{k} \Big\}_{k \in \Z,\om \in \R} .
\]
It is indexed by $v = (\om,k)$. 
We show that the covariance of $\what \HH (\F_u X)$ is sparse but not
necessarily diagonal.
Section \ref{psdofnsodfs} explains that if $X$ is stationary
then $\what X(\om)$ and $\what X(\om')$ are not correlated if 
$\om \neq \om'$ but
$\what X(\om)^k$ and $\what X(\om')^{k'}$ may be correlated if $X$ is not
Gaussian. Similarly we show that the covariance of
$[\what X(\om)]^k$ and $[\what X(\om')]^{k'}$ may become
non-zero for $\om \neq \om'$, which reveals non-Gaussian properties.

Since $\what \HH$ is bi-Lipschitz and the Fourier transform
is unitary up to a factor $d$, it results that
$\UPhi = \what \HH \F_u$ is bi-Lipschitz with
Lipschitz constants $A_\UPhi = d A_{\HH}$ and $B_\UPhi = d B_{\HH}$.
Proposition \ref{bioLihscionfth} implies that 
\begin{equation}
\label{traninequali22}
d A_{\HH}\, \sigma^2(X) \leq   \sigma^2(\what \HH( X))\leq  d B_{\HH} \,\sigma^2(X).
\end{equation}

\paragraph{Sparse phase harmonic covariance}
Phase harmonic covariance coefficients are
\[
K_{\what \HH \F_u }
(\om, k,\om',k') = \what h(k) \, \what h(k')^* \,
\Cov ([\what X (\om)]^{k}\,,\,[\what X(\om')]^{k'}) .
\]
If $\what h$ is compactly supported in $[0,k_{\max}]$, since
$X$ is of dimension $d$, the covariance
$K_{{\what \HH}\F}$ has $(k_{\max}+1)^2 d^2$ coefficients.
The following proposition proves that coefficients of
$K_{{\what \HH}\F}$ are mostly zero and it is 
diagonal if $X$ is Gaussian. Diagonal values
are specified.

\begin{theorem}
\label{SparseSpecMat}
If $X$ is real stationary then
\begin{equation}
\label{densfi8sdfs}
\Cov ([\what X(\om)]^k,[\what X(\om')]^{k'}) = 0 ~~
\mbox{if}~~k \om \neq k' \om' .
\end{equation}
Along the diagonal, 
if $\om\neq 0$ and $k \neq 0$ or if $\om = 0$ and $k$ is odd then
\begin{equation}
\label{densfi8sdfs7}
\Cov ([\what X(\om)]^k,[\what X(\om)]^{k}) = {\Cov (\what X(\om),\what X(\om))}~.
\end{equation}
If $\om = 0$ and $k$ is even then
\begin{equation}
\label{densfi8sdfs000}
\Cov ([\what X(0)]^k,[\what X(0)]^{k}) = {\Cov (|\what X(0)|,|\what X(0)|)}~.
\end{equation}
For all $\om \neq 0$
\begin{equation}
\label{densfi8sdfs8}
\frac{\Cov (|\what X(\om)|,|\what X(\om)|)}
{\Cov (\what X(\om),\what X(\om))} =
1 - \frac {\E(|\what X(\om)|)^2}{\E(|\what X(\om)|^2)} .
\end{equation}
If $X$ is Gaussian then $K_{\what \HH \F}$ is diagonal and
${\E(|\what X(\om)|)^2}/{\E(|\what X(\om)|^2)} = \pi/4$ 
if $\om \neq 0$.
\end{theorem}

The proof is in Appendix \ref{covarianceFourier}. All equalities or
non-equalities on frequencies must be understood modulo $2 \pi$ in dimension $r$.
Property (\ref{densfi8sdfs}) 
proves that $K_{\what \HH \F_u}$ is 
highly sparse. If $X$ is not Gaussian then off-diagonal coefficients are
typically not zero when $k \om = k' \om'$.
For $k = k' = 0$, we get $d^2$ modulus covariances 
$\Cov (|\what X(\om)|,|\what X(\om')|)$ which are a priori non-zero for 
all $(\om,\om')$. It provides no 
information on the phase of $\what X(\om)$ and $\what X(\om')$.
Phase correlations are captured when $k \neq 0$.
If $\om \neq 0$, it results from (\ref{densfi8sdfs}) that
$\Cov ([\what X(\om)]^k,[\what X(\om')]^{k'}) \neq 0$ only
if $\om'$ is colinear with $\om$. For a grid $\Lambda_d$ of size
$d$, there are $O(d^{1/r})$ such frequencies $\om'$. 
The total number of non-zero off-diagonal covariance
coefficients is thus at most $d^2 + O(d^{1+1/r})$. 

For a non-Gaussian process, 
off-diagonal coefficients with $k \om = k' \om'$ 
are typically non-zero. This is illustrated with 
a  random shift process $X(u) = x(u-S)$ where $S$ is uniformly
distributed in $\Lambda_d$. If $k \om = k' \om'$ then one can verify that
\[
\Cov(\what X(\om)^k, \what X(\om')^{k'}) = [\what x(\om)]^k\, [\what x(\om')]^{-k'} ~~\mbox{if}~~\om \neq 0 , \om' \neq 0.
\]
If $\what x(\om)$ does not vanish then all these coefficients are non-zero.

Along the diagonal, 
(\ref{densfi8sdfs7}) and (\ref{densfi8sdfs000})
prove that all coefficients are either equal to
$\Cov(\what X(\om),\what X(\om))$ or to $\Cov(|\what X(\om)|,|\what X(\om)|)$. Property (\ref{densfi8sdfs8}) also proves that the ratio between these
covariance values depend upon the ratio
${\E(|\what X(\om)|)^2}/{\E(|\what X(\om)|^2)}$. 
This ratio measures the sparsity of multiple realizations of
the random variable $\what X(\om)$. Indeed, expected values are estimated by
normalized sums, so ratio of expected values can be interpreted as ratios
of $\ell_1$ over $\ell_2$ norms, which is a measure or sparsity.
For all Gaussian random variables, this ratio is equal to $\pi/4$.
If the ratio is smaller than $\pi/4$ then
$\what X(\om)$ has a high probability to be
relatively small and it has large amplitude outliers of low probability.
The probability distribution of $\what X(\om)$ is then highly peaked in
zero with a long tail. The smaller this ratio the lower the entropy of
this marginal probability distribution.

We saw in (\ref{KLdivers}) that a maximum entropy model $\wtilde X$ 
introduces errors if its entropy is much larger than the entropy
of $X$. The entropy $ H(\tilde p)$ is bounded by the sum of
the entropy of each random variable $\what X(\om)$, which is conditioned
by  $\Cov(\what X(\om),\what X(\om)$ and 
$\Cov(|\what X(\om)|,|\what X(\om)|)$. It gives an upper bound on
the entropy $H(\tilde p)$, which decreases if the sparsity of 
$\what X(\om)$ increases.
However, if $X(u)$ has sharp localized
transitions then its Fourier coefficients have typically a large 
amplitude across most frequencies and are not sparse. On the 
contrary, wavelet coefficients may be sparse. In this case,
we shall capture this sparsity and thus compute
maximum entropy models of lower entropy, by replacing the
Fourier transform by a wavelet transform.

\section{Wavelet Phase Harmonics}
\label{s:wavecov}

To model random vectors $X$ whose realizations include singularities and sharp
transitions, we define a phase harmonic
representation $\UPhi = \what \HH\,L$ where
$L$ is a wavelet transform as opposed to a Fourier transform.
Indeed, wavelet transform provides sparse representations of such signals
and Section \ref{maxentsec} explains that imposing sparsity constraints 
typically improves the accuracy of maximum entropy models. 
We concentrate on two-dimensional image applications. The same approach
applies to signals of any dimension, with an appropriate
wavelet transform.
Section \ref{s:wavecov1} 
reviews the construction of complex 
steerable wavelet frames and the properties of their
covariance matrices. Section \ref{phaharmcov} studies the
covariance of wavelet phase harmonics.

\subsection{Steerable Wavelet Frame Covariance}
\label{s:wavecov1}

Complex steerable wavelet frames were introduced in \cite{steerableSimoncelli} and are further studied in \cite{unseral_steerable13}, to easily compute
wavelet coefficients of rotated images. 
For simplicity, we begin by
introducing wavelets as a localized functions $\psi(u)$ 
for $u \in \R^2$, with $\int \psi(u)\, du = 0$.
Complex steerable wavelets have a Fourier 
$\what \psi(\om)$ concentrated over one-half of the
Fourier domain. We impose that
$\psi(-u) = \psi^* (u)$ so that $\what \psi(\om)$ is real.
This Fourier transform is centered
at a frequency $\xi \in \R^2$ and is non negligible for 
$\om \in \R^2$ such that 
$|\om - \xi | \leq C' |\xi|$ for some $C' > 0$. 
Figure \ref{waveletfig} gives an example 
of such a wavelet, which is specified in Appendix \ref{appendixcomplexsteer}.

Let $r_{\ell}$ be a rotation by an angle $2 \ell \pi/Q$. 
Multiscale steerable wavelets are derived from $\psi$ with dilations
by $2^j$ for $j \in \Z$, and rotations over
$Q$ angles $\theta = 2 \ell \pi / Q$ for $0 \leq \ell < Q$
\begin{equation}
\label{steerable}
\psi_{\la} (u) =  2^{-j} \psi (2^{-j} r_{  - \ell  } u ) ~~\Rightarrow~~
\what \psi_{\la} (\om) =  2^{j} \what \psi (2^{j} r_{  \ell  }  \om ) ~~\mbox{with}~~\la = 2^{-j} r_{-\ell}\, \xi.
\end{equation}
Since $\what \psi(\om)$ is 
non negligible for $|\om - \xi | \leq C |\xi|$ it results that
$\what \psi_\la(\om)$ is centered at $\la$ and
non negligible for $|\om - \la| \leq  C |\la|$.
In space, $\psi_\la(u)$ is non-negligible for $|u| \leq C' |\la|^{-1}$.
We limit the scale $2^j$ to a maximum $2^{\jmax}$. The lowest frequencies
are captured by a wavelet centered at $\la = 0$.
It is computed by dilating a 
function $\phi(u)$ such that $\int \phi(u)\,du = 1$:
\begin{equation}
\label{steerable0}
\psi_{0} (u) =  2^{-\jmax} \phi (2^{-\jmax} u )  ~~\Rightarrow~~
\what \psi_{0} (\om) =  2^{\jmax} \what \phi (2^{\jmax}    \om ) .
\end{equation}

\begin{figure}
\centering
\begin{subfigure}[t]{4cm}
	\caption{}
	\includegraphics[height=4cm]{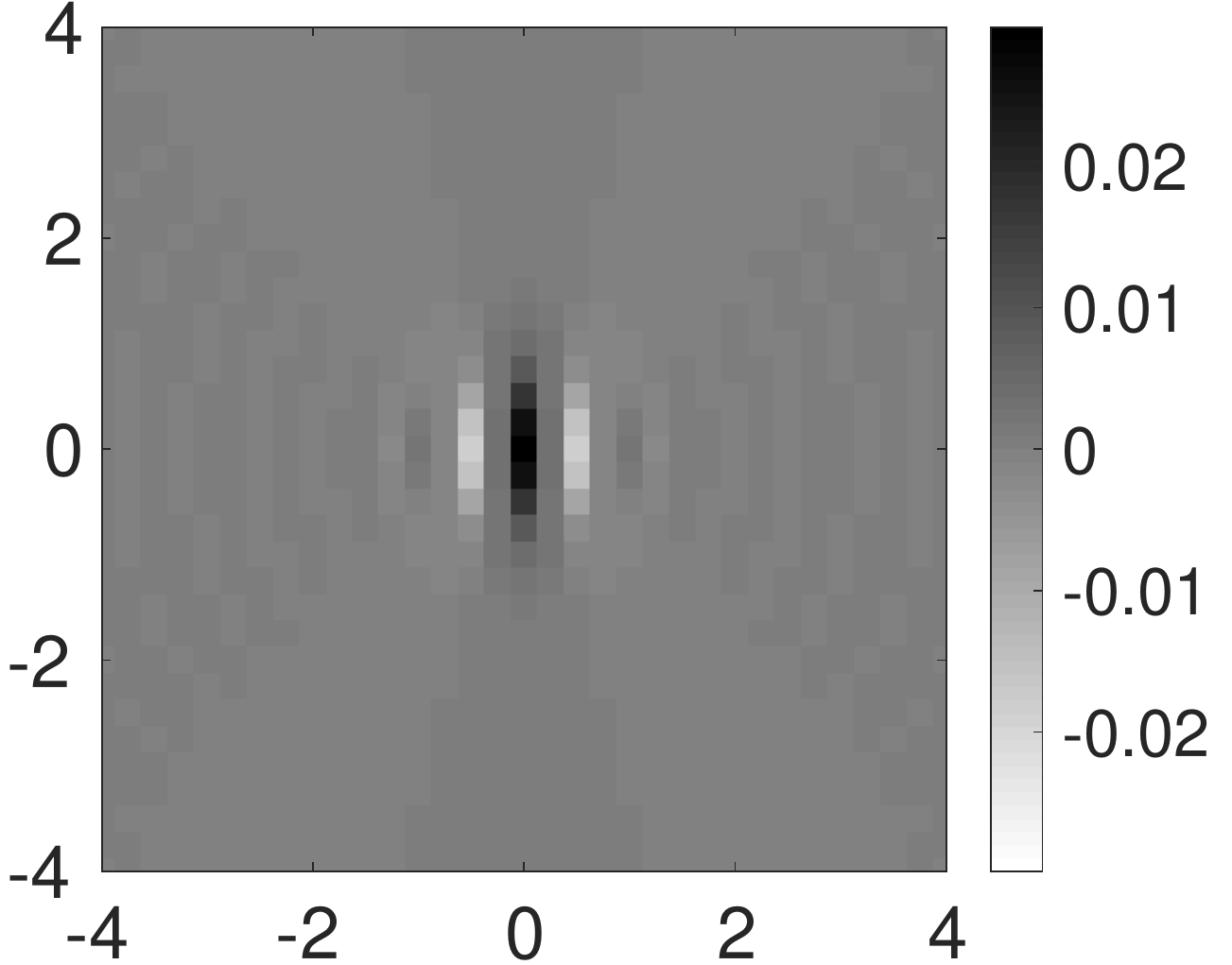}
\end{subfigure}
\quad \quad \quad \quad 
\begin{subfigure}[t]{4cm}
	\caption{}
	\includegraphics[height=4cm]{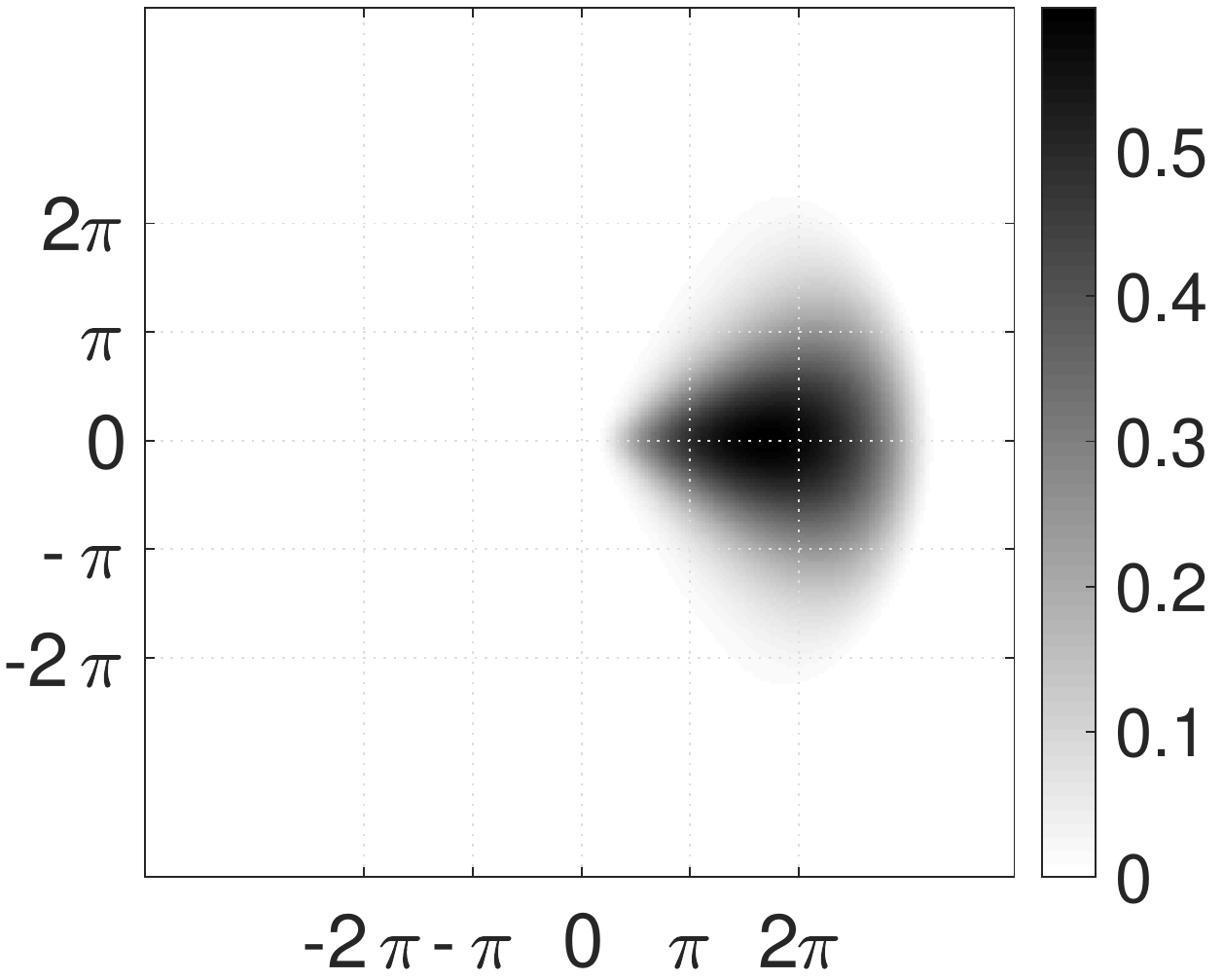}
\end{subfigure}
\caption{(a): Real-part of steerable bump wavelet $\psi(u)$. (b): Fourier transform $\what \psi(\om)$.}
\label{waveletfig} 
\end{figure}

A wavelet frame is constructed by translating
each $\psi_\la$ for $\la \neq 0$ 
by $u = 2^{j-1} n$ and $\psi_0$ by
$u = 2^{\jmax -1} n$ for all $n \in \Z^2$.
It introduces a factor $2$ oversampling relatively to a wavelet
orthonormal basis \cite{mallatbook}, which creates some redundancy. 
The wavelet transform of $x \in \Ltwo$ is defined by
\[
\W x= \{ x \star  \psi_{\la}(u)  \}_{(\la,u) \in \Gaw}~~
\]
where $\Gaw$ is a frequency-space index set with
$(\la,u) = (2^{-j} r_{-\ell} \xi, 2^{j-1} n)$ for
$1 \leq j \leq \jmax$, $0 \leq \ell < Q$, $n \in \Z$, 
or $(\la,u) = (0, 2^{\jmax -1}n )$. 

Under appropriate conditions on 
$\psi$, the wavelet
family $\{ \psi_{\la}(\cdot -u) \}_{(\la,u) \in \Gaw}$ is a
frame of $\Ltwo$ \cite{unseral_steerable13}.
This means that there exists $0 < A_{\W} \leq B_{\W} $ such that 
for any $x \in \Ltwo$ 
\begin{equation}
\label{frames}
A_{\W}\, 
\|x\|^2 \leq \| \W x \|^2 \leq 
B_{\W}\,  \|x\|^2
\end{equation}
with $\| \W x \|^2 = \sum_{(\la,u) \in \Gaw} |x \star \psi_{{\la}}(u) |^2$.

The wavelet transform can be redefined over discrete
images $x$ of $d$ pixels supported
in a two-dimensional square grid $\Lambda_d$,
uniformly sampled at intervals $1$ in  $[1,d^{1/2}]^2$.
It requires to discretize and modify ``boundary wavelets'' whose supports
intersect image boundaries. This can be done
over steerable wavelets 
\cite{steerableSimoncelli,unseral_steerable13}, 
while preserving the frame constants $A_{\W}$ and $B_{\W}$. 
The resulting wavelet $\psi_{\la}(\cdot - u)$ are supported in $\Lambda_d$.
They are still indexed by $(\la,u) \in \Gaw$. If
$\la = r_{-\ell} \xi \neq 0$ for
$1 \leq j \leq \jmax$, $0 \leq \ell < Q$ then
$\sum_{u \in \Lambda_d} \psi_{\la}(u) = 0$. 
If $\la = 0$ then $\sum_{u \in \Lambda_d} \psi_{0}(u) = 2^{\jmax}$.
Since $\jmax \leq (\log_2 d)/2$, there are at most
$ Q (\log_2 d)/2  + 1$ different frequency channels $\la$. 
For $\la = r_{-\ell} \xi \neq 0$, $\psi_{\la}$
is translated by $u = 2^{j-1} n \in \Lambda_d$ which yields
$2^{-j+1} d$ wavelet coefficients. The total number of wavelets
coefficients is about $4 Q d / 3$  if $\jmax = (\log_2 d)/2$.

\begin{figure}	
\centering
	\begin{subfigure}[t]{4.5cm}
		%\caption{}
		\includegraphics[height=4.5cm]{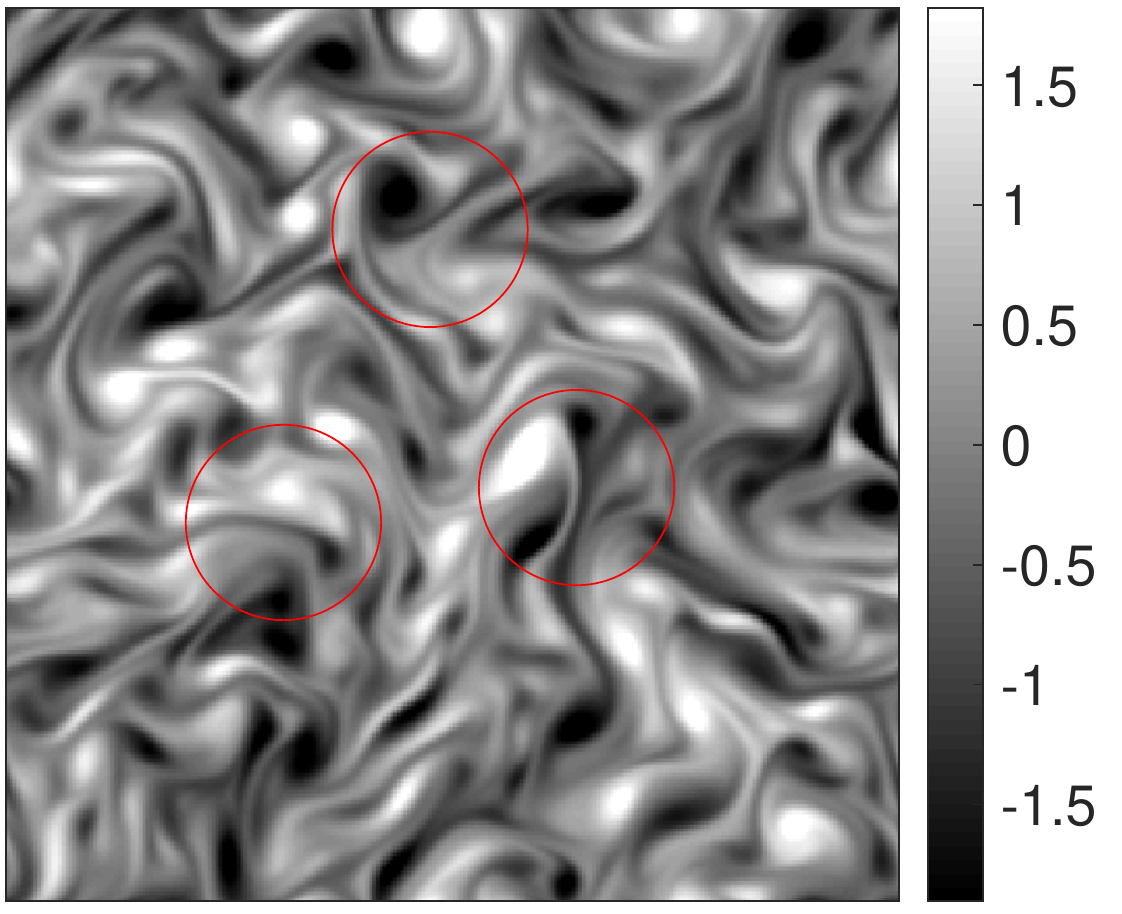}
	\end{subfigure} \\
	\begin{subfigure}[t]{4.5cm}
		\caption{}
		\includegraphics[height=4.5cm]{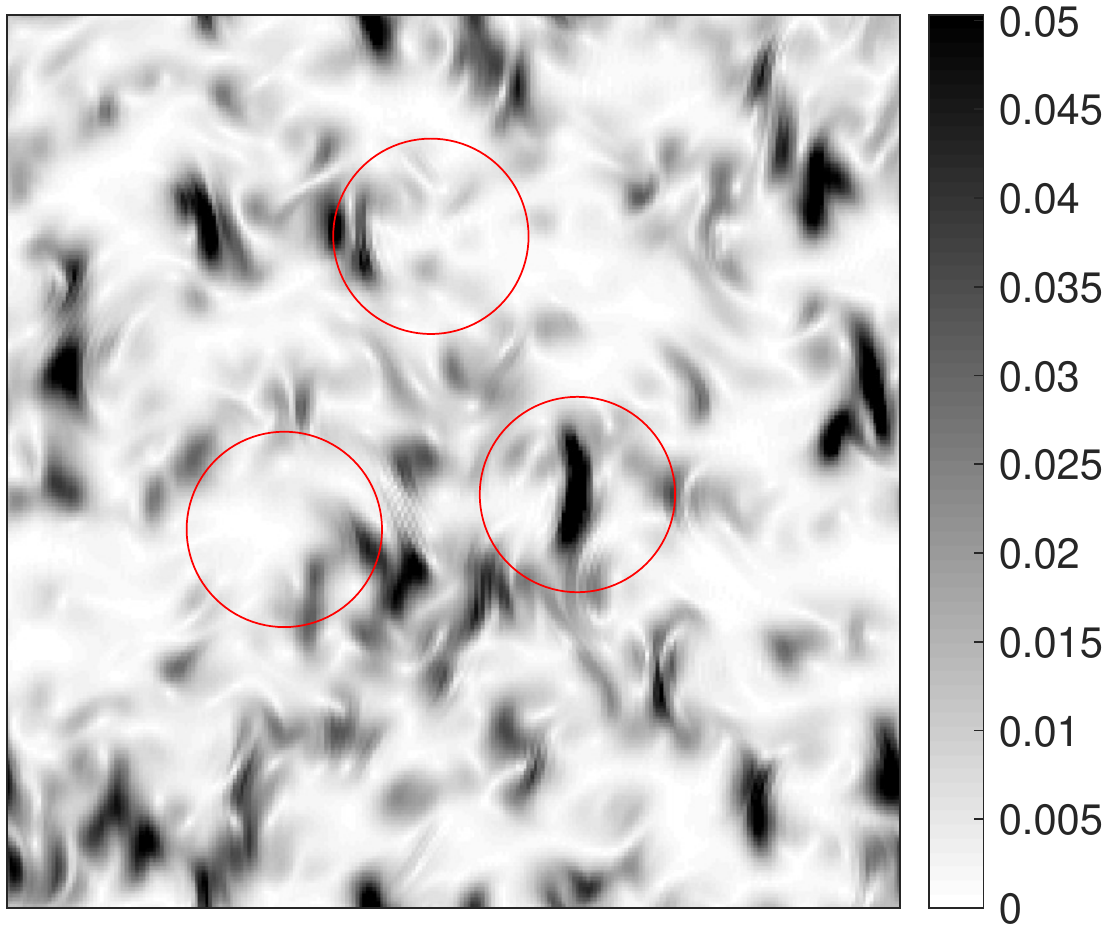}
		\includegraphics[height=4.5cm]{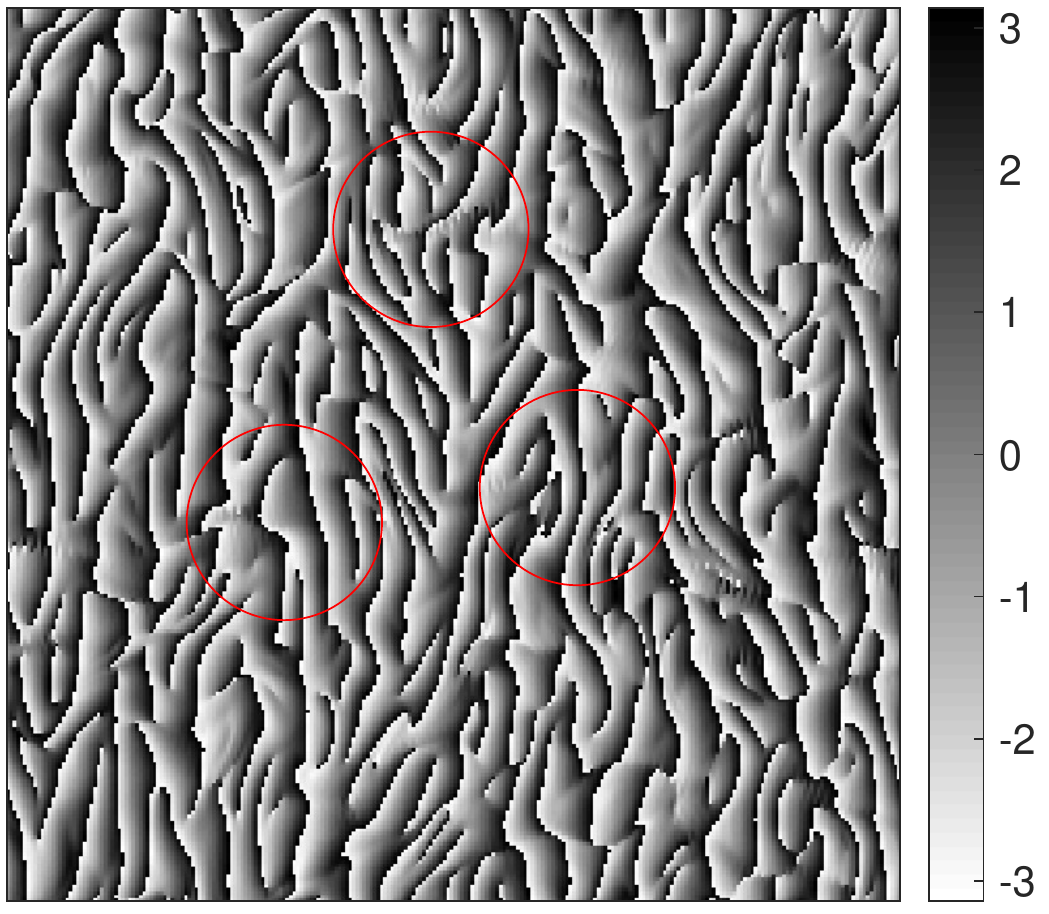}
	\end{subfigure} $\quad \quad$
	\begin{subfigure}[t]{2.25cm}
		\caption{}
		\includegraphics[height=2.25cm]{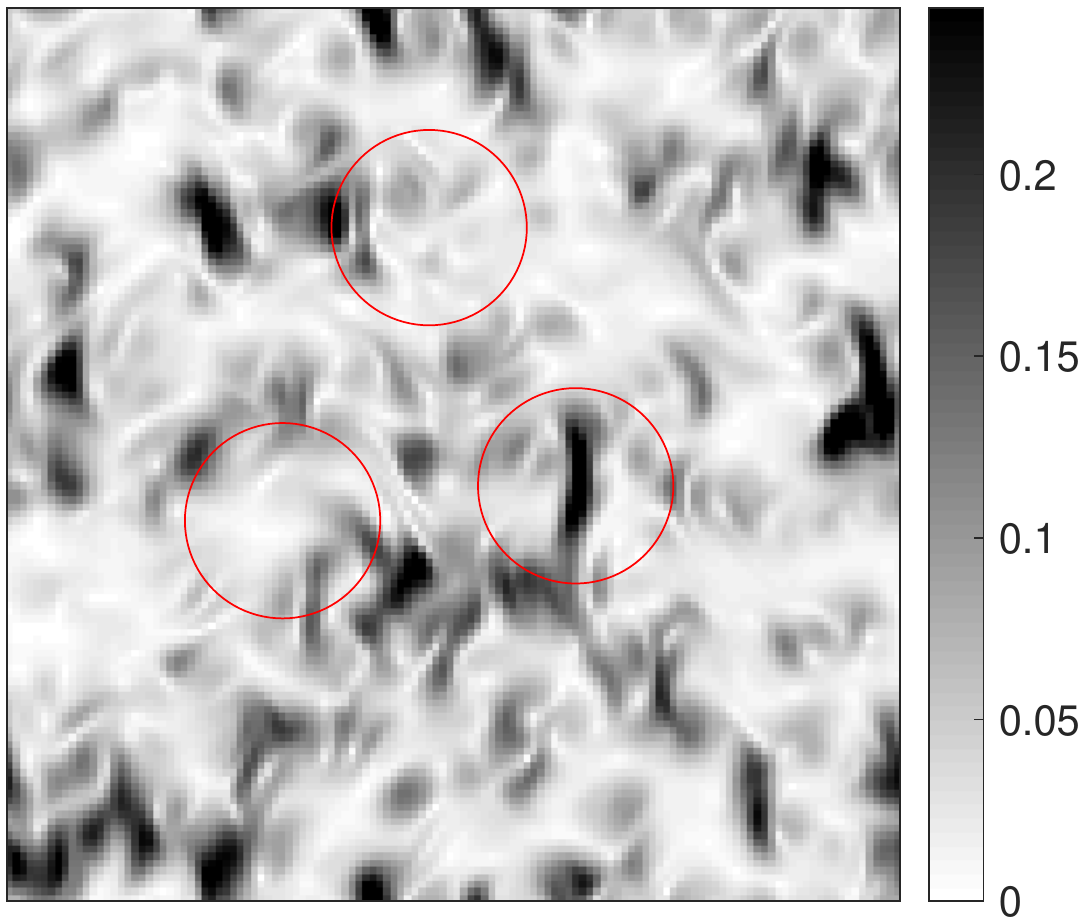} 
		\includegraphics[height=2.25cm,trim={0.2cm 0.2cm 0.2cm 0.2cm},clip]{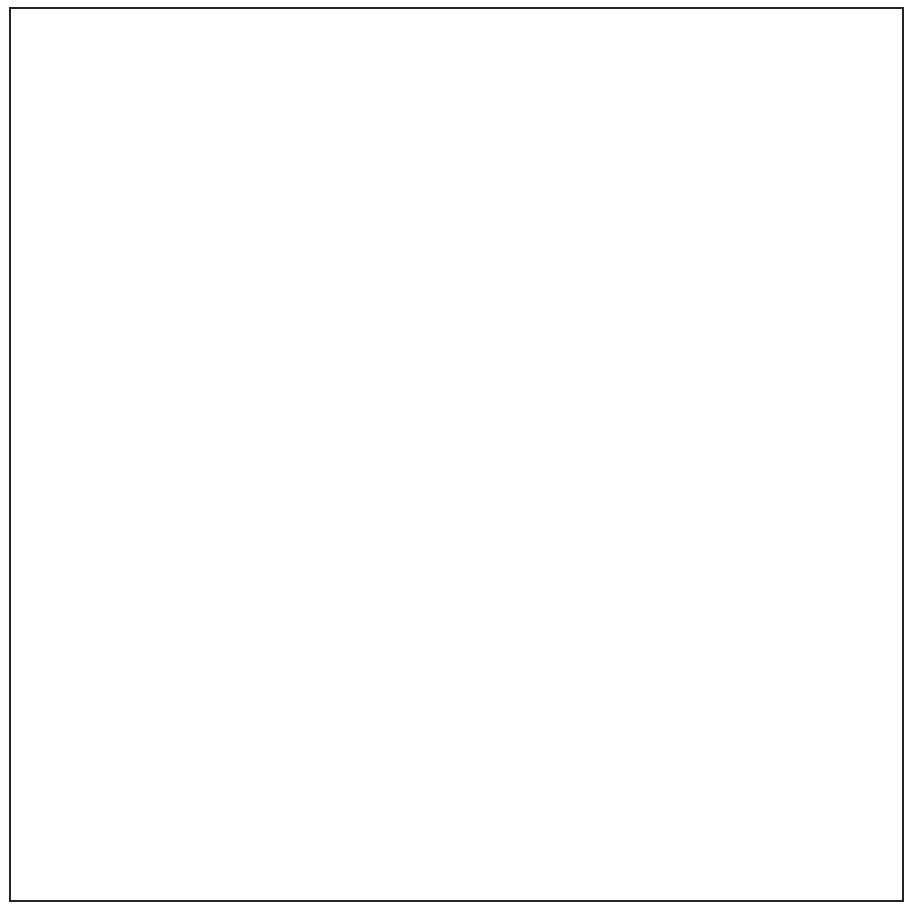}
		\includegraphics[height=2.25cm]{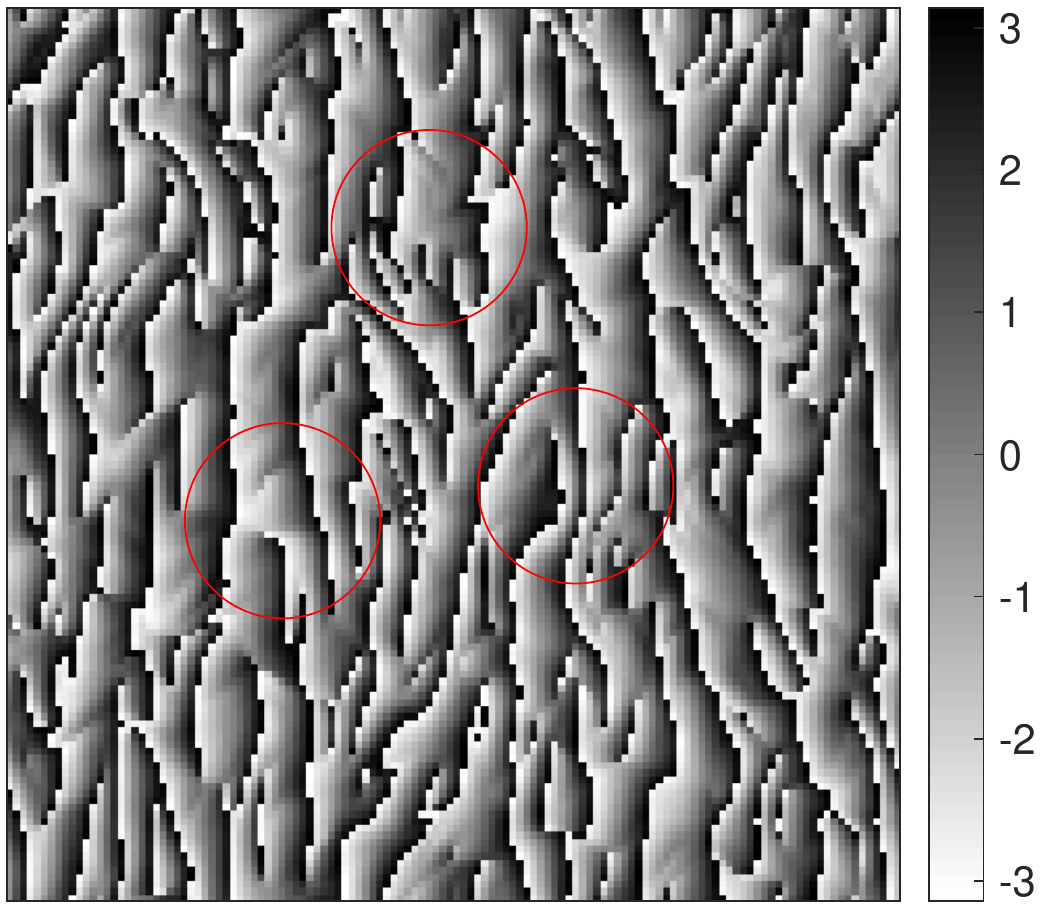}
	\end{subfigure} $\quad $
	\begin{subfigure}[t]{4.5cm}
		\caption{}
		\includegraphics[height=4.5cm]{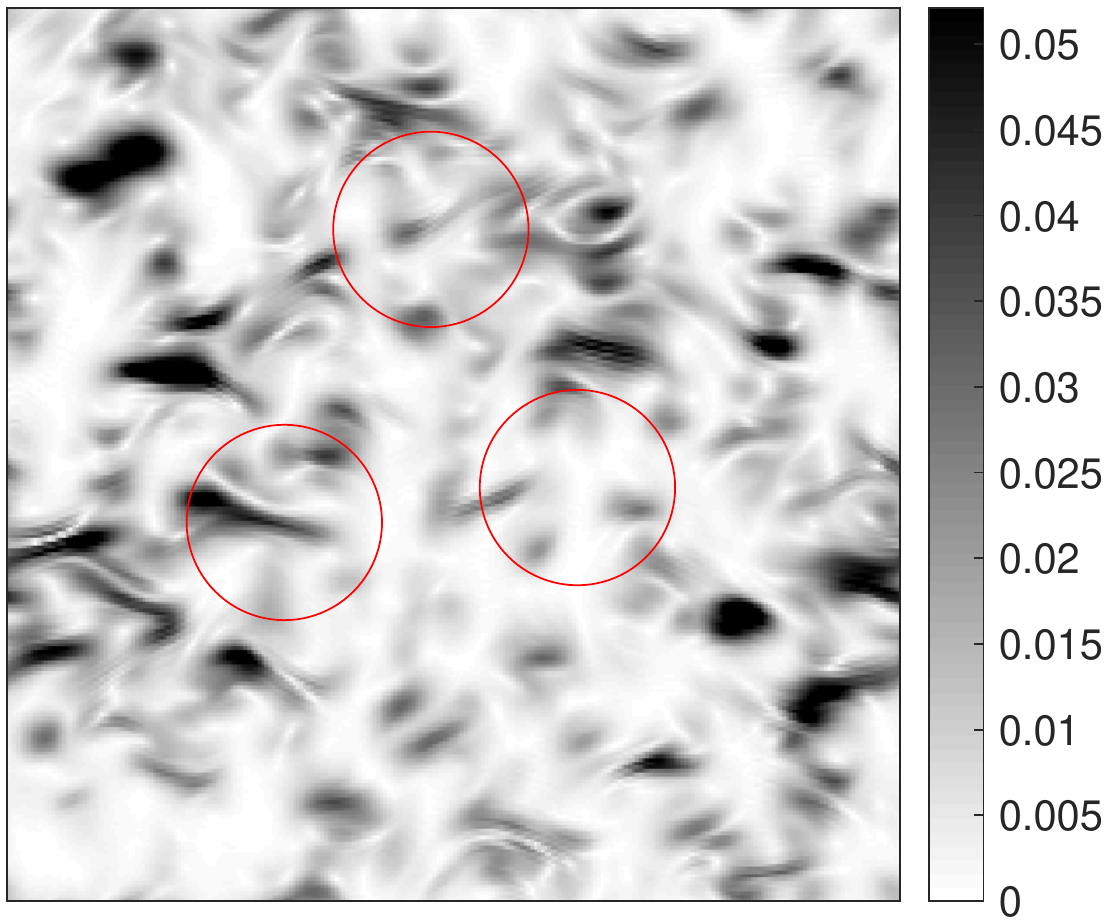}
		\includegraphics[height=4.5cm]{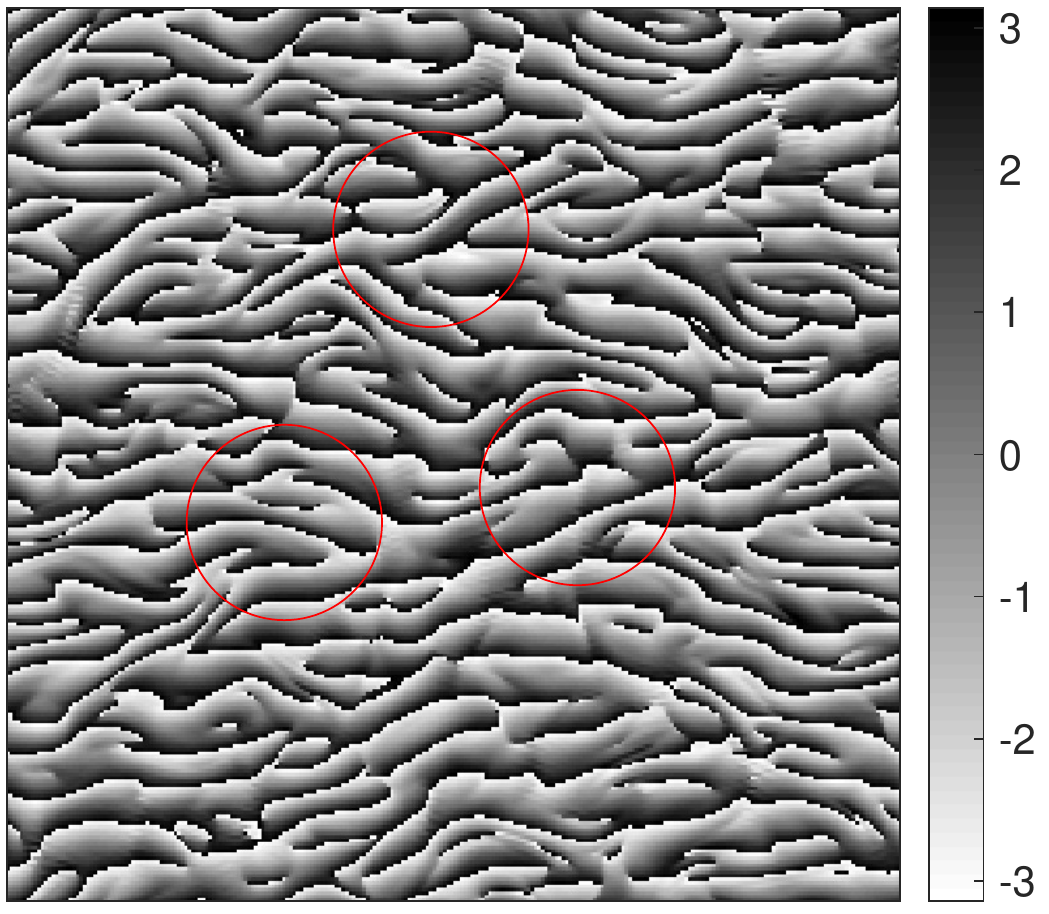}
	\end{subfigure} $\quad \quad$
	\begin{subfigure}[t]{2.25cm}
		\caption{}
		\includegraphics[height=2.25cm]{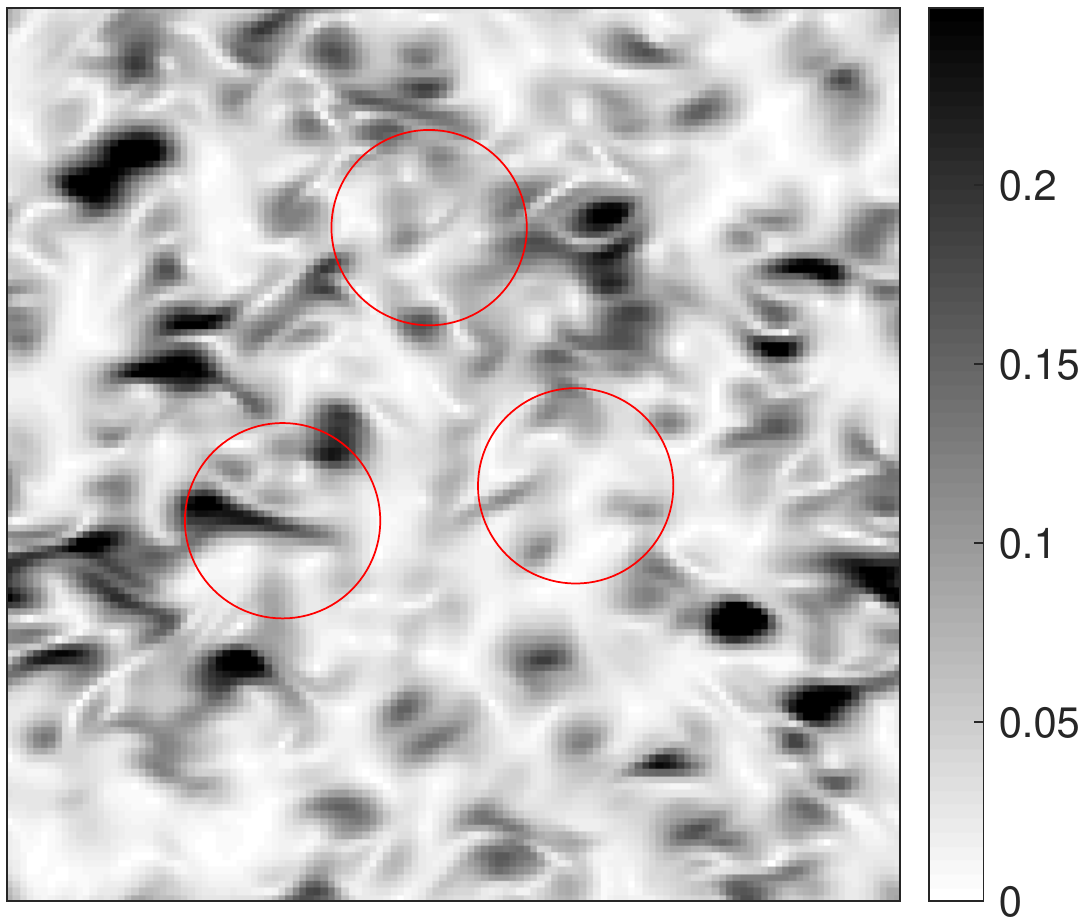}
		\includegraphics[height=2.25cm,trim={0.2cm 0.2cm 0.2cm 0.2cm},clip]{./figs/empty.pdf}
		\includegraphics[height=2.25cm]{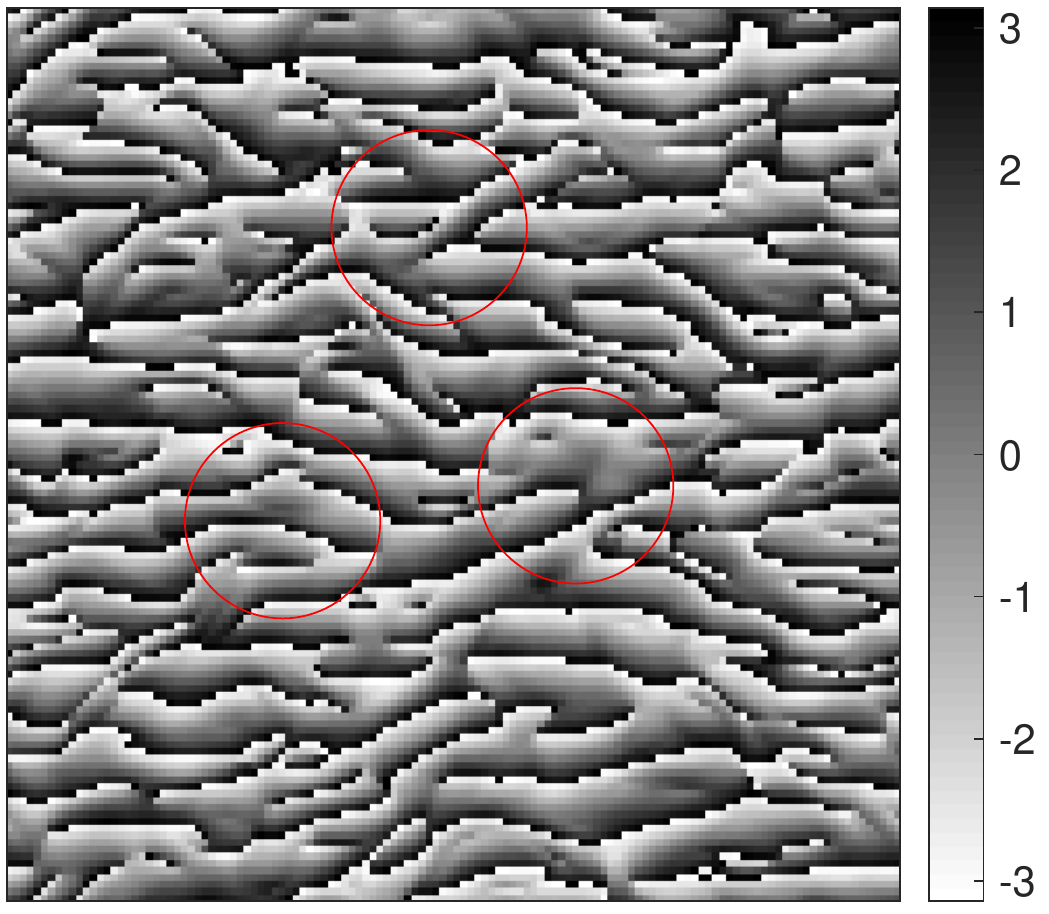}
	\end{subfigure} 
	\caption{Top: turbulent vorticity field.
Bottom: Each image gives the modulus (above) or the phase (below)
of wavelet coefficients $x \star \psi_\la (u)$ for different frequency
channels $\la = 2^{-j} r_{-\ell} \xi$.
Large modulus coefficients are shown in black. The columns correspond to
different scales and angles $(j,\ell)$.
(a): $(j,\ell)=(1,0)$. (b): $(j,\ell)=(2,0)$. (c): $(j,\ell)=(1,Q/4)$. (d): $(j,\ell)=(2,Q/4)$.}
	\label{fsafsag}
\end{figure}

The mother wavelet $\psi$ is chosen in order
to obtain a sparse wavelet representation of realizations
of $X$, with few large amplitude
wavelet coefficients. This sparsity highlights non-Gaussian properties.
Figure \ref{fsafsag} displays the modulus and phase of wavelet coefficients of the vorticity field of a turbulent flow.  
This flow is
obtained by running the 2D Navier Stokes equation with
periodic boundary conditions, initialized with a
random Gaussian field \cite{schneider2006coherent}. After a fixed time, 
it defines a stationary but non-Gaussian random process.
For each scale and orientation, large amplitude modulus 
coefficients are located at positions where the image has sharp transitions,
and the phase depends upon the position of these sharp
transitions. 

\paragraph{Wavelet covariance}
A wavelet transform defines a linear representation
$\UPhi = \W$ indexed by $v = (\la,u)$. 
Similarly to Fourier coefficients, 
we show that wavelet coefficients have a covariance which nearly
vanish at different frequencies.

Similarly to Fourier coefficients, wavelet coefficients have a zero mean
at non-zero frequencies. 
If $\la \neq 0$ then
$\E(X \star \psi_\la (u)) = 0$
because $\sum_u \psi_{\la} (u) = 0$. 
If $\la = 0$ then
$\sum_u \psi_{0}(u) = 2^{J}$ so  $\E(X \star \psi_0 (u)) =  
2^{J} \E(X(u))$.
The covariance at $v = (\la,u)$ and 
$v'=(\la',u')$ is
\[
K_\W (v,v') = 
\Cov (X \star \psi_\la (u)\,,\,X \star \psi_{\la'} (u'))~.
\]
It depends on $u-u'$ because $X$ is stationary.
Let $\what \psi_{\la} (\om)$ be the discrete Fourier transform
of $\psi_{\la}(u)$ defined in (\ref{dionsdfsdf}).
Since wavelet coefficients are convolutions,  
covariance values can be rewritten from the power spectrum
$\what K(\om) = \frac{1}{d} \Cov(\what X(\om),\what X(\om))$ of $X$
\begin{equation}
\label{Fourier}
K_\W  (v,v') = 
\sum_{\om \in \Lambda_d} \what K (\om)\, \what \psi_{{\la'}}  (\om)\, 
\what \psi^*_{{\la} } (\om) \,e^{i (u-u'). \om}. 
\end{equation}
It results that
$K_\W (v,v') = 0$
if $\what \psi_{\la} (\om)\, \what \psi_{\la'} (\om) = 0$ for all $\om$.
Since $\what \psi_{\la}(\om)$ is non-negligible only if
$|\om - \la| \leq C |\la|$, the covariance
$K_\W  (v,v')$ is non-negligible for $\la \neq \la'$ only if 
\begin{equation}
\label{spansdfoius0}
\frac{|\la - \la'|} {|\la| + |\la'|} \leq C\,.
\end{equation}
It shows that similarly to Fourier coefficients,
wavelet covariances are negligible across frequencies 
which are sufficiently far apart. 

\begin{figure}
\centering
	\begin{subfigure}[t]{4.5cm}
		\caption{}
		\includegraphics[height=4.5cm]{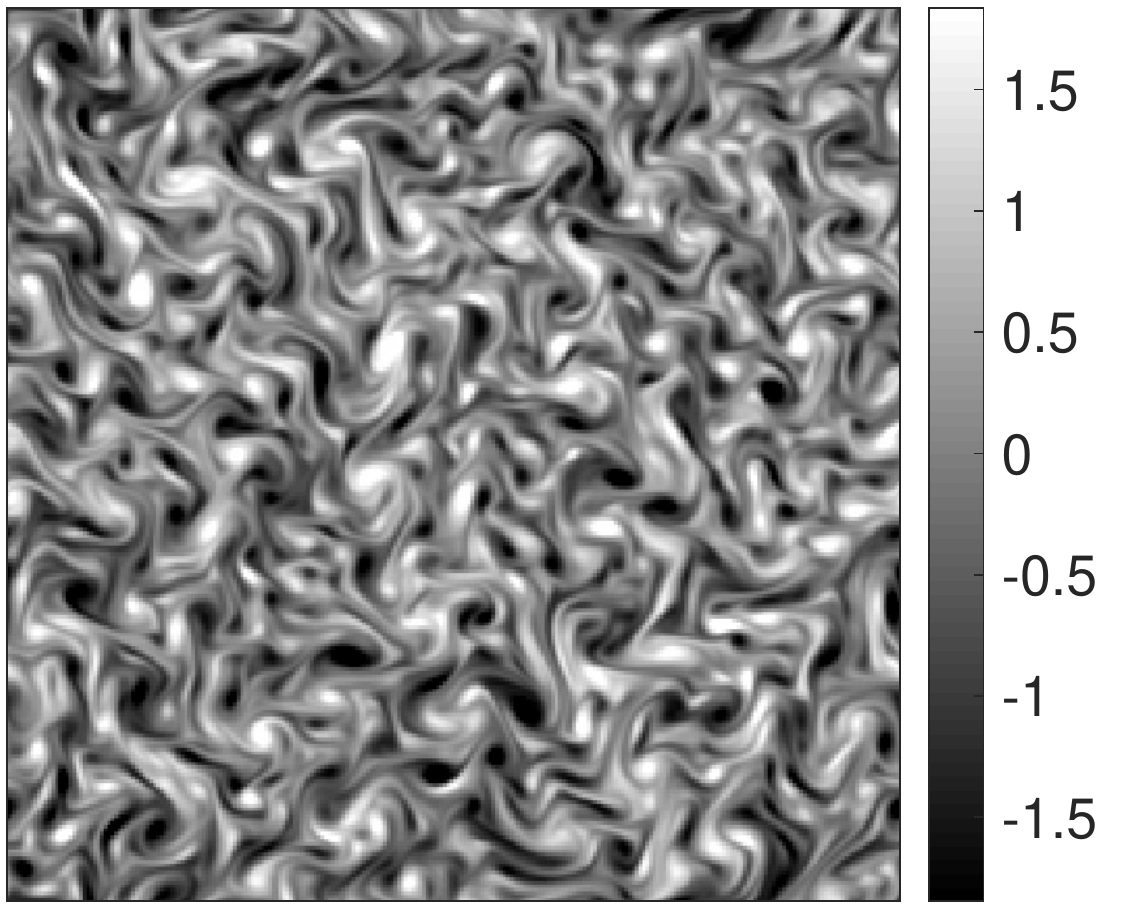}
	\end{subfigure} $\quad \quad$
	\begin{subfigure}[t]{4.5cm}
		\caption{}
		\includegraphics[height=4.5cm]{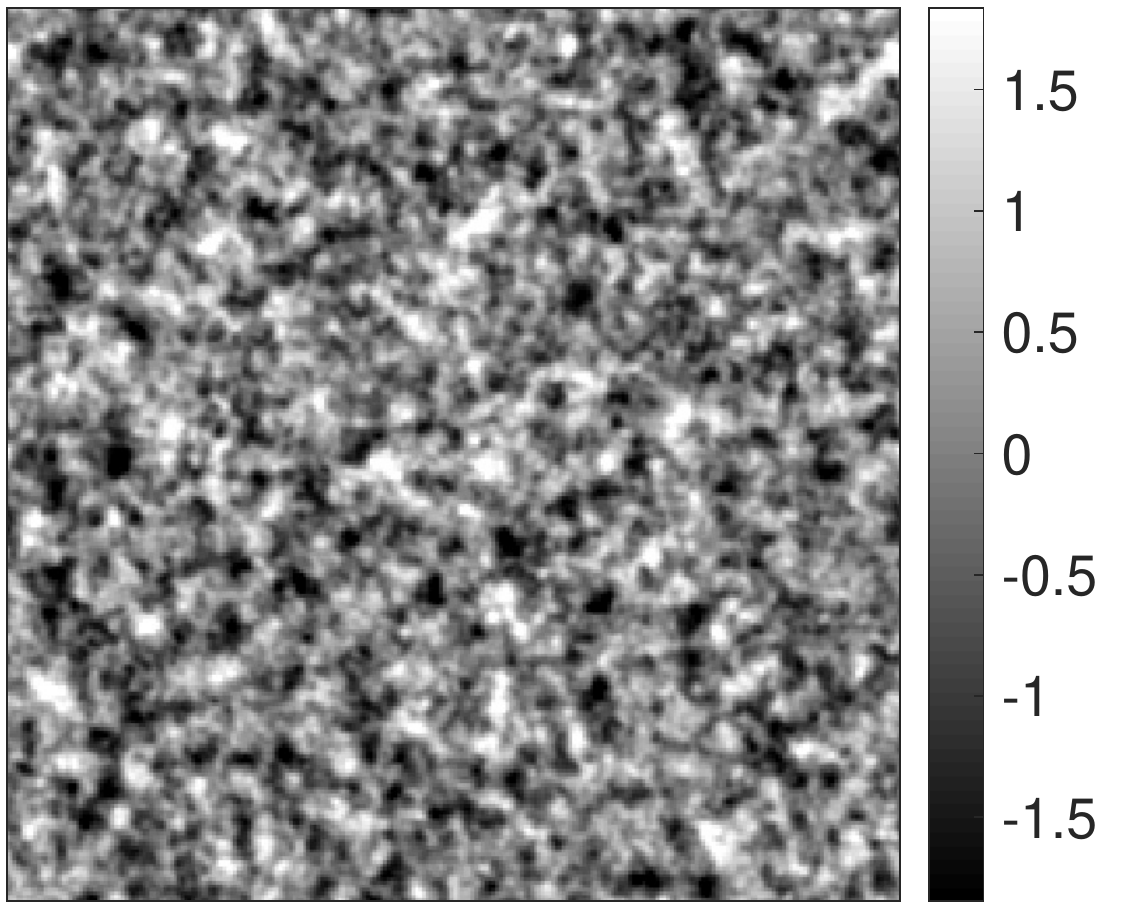}
	\end{subfigure} $\quad \quad $
	\begin{subfigure}[t]{4.5cm}
		\caption{}
		\includegraphics[height=4.5cm]{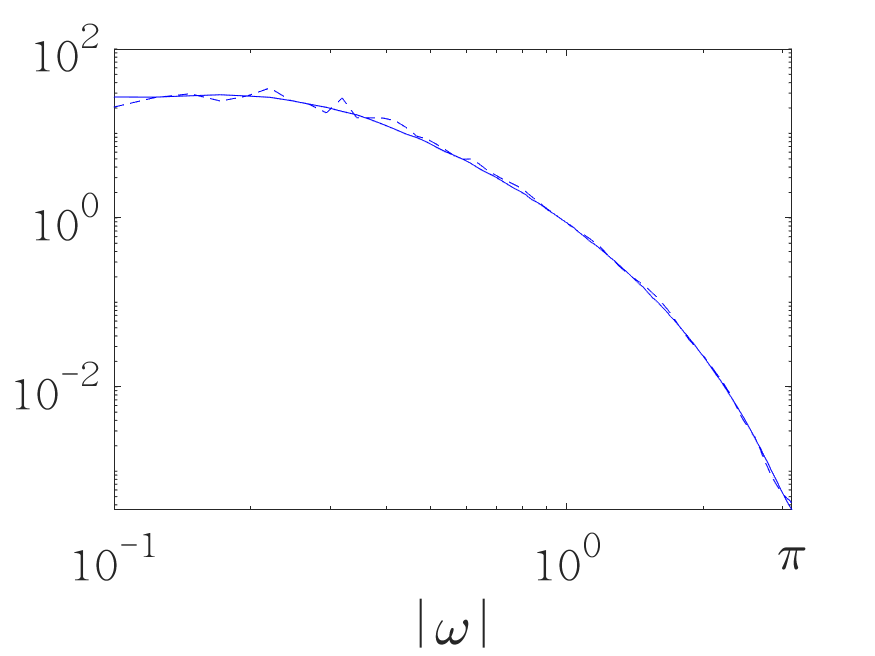}
	\end{subfigure}	
	\quad \quad
\caption{(a): Realization of a stationary
turbulent vorticity field $X$. 
(b): Realization of a Gaussian maximum entropy model $\widetilde X$
calculated from wavelet covariances estimated on a foveal graph.
(c): The full line and dashed lines are the power spectrum of $X$ and the power spectrum of $\widetilde X$ respectively, as a function of the radial frequency $|\om|$ (shown in log-log scale).
}
\label{Gausssynthesisfig} 
\end{figure}

\paragraph{Maximum entropy wavelet graph model}
A maximum entropy model conditioned by wavelet covariances
is Gaussian because the wavelet transform is linear.
Figure \ref{Gausssynthesisfig}(a) gives a realization
of a stationary turbulent flow $X$. 
A low-dimensional maximum entropy model is defined on a graph of
covariance coefficients $K_\W(v,v')$ 
for $v'=(\la',u')$ in a small neighborhood of $v = (\la,u)$.
Since wavelet coefficients are nearly decorrelated across frequencies,
at each scale $2^j$, the neighborhood of $v = (\la,u)$ is defined as
the set of $v'=(u',\la')$ such that $\la' = \la$ and
$|u-u'| \leq 2^{j-1} \Delta$ for a fixed $\Delta$. Covariances are thus
specified over a spatial range proportional to the scale. 
This is a foveal neighborhood which is sufficient
to approximate the
covariances of large classes of random processes such as
fractional Brownian motions \cite{Mathieuthesis}.
Since $(u,u') = 2^{j-1} (n,n')$, the neighborhoods of all $v$ have
the same size, which is  smaller than
$(2 \Delta +1)^2$.

Since $X$ is stationary, its probability distribution
is invariant to the group $G$ of translations.
The number of covariance coefficients that must be estimated
is equal to the number $|E_G|$ of edges in the graph
modulo translations. It is equal to the number of wavelet
frequencies $\la$ multiplied by the size of each neighborhood, and 
thus bounded by $(\jmax Q + 1) (2 \Delta + 1)^2 = O(\log_2 d)$.
This model size is much smaller than the image size $d$ when $d$ is large.

Figure \ref{Gausssynthesisfig}(b) shows a realization of
the maximum entropy Gaussian model $\widetilde X$.
It is conditioned by wavelet
covariances on a foveal graph with $\Delta = 2$. 
The wavelet transform is computed with a bump wavelet specified 
in Appendix \ref{appendixcomplexsteer},
with $ J=5$, $Q=16$ and $d = 256^2$.
In this case $|E_G| / d = 3.6 \, 10^{-2}$. 
The covariances
are estimated from a single realization of $X$.
The calculation of the Lagrange multipliers in (\ref{macrosndsf}) 
is explained in Appendix \ref{appendixmaxent}. To measure the accuracy
of this model we compare the power spectrum of $X$ and the Gaussian
model $\wtilde X$. Both processes are isotropic so 
Figure \ref{Gausssynthesisfig}(c) gives the radial
log power spectrum of $X$ and $\wtilde X$ as a function of $|\om|$.
These power spectrum
are nearly the same, which means that the
wavelet covariance graph gives an accurate estimation of
the second order moments of $X$ from a single realization. 

The realization of the Gaussian model in
Figure \ref{Gausssynthesisfig}(b) has a geometry which is
very different from the turbulence flow $X$ in
Figure \ref{Gausssynthesisfig}(a). It shows that $X$
is highly non-Gaussian, which also appears in Figure \ref{fsafsag}.
The wavelet coefficients of a stationary process $X$ are not correlated
at different scales and angles, which would imply that they are
independent if $X$ was Gaussian. On the contrary,
Figure \ref{fsafsag} shows that the 
the modulus and phases of wavelet coefficients of $X$
are strongly dependent across scales and angles. 
High amplitude modulus coefficients are located in the same spatial 
neighborhoods because they are produced by the same 
sharp transitions of the flow. 
Next section explains how to capture this dependence with phase
harmonics.

\subsection{Wavelet Phase Harmonic Covariance}  
\label{phaharmcov}

A wavelet phase harmonic representation $\Phi = \what \HH \, L$ is
computed with a linear wavelet transform $L = \W$. 
As in the Fourier case, the covariance of $\Phi(X) = \what \HH(\W X)$
has non-zero covariance
coefficients across different frequency bands. We study
the properties of the resulting covariance matrix, and the role of
sparsity.

\paragraph{Wavelet phase harmonics}
To specify the dependence across frequencies,
we apply a phase harmonic operator to wavelet coefficients:
\[
\what \HH (\W x) = 
\Big\{ \what{h}(k)\, [x \star \psi_{{\la}}(u)]^{k} \Big\}_{(\la,u) \in \Gaw,k \in \Z}~.
\]
The coefficients of 
$\UPhi = \what \HH \W$ are indexed by $v = (\la,k,u)$. 
The covariance coefficients of $\what \HH (\W X)$ are
\[
K_{{\what \HH}\W} (v,v') = \what h(k) \, \what h(k')^* \,
\Cov ([X \star \psi_\la (u)]^{k}\,,\,[X \star \psi_{\la'}(u')]^{k'}) .
\]
Since $X$ is stationary, it only depends on $u-u'$.
Wavelet harmonic covariances only need to 
be calculated for $k \geq 0$ and $k' \geq 0$. Indeed,
since $X$ is real and $\psi(-u) = \psi^* (u)$ one can verify that
$K_{\what \HH \W}(v,v')$ does not change its value
if $(\la,k)$ becomes $(-\la,-k)$ or if $(k,k')$ becomes $(-k,-k')$.
Such wavelet harmonic covariances have first been computed by
Portilla and Simoncelli \cite{portilla2000parametric}
to characterize the statistics of image textures. Their representation
correspond to $(k,k')$ equal to $(0,0)$, $(1,1)$, $(1,2)$, which
amounts to choosing $\hat h(k) = 1_{[0,2]}(k)$.

Since $\what \HH$ and $\W$ are bi-Lipschitz the operator
$\what \HH \W$ is also bi-Lipschitz, 
with lower and upper bounds $A_{\HH}\ A_\W$ and $B_{\HH}\, B_\W$.
Proposition \ref{bioLihscionfth} implies that
\begin{equation}
\label{traninequali2}
A_{\HH}\, A_\W\, \sigma^2(X) \leq \sigma^2(\what \HH( \W X))\leq  B_{\HH} \,B_\W\,
\sigma^2(X),
\end{equation}
which controls the variance of wavelet harmonic coefficients.

\paragraph{Rectified neural network coefficients}
Ustyuzhaninov et. al. in \cite{Ustyuzhaninov2017a} have
shown that one can get good texture synthesis
from the covariance of a one-layer convolutional neural network,
computed with a rectifier.
In the following we show that these
statistics are equivalent to 
phase harmonic covariances, computed with a rectifier
phase window $h(\alpha)$. 

Section \ref{waveharmosec} proves that
$\what \HH = \F_\alpha \HH$, where $\HH$ computes
a phase windowing of wavelet coefficients
\[
\HH (\W x) = 
\Big\{ |x \star \psi_\la (u)|\,
h(\varphi(x \star \psi_\la (u)) + \alpha) \}_{(\la,u) \in \Gaw,\alpha \in [0,2\pi]}.
\]
The covariance of $\what \HH (\W X)$ and 
$\HH (\W X)$ thus satisfy
$K_{{\what \HH}\W} = \F_\alpha \, K_{{ \HH}\W}  \, \F_\alpha^{-1}$.
The following proposition proves that $K_{\HH \W}$ gives the
covariance of rectified wavelet coefficients if $h$ is a
rectifier phase window.

\begin{proposition}
\label{ReluRepres}
Let $v = (\la,\alpha,u)$ and $v' = (\la',\alpha',u')$.
For a rectifier phase window $h(\alpha) = \rho(\cos \alpha)$
\begin{equation}
\label{confrect}
K_{{\HH}\W} (v,v') = 
\Cov \Big(\rho( X \star \psi_{\la,\alpha}(u)) \,,\,\rho( X \star 
\psi_{\la',\alpha'}(u') )\Big) 
\end{equation}
with  $\psi_{\la,\alpha} (u) = {\rm \Real} (e^{-i \alpha} \psi_{\la} (u))$.
\end{proposition}

{\it Proof:}
We proved in (\ref{phase-shift-filt}) that if 
$h(\alpha) = \rho(\cos \alpha)$ then
\[
\HH (z) = \{ \rho( {\rm Real}( e^{i \alpha} z )) \}_{\alpha \in [0,2\pi]} . 
\]
It results that
\[
K_{{\HH}\W} (v,v') = \Cov(
\rho( {\rm Real}(e^{i \alpha}  X \star \psi_{\la}(u) )) , 
\rho( {\rm Real}(e^{i \alpha'}  X \star \psi_{\la'}(u'))) \rb ,
\]
which proves (\ref{confrect}). $\Box$

Rectified wavelet coefficients
$\rho( X \star \psi_{\la,\alpha}(u))$ can be interpreted as
one-layer convolutional network coefficients, computed with 
wavelet filters $\psi_{\la,\alpha}$ of different frequencies $\la$
and different phases $\alpha$. This result applies to any linear
operator $L$, not necessarily a wavelet transform.
The statistics used by
Ustyuzhaninov et. al. in \cite{Ustyuzhaninov2017a} 
use local cosine or random filters
in their network as opposed to steerable wavelets. It corresponds
to convolutional operators $L$ that are not wavelet transforms.

\paragraph{Sparse harmonic covariance} 
Proposition \ref{SparseSpecMat} specifies the properties of 
a Fourier phase harmonic covariance. 
We qualitatively explain, without proof, why 
$\Cov([X \star \psi_\la (u) ]^k,[X \star \psi_{\la'} (u')]^{k'})$
has similar sparsity properties.

For a fixed $\la$ and $k$,  $[X \star \psi_\la (u) ]^k$ is
a stationary random vector in $u$. 
The covariance 
of $[ X \star \psi_{\la}(u)]^k$ and 
$[X \star  \psi_{\la'}(u')]^{k'}$ is non-zero only if their
power-spectrum have a support which overlap. We give a necessary
condition by approximating these spectrum supports. 

\begin{figure}
	\centering
	\begin{subfigure}[t]{4cm}
		\caption{}
		\includegraphics[height=4cm]{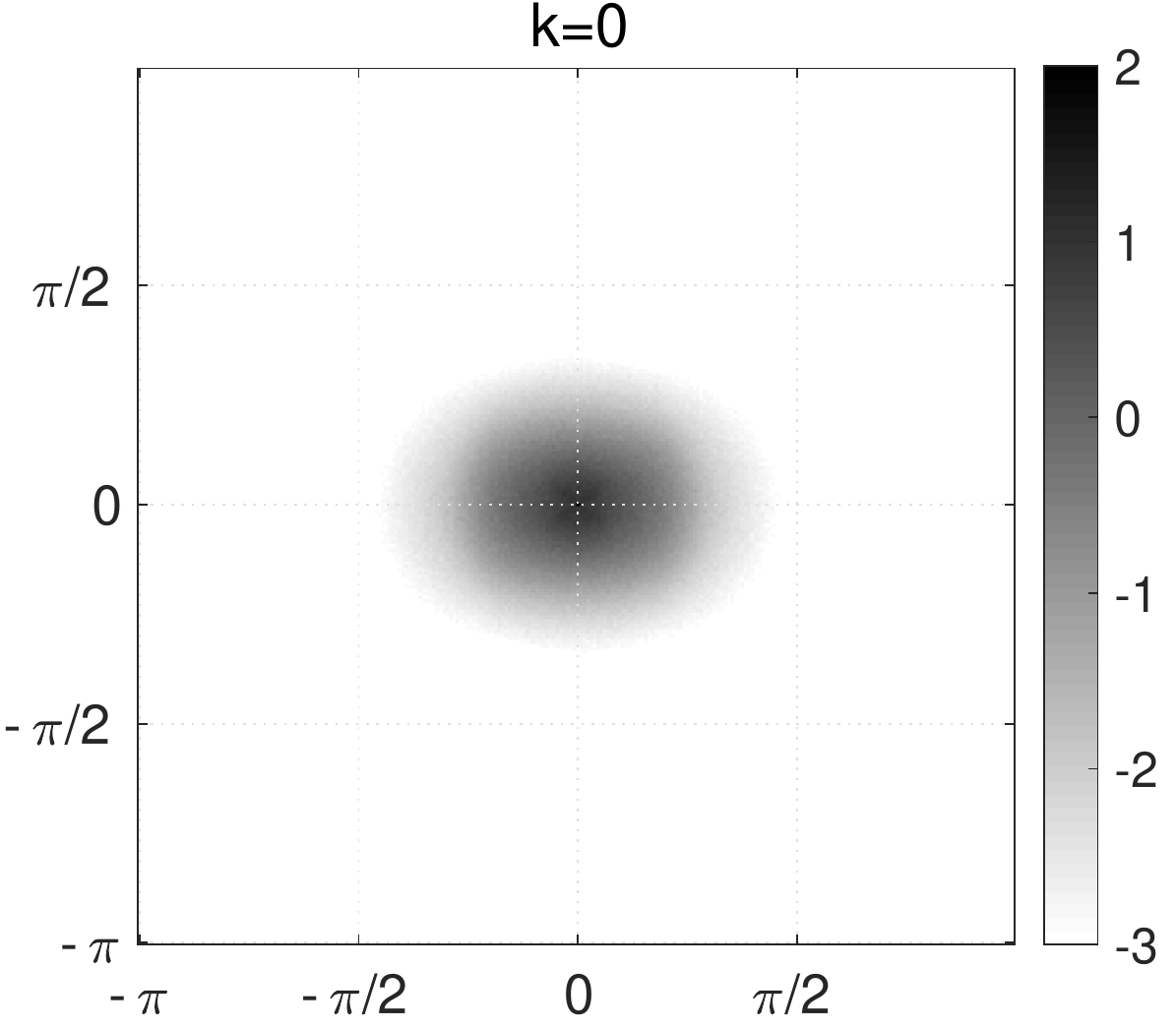}
	\end{subfigure}
	\quad \quad
	\begin{subfigure}[t]{4cm}
		\caption{}
		\includegraphics[height=4cm]{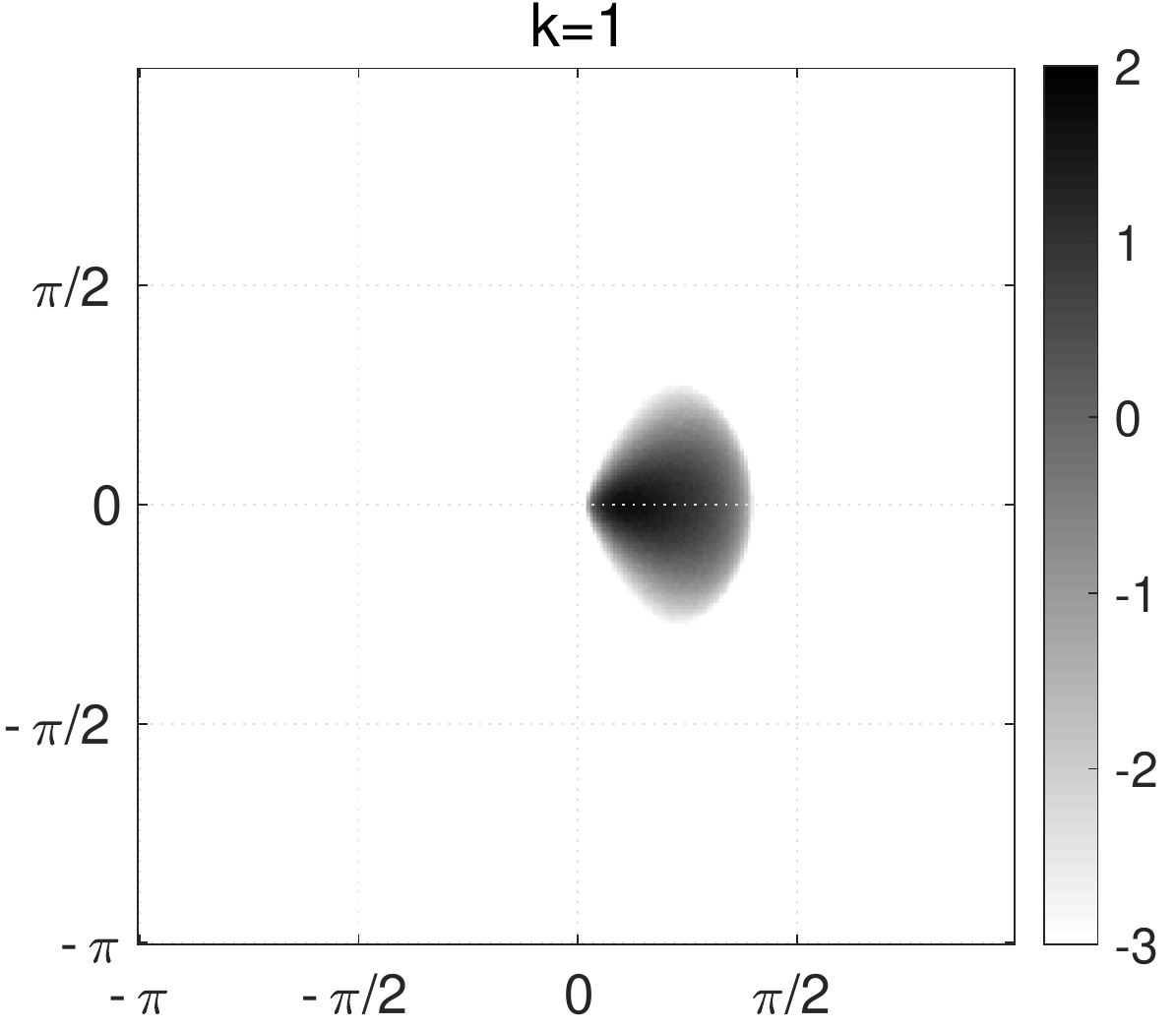}
	\end{subfigure}
	\quad 	\quad
	\begin{subfigure}[t]{4cm}
		\caption{}
		\includegraphics[height=4cm]{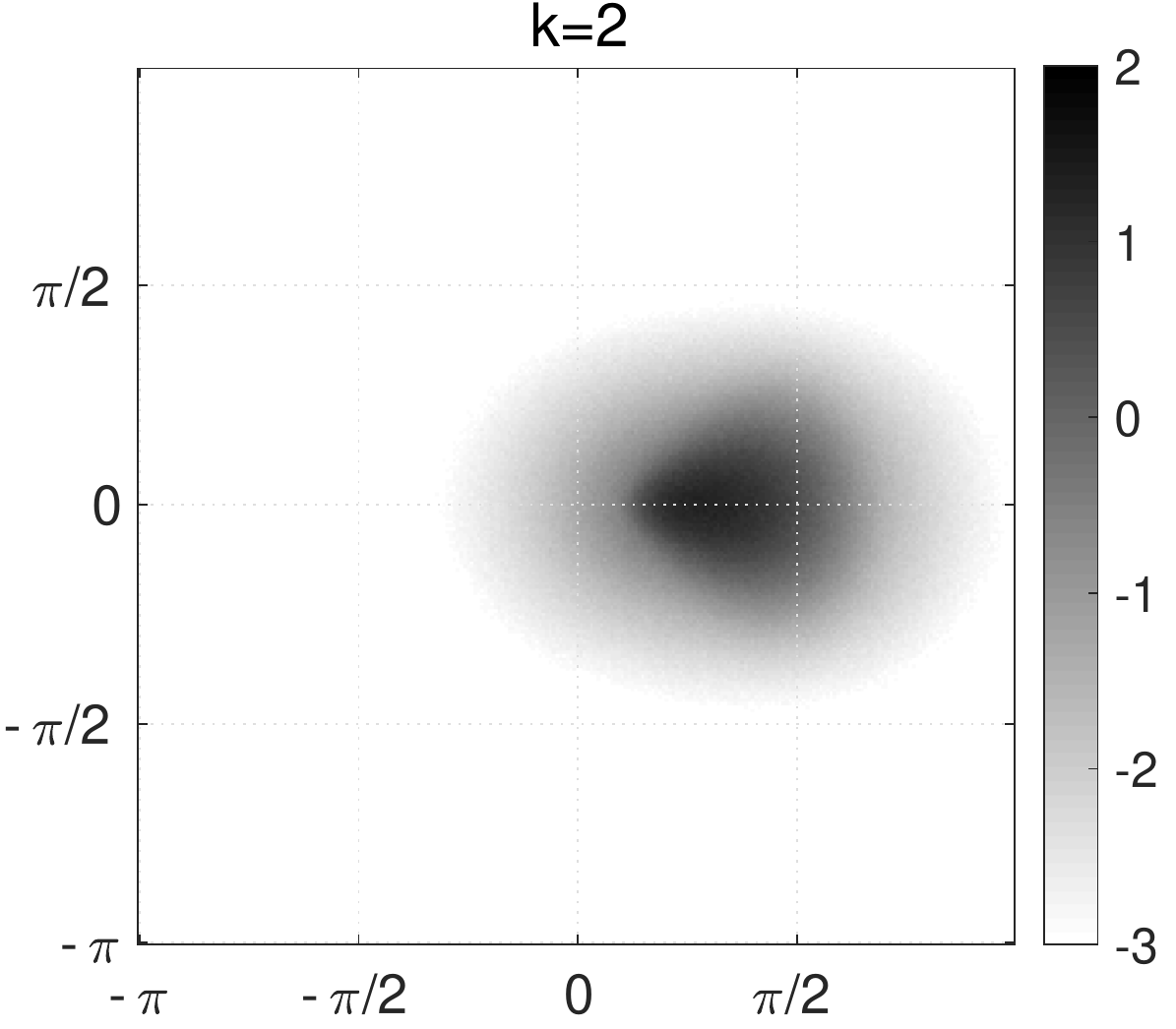}
	\end{subfigure}
\caption{
Power spectrum of $[X \star \psi_\la (u) ]^k$ for a fixed
$\la$,
for the turbulent vorticity flow at the top of 
Figure \ref{fig:synthesis2}(a),
and different $k$. The power spectrum 
is shown in log base 10.
(a): $k=0$. (b): $k=1$. (c): $k=2$. 
}
\label{fig:freqtrans}
\end{figure}

For $k=1$, the spectrum of 
$ X \star \psi_{\la}(u) $ has an energy concentrated 
at frequencies where the Fourier transform of
$\what \psi_{\la}$ is concentrated, which is included
in a ball centered at $\la$ of radius $C |\la|$.
For $|k| > 1$, \cite{mzr-18} explains that the Fourier transform
of $[X \star   \psi_\la(u) ]^{k}$ and hence its power spectrum is
concentrated in a ball centered at $k \la$ of
radius $|k| C  |\la|$.
If $k = 0$ then the spectrum of 
$|X \star \psi_{\la}(u) |$ is concentrated 
in a ball centered at the $0$ frequency, of radius $ C |\la|$.
This is illustrated numerically in Figure \ref{fig:freqtrans}
which displays the power spectrum
of $[X \star \psi_\la (u)]^k$ for $k = 0,1,2$. In this case, 
$X$ is a stationary vorticity field shown
at the top of Figure \ref{Gausssynthesisfig}(a).
These power
spectrum are estimated for a fixed $\la$ from $100$ independent
realizations of $X$. For $k = 1$, the power spectrum
is supported over the Fourier
support of $\what \psi_{\la}$. For $k = 2$ its support is approximately
dilated by $2$, whereas for $k = 0$ it is centered at the zero frequency.

It results that
the spectrum of $ [X \star \psi_{\la}(u)]^k$ and 
$[X \star \psi_{\la'}(u')]^{k'}$ have a support which overlap only if
\begin{equation}
\label{freqeconsfd}
|k \la - k' \la'| \leq C (\max(|k|,1) \,|\la| 
+ \max(|k'|,1)\,|\la'| )~.
\end{equation}
Similarly to the Fourier case, if $k = k' = 0$ then
this condition is satisfied for any pair of frequencies
$(\la,\la')$. It corresponds to covariances of modulus coefficients
$\Cov (|X \star \psi_\la (u)|\,,\,|X \star \psi_{\la'}(u') |)$,
which are typically non-zero when the process is non-Gaussian.

If $k \neq 0$ and $k' \neq 0$ 
then non-negligible values of
$\Cov([X \star \psi_\la (u) ]^k,[X \star \psi_{\la'} (u')]^{k'})$
occur for
$k \la \approx k' \la'$.
This is similar to the vanishing property (\ref{densfi8sdfs})
of Fourier harmonic coefficients, at frequencies which are colinear.
Since $\la = 2^{-j} r_{-\ell} \xi$ and $\la' = 2^{-j'} r_{-\ell'} \xi$, it
requires that $2^{j-j'} \approx |k'|/|k|$ and that $\ell \approx \ell'$.

Diagonal covariance coefficients have 
similar properties as Fourier coefficients. 
They are specified by 
$\Cov(|X \star \psi_\la (u)| ,|X \star \psi_\la (u)| )$
and $\Cov(X \star \psi_\la (u) ,X \star \psi_\la (u) )$,
which do not depend upon $u$. Moreover when $\la \neq 0$, 
\[
\frac{\Cov(|X \star \psi_\la (u)| ,|X \star \psi_\la (u)| )}
{\Cov(X \star \psi_\la (u) ,X \star \psi_\la (u) )}=
1 - \frac{\E(| X \star \psi_{\la}(u)|)^2}
{\E(| X \star \psi_{\la}(u)|^2)} .
\]
The ratio ${\E(| X \star \psi_{\la}(u)|)^2}/
{\E(| X \star \psi_{\la}(u)|^2)}$ 
measures the sparsity of wavelet
coefficients $ X \star \psi_{\la}(u)$. Indeed, the sparsity of a time series
is usually quantified by its  $\ell_1$ norm for a fixed $\ell_2$ norm
\cite{mallatbook}, which amounts to compute their ratios. Since
$ X \star \psi_{\la}(u)$ is stationary, spatial sums over $u$ converge to expected values. It results that a ratio of an $\ell_1$ over an $\ell_2$ norm
$ (\sum_u | X \star \psi_{\la}(u) | )^2 / (\sum_u | X \star \psi_{\la}(u) |^2 ) $
is an estimator of the ratio of expected values.
If $X$ is Gaussian then $X \star \psi_\la (u)$ 
is a complex Gaussian random variable and one can verify that
this ratio is $\frac \pi 4$ for all 
$\la$. A ratio smaller than $\frac \pi 4$ implies
that wavelet coefficients $ X \star \psi_\la (u)$ 
are sparse and hence that $X$ is non-Gaussian.

To increase the accuracy of a maximum entropy
model $\widetilde X$, 
the Kullback-Leibler divergence (\ref{KLdivers}) shows that
we must minimize its entropy. As in the Fourier case, an upper bound
of the entropy is obtained as a sum of the entropy of the marginals
of wavelet coefficients. The marginal entropies
get smaller by reducing the sparsity ratio
${\E(| X \star \psi_{\la}(u)|)^2}/{\E(| X \star \psi_{\la}(u)|^2)}$.
To minimize this upper bound of the model entropy, this suggests
by finding a mother wavelet $\psi$
which yields sparse coefficients.

\paragraph{Gaussianity test}
Even though $\psi_{\la}$ and $\psi_{{\la'}}$ may have disjoint
Fourier supports, the previous analysis showed that one
can find $k$ and $k'$ such that the spectrum of
$[X \star \psi_\la (u)]^{k}$ and $[X \star \psi_{\la'}(u')]^{k'}$ overlap,
for example with $k = k' = 0$. A priori, the 
covariance $K_{\wH \W} (v,v')$ is then non-zero,
unless $X$ is Gaussian in which case these coefficients vanish, 
as proved by the following theorem.

\begin{proposition}
Let $(\la,\la')$ be such that
$\what{\psi}_{\la}  \, \widehat{\psi}_{{\la'}}  = 0$.
If $X$ is Gaussian and stationary 
then $K_{\wH \W} (v,v') = 0$ if $v=(\la,k,u)$
and $v'=(\la',k',u')$, for any $(k,u,k',u')$.
\end{proposition}

{\it Proof:} 
If $X$ is Gaussian then $X \star \psi_\la (u)$ and
$ X \star \psi_{{\la'}}(u')$ are jointly Gaussian random variables. 
Equation (\ref{Fourier}) proves that 
if $\what{\psi}_{\la}\,  \widehat{\psi}_{{\la'}} = 0$ then
$\Cov( X \star \psi_\la (u) , X \star \psi_{\la'}(u')) = 0$. 
These Gaussian random variables are
uncorrelated and therefore independent.
It results that
$[X \star \psi_\la (u)]^{k}$ and $[X \star \psi_{\la'}(u')]^{k'}$ 
are also independent and thus have a covariance which is zero.$\Box$

This proposition gives a test of Gaussianity by considering 
two frequencies $\la$ and
$\la'$ where $|\la - \la'|$ is sufficiently large so that the
support of 
$\widehat \psi_{\la}$ and $\widehat \psi_{\la'}$ do not overlap.
If $k\la  \approx k' \la'$ then 
wavelet harmonic covariances are zero if $X$ is Gaussian but it
is typically non-zero if $X$ is not Gaussian.

\section{Microcanonical Models}
\label{s:maxent}

Section \ref{maxentsec} introduces 
macrocanonical maximum entropy models conditioned by 
covariance coefficients of a representation $\UPhi(X)$. 
The resulting maximum entropy distribution $\tilde p$ depend upon
Lagrange coefficients which are computationally very expensive 
to calculate, despite the development of efficient algorithms
\cite{Bortolli}. Section \ref{microsec}
reviews maximum entropy microcanonical models which
avoid computing these Lagrange multipliers.
Section \ref{microsec2} 
gives an alternative microcanonical model which is calculated with
a faster algorithm, which transports a maximum entropy measure by
gradient descent.

\subsection{Maximum Entropy Microcanonical Models}
\label{microsec}
Microcanonical models avoid the calculation of Lagrange multipliers and
are guaranteed to exist as opposed to macrocanonical models. 
We first specify covariance estimators and 
then briefly review
the properties of these microcanonical models.

Microcanonical models
rely on an ergodicity property which insures that
covariance estimations 
concentrate near the true covariance  $K_{\UPhi}$
when $d$ is sufficiently large. Let $G$ be a known group of
linear unitary symmetries of the density $p$ of $X$. It includes
translations because $X$ is stationary.
An estimation of the covariance $K_{\UPhi}$ over the edge set $E_G$ is
computed in (\ref{phiFouriensf6}) from a 
single realization $\bar x$ of $X$, by transforming $\bar x$
with all $g \in G$. 
To differentiate realizations
of $X$ from 
other $x \in R^d$, we associate a similar covariance estimation 
to any $x \in \R^d$. This covariance is
centered on the empirical mean $\widetilde M_{\UPhi}$
computed in (\ref{phiFouriensfmean6}) and defined by
\begin{equation}
\label{phiFouriensf666}
\widetilde K_{\UPhi x} (\ga,\ga') = 
\frac {1} {|G|} \sum_{g \in G} 
\Big( \UPhi_\ga (g.  x) - \widetilde M_{\UPhi} (\ga) \Big)\,
\Big( \UPhi_{\ga'} (g.  x) - \widetilde M_{\UPhi} (\ga') \Big)^*\,.
\end{equation}
It is invariant to 
the action of any $g \in G$ on $x$.

We denote 
$\|K_{\UPhi} \|_\LaG^2 = \sum_{(\ga,\ga') \in \LaG} |K_{\UPhi} (\ga,\ga')|^2$. 
We shall suppose that $X$ satisfies the following covariance 
ergodicity property over $\LaG$:
\begin{equation}
	\forall \epsilon > 0, \quad  
\lim_{d \to  \infty} \mbox{Prob} ( \| \widetilde K_{\UPhi X} - K_{\UPhi}  \|_\LaG \leq \epsilon ) = 1.  
\label{eq:erg}
\end{equation}
To satisfy this property the number $|\LaG|$ 
of covariance moments
must be small compared to the dimension $d$ of $X$.
The ergodicity hypothesis implies that 
when $d$ is sufficiently large, with high probability, 
the empirical covariance
$\widetilde K_{\UPhi \bar{x}}$ of a realization $\bar x$ of $X$ 
is close to the true covariance $K_{\UPhi}$ over $\LaG$.

Given a realization $\bar x$ of $X$,
a microcanonical set of width $\epsilon$ 
is the set of all $x \in \R^d$ which have nearly the same covariance estimations as $\bar{x}$:
\begin{equation}
\label{mcifsdfnsdf}
	\Omega_{\epsilon}  = \{ x \in \R^d: \| \widetilde K_{\UPhi x} - \widetilde K_{\UPhi \bar{x}} \|_\LaG \leq  \epsilon \} . 
\end{equation}
A maximum entropy microcanonical model 
has a probability distribution of maximum entropy supported in 
$\Omega_{\epsilon}$. 
The set 
$\Omega_{\epsilon}$ is bounded. Indeed, if 
$x \in \Omega_{\epsilon}$ then $\|\widetilde K_{\UPhi x} \|_\LaG \leq 
\| \widetilde K_{\UPhi \bar{x}} \|_\LaG +  \epsilon$. 
Since $\LaG$ includes all diagonal coefficients $(\ga,\ga)$ 
for $\ga \in \Ga$, one can derive an upper bound for $\|{\UPhi x} \|$.
Since $\UPhi$ is bi-Lipschitz, we also obtain an upper bound for $\|x\|$.
Since $\epsilon>0$, 
the Lebesgue measure of $	\Omega_{\epsilon} $ is non-zero, therefore
the maximum entropy distribution is
uniform in $\Omega_{\epsilon}$ relatively to the Lebesgue measure.
A major issue in statistical physics is to find sufficient 
conditions to prove the Boltzmann equivalence principle
which guarantees the convergence of
microcanonical and macrocanonical models towards the same Gibbs measures
when $d$ goes to $\infty$ \cite{ZeitouniBook}. This involves the proof of
a large deviation principle which expresses concentration properties
of the covariance of $\UPhi(X)$. If $\UPhi$ is continuous and bounded 
so that the interaction potential
$(\UPhi  - M_\UPhi )( \UPhi - M_\UPhi )^*$ is also continuous and bounded,
and if 
there is no phase transition, which means that the limit is a unique
Gibbs measure, then one can prove that microcanonical and macrocanonical measures converge to 
the same limit for an appropriate topology \cite{stroock,DiGorggi}.
The bounded hypothesis is not necessary and may not be satisfied.

The ergodicity property (\ref{eq:erg}) guarantees that
$X$ concentrates
in $\Omega_{\epsilon}$ with a high probability when $d$ is sufficiently
large. The microcanonical set $\Omega_\epsilon$ may however be much larger
than the set where $X$ concentrates, which means that the maximum
entropy microcanonical distribution may have a much
larger entropy than the entropy of $X$. As in the macrocanonical case,
the representation $\UPhi$ must be optimized in order to reduce the
maximum entropy, which motives the use of sparse representations.

\subsection{Gradient-Descent Microcanonical Models}
\label{microsec2}

Sampling a maximum entropy
microcanonical set requires to use Monte Carlo algorithms.
They are computationally very expensive when the number
of moments $|\LaG|$ and the dimension $d$ are large,
because their mixing time become prohibitive
 \cite{doi:10.1063/1.477552}. 
Following the approach in \cite{bruna2018multiscale}, 
we approximate these microcanonical
models with a gradient descent algorithm.

Gradient-descent microcanonical models 
are computed 
by transporting an initial Gaussian white noise measure
into the microcanonical set $\Omega_{\epsilon}$. 
This transport is calculated with a gradient descent
which progressively minimizes
\begin{equation}
f(x) = \| \widetilde K_{\UPhi x} - \widetilde K_{\UPhi \bar{x}} \|_\LaG^2 ,
\label{eq:microobj}
\end{equation}
with a sufficiently large number of
iterations so that $f(x) < \epsilon$ and hence $x \in \Omega_\epsilon$.

The initial 
Gaussian white noise is a maximum entropy distribution conditioned by
a variance $\sigma^2$. This variance
is chosen to be an upper bound of the empirical variance
of $\bar x(u)$ along $u$, calculated from diagonal coefficients of
$\widetilde K_{\UPhi \bar x}$. It guarantees that the resulting white
noise has an entropy larger than the microcanonical model.
The gradient
descent progressively reduces this entropy while transporting the measure
towards $\Omega_\epsilon$. Entropy reduction bounds are computed 
in \cite{bruna2018multiscale}. 

The initial $x_0$ is Gaussian white noise. 
For all $t \geq 0$, the gradient descent 
iteratively computes
\[
x_{t+1} = x_t - \eta\, \nabla f (x_t)~.
\]
This operation transports the
probability measure $\mu_t$ of $x_t$ into
a measure $\mu_{t+1}$ of $x_{t+1}$.
We stop the algorithm at a time $t = T$ which
is large enough so that $f(x_T) < \epsilon$ with a high probability.
Since $f(x)$ is not convex, 
there is no guarantee that $f(x_T) < \epsilon$
even for large $T$. 
In numerical calculations in Section \ref{approxnsec}, 
we may use several initializations, typically $10$,
to evaluate the model. 
The differentiability of $f$ relies on 
the smoothness of the function $ z \mapsto [z]^k$, 
which is differentiable at $z \neq 0$. 
The implementation uses an analytic formula of its derivative
at $z \neq 0$ and applies the same formula at $z=0$.

The empirical covariance 
$\wtilde K_{\UPhi x}$ is computed in (\ref{phiFouriensf666})
with an 
average over all symmetries of a known group $G$ of linear
unitary operators in $\R^d$. The following theorem proves that
$G$ is also a group of symmetries of the probability measures
obtained by gradient descent.

\begin{theorem}
\label{thm:cgsym}
For any $t \geq 0$, the probability measure $\mu_t$ of $x_t$
is invariant to the action of $G$.
\end{theorem}

The proof is in Appendix \ref{algosym}. 
It is a minor modification of the proof
in \cite{bruna2018multiscale}.
We replace the gradient descent method by 
L-BFGS algorithm with line search \cite{OptimBook}.
It has a faster and more accurate numerical convergence.
Since $G$ includes translations, this theorem implies that
each $x_t$ is stationary.
It is shown in \cite{bruna2018multiscale} that
the transported density $\mu_T$ supported in
$\Omega_\epsilon$ may be different from the
maximum entropy density in $\Omega_\epsilon$ because the gradient
descent reduces too much the entropy. In general, 
the precision of these
microcanonical gradient descent models are not well understood.
However, this theorem proves that the gradient descent 
preserves all known symmetries of the distribution
of $X$, which is also true for a maximum entropy measure.
The sampling algorithm in \cite{portilla2000parametric} is different since it relies on iterative gradient projections for $x$ to satisfy constraints of various nature, which is not guaranteed to preserve symmetries.

The gradient descent may converge faster 
by preconditioning $f(x)$. This is done by 
replacing estimated covariances $\wtilde K_{\UPhi x} (\ga,\ga')$ 
by normalized correlation coefficients 
\begin{equation}
\label{cofrrelasts}
\frac{\wtilde K_{\UPhi x} (\ga,\ga')}
{\wtilde K_{\UPhi \bar x}(\ga,\ga)^{1/2}\,\wtilde K_{\UPhi \bar x}(\ga',\ga')^{1/2}} ~.
\end{equation}
Any $g \in G$ is a symmetry of $\wtilde K_{\UPhi x}$ and is therefore
a symmetry of normalized correlation coefficients, so 
Theorem \ref{thm:cgsym} remains valid with this preconditioning.

\section{Foveal Wavelet Harmonic Covariance Models}
\label{approxnsec}
We study microcanonical models conditioned by wavelet harmonic 
covariances, computed
with different symmetry groups $G$, and different
sufficient statistics set $\LaG$.
Section \ref{microsfoveal} introduces foveal models which limit the
multiscale spatial range of coefficients in $\LaG$.
Section \ref{evalucov} gives a methodology to evaluate the precision of different
foveal models, with numerical results.

\subsection{Rotation and Reflection Symmetries}

The covariance $K_\UPhi$ is estimated from a single realization of $X$,
by taking advantage of a known group of symmetries $G$ of 
the probability distribution of $X$. 
For a wavelet phase harmonic representation $\UPhi = \what \HH \W$,
we specify the properties of $K_\UPhi$ when $G$ includes
sign changes, reflections and rotations.

The following proposition proves that 
some covariance coefficients vanish when symmetries include
a sign change or a central reflection. These covariances thus
do not need to be included in the sufficient statistics set $\LaG$.
The proof is in Appendix \ref{proofcovsign}.

\begin{proposition}
\label{prop:covsign}
Let $v = (\la, k ,u)$ and $v' = (\la', k' ,u')$.\\ 
(i) If the sign change $g. x = -x$ is a symmetry of the probability distribution
of $X$
then $K_{\wH \W} (v,v') = \wtilde K_{\wH \W x}(v,v')   = 0$ if
$k + k'$ is odd. \\
(ii) If the central reflection
$g.x (u) = x(-u)$ is a symmetry of the probability distribution
of $X$ and $\what h$ is real
then $K_{\wH \W} (v,v')$, and 
$\wtilde K_{\wH \W x} (v,v')$ are real.
\end{proposition}

\paragraph{Isotropic models}
An isotropic random process $X$
has a probability distribution which is invariant by 
rotations. The group $G$ then includes all translations and rotations. 
We show that $K_{\what \HH \W}$ becomes sparse after applying
a Fourier transform on rotation angles.

If $g = r_\eta$ is a rotation by $2 \pi \eta/Q$ for $0 \leq \eta < Q$ 
then 
$ (r_\eta . x) \star \psi_{\la} (u) = x \star \psi_{r_\eta \la}(r_{-\eta} u)$
where $r_\eta \la = 2^{-j} r_{\eta-\ell} \xi $ 
is the rotation of $\la = 2^{-j} r_{-\ell} \xi$. 
To eliminate the effect of the change of position $r_{-\eta} u$ on 
covariance coefficients, we only keep wavelet harmonic covariances
at a same spatial position. This means that $v'= (\la',k',u')$ is
a neighbor of $v= (\la,k,u)$ in the covariance graph model only if 
$u = u'$. It implies that $(v,v') \in \LaG$ only if $v$ and $v'$
have the same spatial position.

Isotropy is a form of stationarity along rotations angles.
To diagonalize angular covariance matrices, we use a 
discrete Fourier transform along rotations written $\F_\ell$. The 
discrete Fourier
transform of $y(\ell)$ for $0 \leq \ell < Q$ 
at a frequency $0 \leq m < Q$ is
\[
(\F_\ell y)(m) = \sum_{\ell = 0}^{Q-1} y(\ell)\, e^{-i 2 m \pi \ell / Q} .
\]
The representation $\UPhi (X) = \F_\ell\,\what \HH  (\W X)$ computes
$\F_\ell ([X \star \psi_{\la}(u)]^{k})$ for 
$\la = 2^{-j} r_{-\ell} \xi$ with $(j,u,k)$ fixed and $\ell$ varying.
It is indexed by $v = (j,m,k,u)$. 
The covariance matrix of $\UPhi (X)$ is
\[
K_{\UPhi} = \F_\ell K_{\what \HH  \W} \, \F_\ell^{-1}.
\]

We consider the restriction of
$K_{\UPhi}$ to
$\LaG$. The next theorem proves that if $X$ is isotropic then
$K_{\Uphi}$ has diagonal angular Fourier matrices. 
Isotropic processes $X$ may also have a probability distribution
which are invariant to  line reflections.
A line reflection of orientation $\eta$ 
computes $g.x(u) = x(u_\eta)$, where $u_\eta$ is 
symmetric to $u$ relatively to a line 
going through the origin in $\R^2$, with an
orientation $\eta \in [0,2\pi]$. If $X$ is isotropic and invariant
to a line reflection for an angle $\eta$ then it is invariant
to line reflections for any $\eta \in [0,2\pi]$.
The following theorem applies to 
the bump steerable wavelets used in numerical experiments.

\begin{theorem}
\label{prop:covrot}
Let $v = (j,m,k,u)$, $v' = (j',m',k',u)$ and $u=(u_1,u_2)$.
If the probability distribution of $X$ stationary and isotropic and 
$\UPhi = \F_\ell\,\what \HH  \W$ then
\begin{equation}
\label{rotation2}
K_{\UPhi} (v,v') = \wtilde K_{\Uphi  x}
(v,v')  = 0~~\mbox{if $m \neq m'$}.
\end{equation}
Furthermore, if the distribution of 
$X$ is invariant to line reflections 
and if the wavelet satisfies
$\psi(u_1,-u_2) = \psi(u_1,u_2)$ and $\phi(u_1,-u_2) = \phi(u_1,u_2)$
then $K_{\Uphi} (v,v')$ and $\wtilde K_{\Uphi  x}(v,v')$ are real if $m = m'$.
\end{theorem}

The proof is in Appendix \ref{proofcovrot}.
This theorem proves that the sufficient statistics set can be
reduced to diagonal angular coefficients
$(v,v') \in \LaG$ with $m = m'$. It reduces its size 
by a factor $Q$. 
Invariance to line reflections implies that 
it is sufficient to
keep the real part of these diagonal values.
The rotation invariance strictly 
applies to the process defined on a continuous variable $u \in \R^2$.
On discrete images it is not valid at the finest scale because of the
square sampling grid, and it is not valid at the largest scale because of their square support. For preconditioning, the covariance of
the isotropic model is also normalized with (\ref{cofrrelasts})
and the Fourier transform $\F_\ell$ is then applied on the normalized
coefficients.

\subsection{Foveal Wavelet Harmonic Covariance}
\label{microsfoveal}

A wavelet harmonic covariance model is defined by the
choice of the harmonic weights $\hat h$, 
by the symmetry group $G$ and neighborhood relations which
defines the edge set $\La$. In the following we
shall impose that $\hat h(k) = 1_{[k_{\min},k_{\max}]} (k)$,
which limits harmonic exponents in the range $[k_{\min},k_{\max}]$.
We define several foveal models which capture different properties.

\paragraph{Foveal models} 
Wavelet harmonic coefficients are indexed by
$v = (\la,k,u)$, with 
$\la = 2^{-j} r_{-\ell} \xi$ and $u = 2^{j-1} n$.
The covariance graph model is specified
by the neighborhoods $\Voi_v$.
A foveal model defines neighborhoods whose size doe not depend
upon $v$. The range of spatial,
scale and angular parameters is limited by three parameters $\Delta_n$,
$\Delta_j$ and $\Delta_\ell$.
A vertex $v'=(\la',k',u')$ with 
$\la' = 2^{-j'} r_{-\ell'} \xi$ and $u' = 2^{j'-1} n'$ is a neighbor of
$v=(\la,k,u)$ only if
\[
|n - n'| \leq \Delta_n~,~|j-j'| \leq \Delta_j~,~|\ell-\ell'| \leq \Delta_\ell~,~(k,k') \in [k_{\min},k_{\max}]^2~.
\]
A foveal model has a spatial range proportional to the scale. 
Long range spatial correlations are partly captured because
$|u-u'| = |n 2^{j-1} - n' 2^{j'-1}|$ become large at large scales.
It provides
high frequency correlations between close points and low-frequency
correlations between far away points. It is similar to a visual
fovea \cite{mallat2003foveal}.
Such foveal models have been
used by Portilla and Simoncelli \cite{portilla2000parametric}
to synthesize image textures.

Because of translation invariance, the sufficient statistics
$E_G$ can be defined by setting $n=0$.
Since there are $Q$ angles $\ell$ and at most $(\log_2 d)/2$ scales $j$,
the size of $\LaG$ is at most
\[
|\LaG| = O(Q \Delta_\ell (k_{\max} - k_{\min}+1)^2(2 \Delta_n+1)^2 \Delta_j \log_2 d ).
\]
Increasing the values
of $k_{\max}, \Delta_j, \Delta_\ell, \Delta_n$ decreases the model bias
but it also increases the size $|\LaG|$ and hence
the variance of the estimation.
To ensure the bi-Lipschitz continuity of $\UPhi = \what \HH \W$, we 
impose that $  k_{\min} \leq 1 \leq k_{\max} $. 
In the following we describe several
models of different sizes, which capture different properties of $X$. Each realization $\bar x$ is an image of $d=256^2$ pixels. We use the bump steerable wavelets of Appendix \ref{appendixcomplexsteer}, 
computed on $J=5$ scales and $Q=16$ angles. 
The maximum scale $2^J$ depends on the integral scale 
of $X$, which is the distance beyond which all coefficients
are nearly independent.
A large number of angles $Q$ gives a finer angular resolution. 
We specify $4$ models corresponding to different choices of neighborhood
parameters. For the first three models, 
the symmetry group $G$ is reduced to translations where as the last
model is also invariant to rotations.

$\bullet$ Model A with $k_{\min} = k_{\max} = 1$. It corresponds to a
Gaussian maximum entropy wavelet model of
Section \ref{s:wavecov1}.
The covariance of wavelet coefficients is neglected
across scales and angles by setting
$\Delta_j = 0$, $\Delta_\ell = 0$ and $\Delta_n = 2$.
The relative dimension of this
model is  $|\LAG^A|/d = 3.6 \, 10^{-2}$.

$\bullet$ Model B with $k_{\min} = 0$ and $k_{\max} = 1$. 
It includes the covariance of the modulus 
of wavelet coefficients across angles. Covariance
across angles at most distant by $\pi/4$ are computed for $k,k' \in \{0,1\}$ 
by setting $\Delta_\ell= Q/4$.  We set $\Delta_j=0$ and $\Delta_n=2$,
and thus only incorporate covariance across 
spatial positions. Compared to Model A, spatial correlations 
are included for $k=k'=1$ but also for $k=k' =0$.
The relative model size is $|\LAG^B|/d = 1.1 \,10^{-1}  $.

$\bullet$ Model C with $k_{\min} = 0$ and $k_{\max} = 2$. 
It incorporates covariances of the modulus and phase
of wavelet coefficients across scales and angles. 
Neighborhoods are limited by
$\Delta_j=1$, $\Delta_\ell = Q/4$ and $\Delta_n = 2$.
Compared to Model B, it incorporates $j'=j+1$ 
to capture scale interactions. The phases of wavelet coefficients
at a scale $2^j$ and $2^{j'}=2^{j+1}$ are correlated with the
harmonic exponents $(k,k') = (1,2)$. The set of $(k,k')$ are
restricted to $k'=0,1,2$ when $k=0$, and $k'=1,2$ when $k=1$.
We use the same spatial correlation range as Model B.
The relative model size is $|\LAG^C|/d = 1.7 \, 10^{-1}  $. 

$\bullet$ Model D with $k_{\min} = 0$ and $k_{\max} =2$
is a rotation invariant version of Model C,
with the same $\Delta_j$, $\Delta_\ell$ and $(k,k')$.
This model sets $\Delta_n=0$ and thus does not capture 
spatial correlations explicitly. 
The symmetry group $G$
is composed of translations and $Q$ rotations by $2 \pi \eta/Q$.
This rotation invariance is represented by
computing a Fourier transform along angles and
setting to zero the covariance coefficients according to 
Theorem \ref{prop:covrot}.
The model size is therefore much smaller with
$|\LAG^D|/d= 1.2 \,  10^{-2}$.
Theorem \ref{thm:cgsym} proves that
the resulting gradient descent microcanonical distribution 
is invariant to these $Q$ rotations.
The Python implementation uses GPU (graphics processing unit) to synthesize
non-Gaussian samples. Each synthesis sample (at $d = 256^2$)
takes $30$min for Model B, $50$min for Model C,
and $8$min for Model D, by running a maximal 5000 number of L-BFGS iterations.
The GPU memory cost is between $3$GB to $6$GB.

\begin{figure}
\scalebox{1}{
\centering
	\begin{subfigure}[t]{4cm}
%		\caption{}
		\includegraphics[height=4cm]{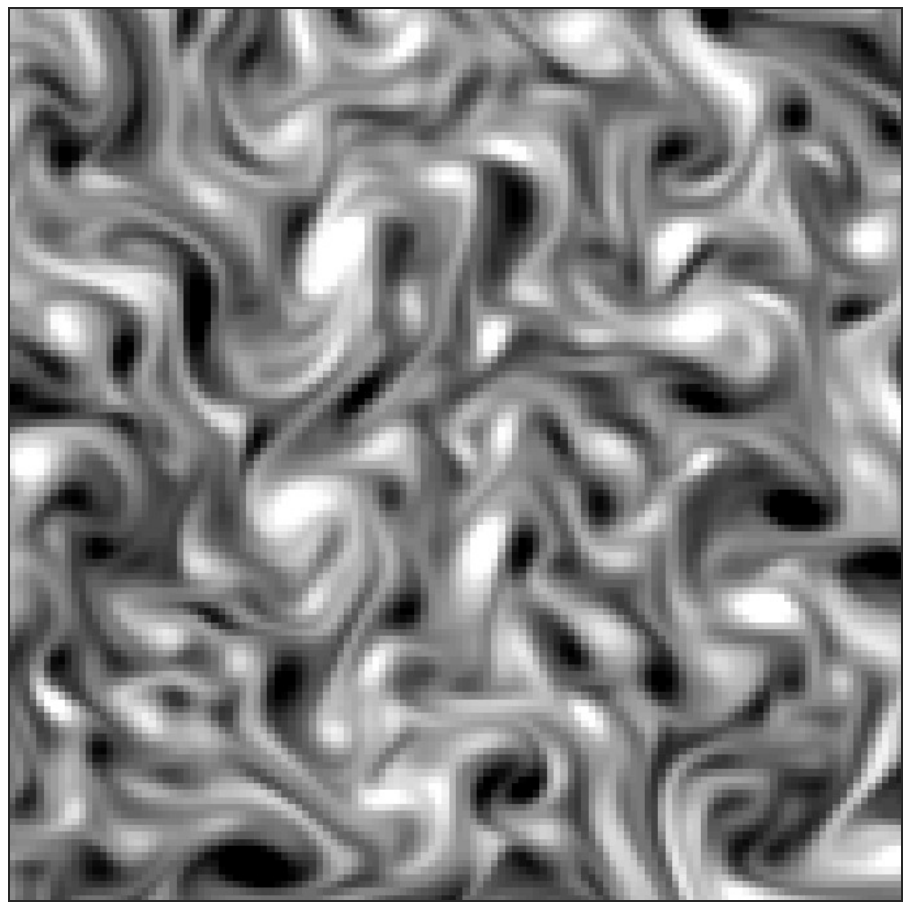}
		\includegraphics[height=4cm]{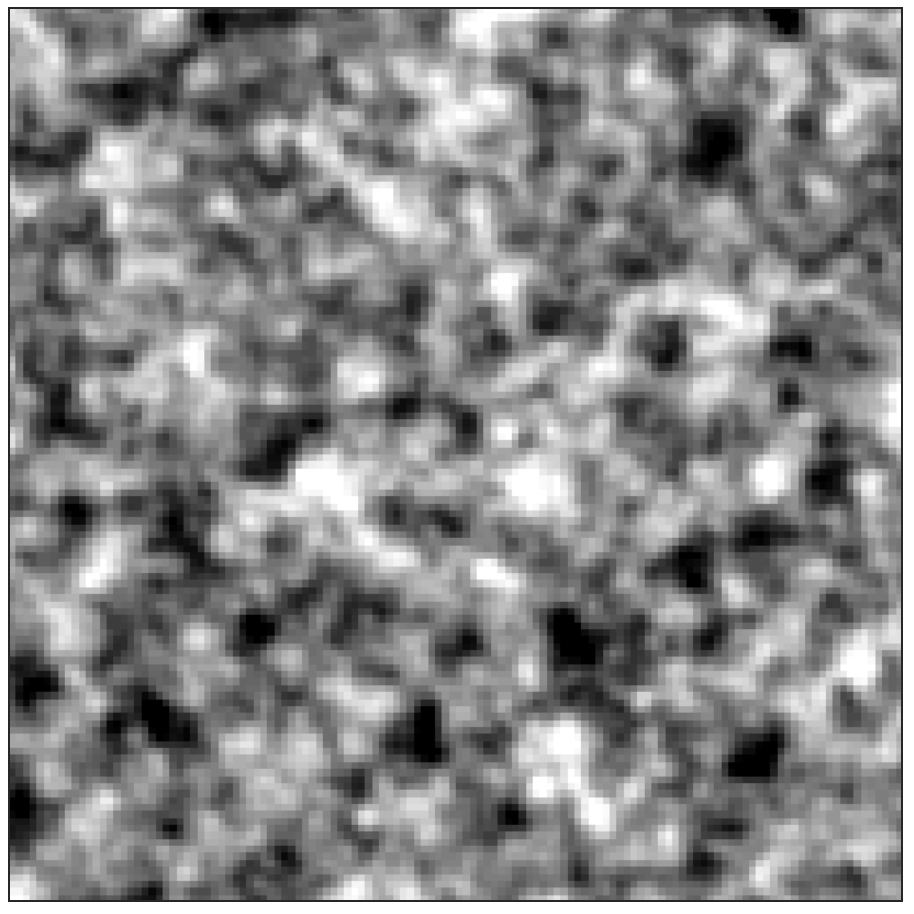}
		\includegraphics[height=4cm]{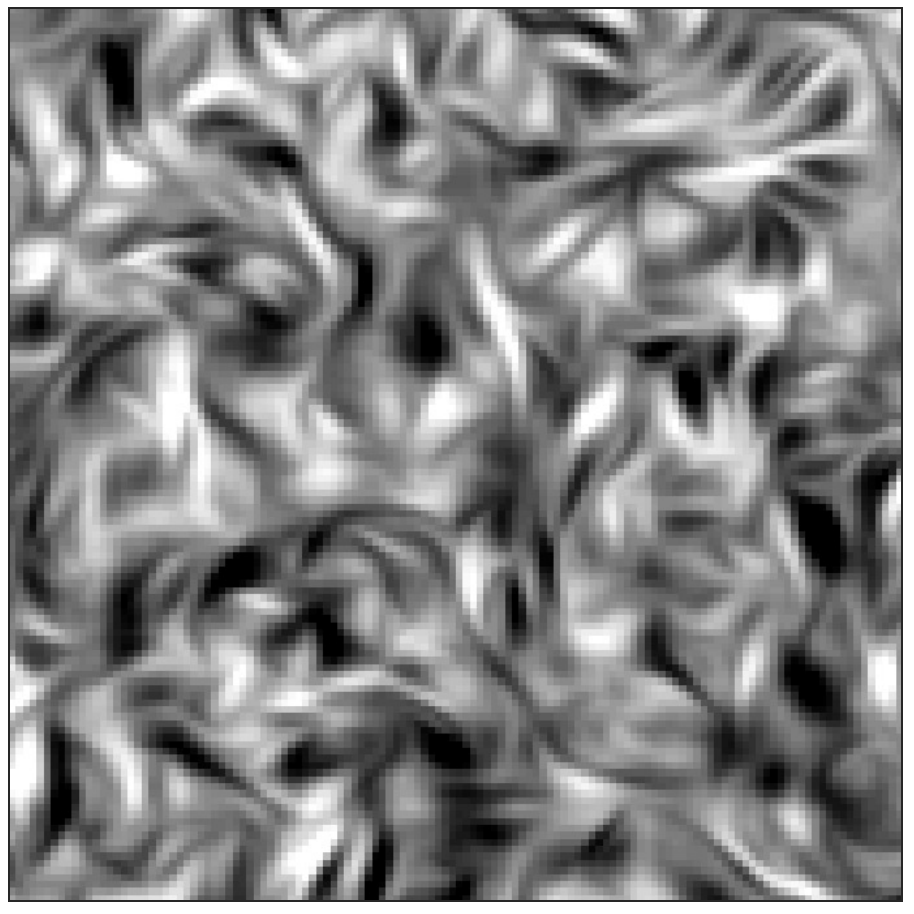}
		\includegraphics[height=4cm]{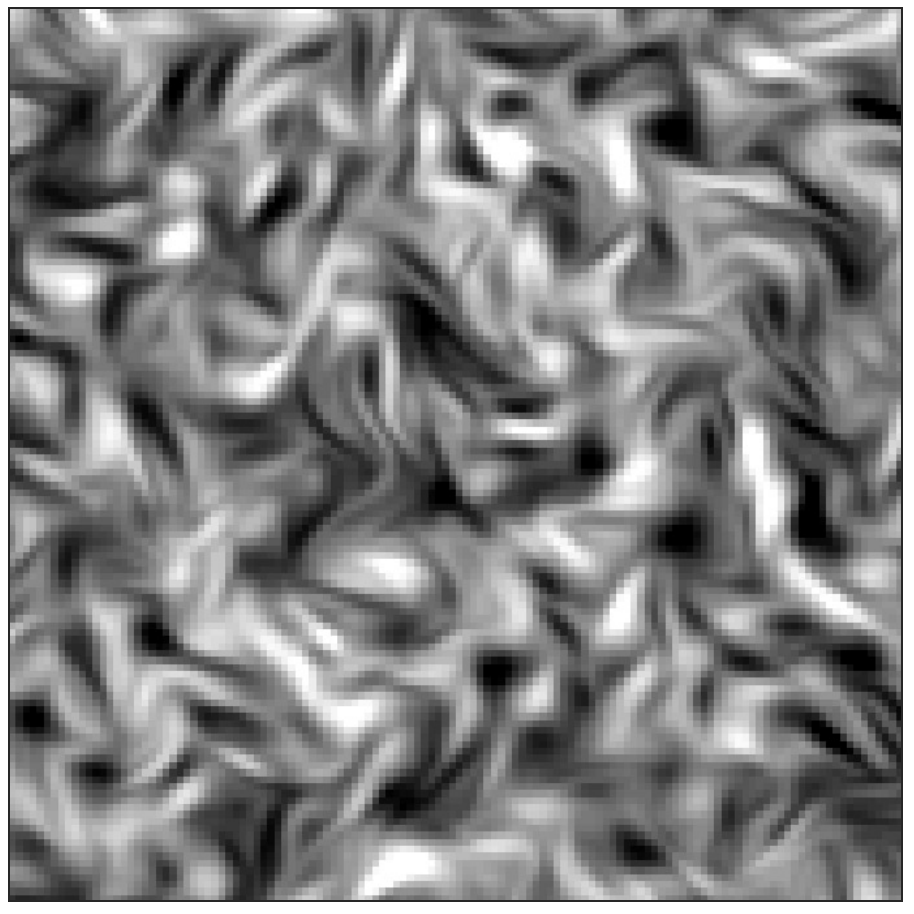}
		\includegraphics[height=4cm]{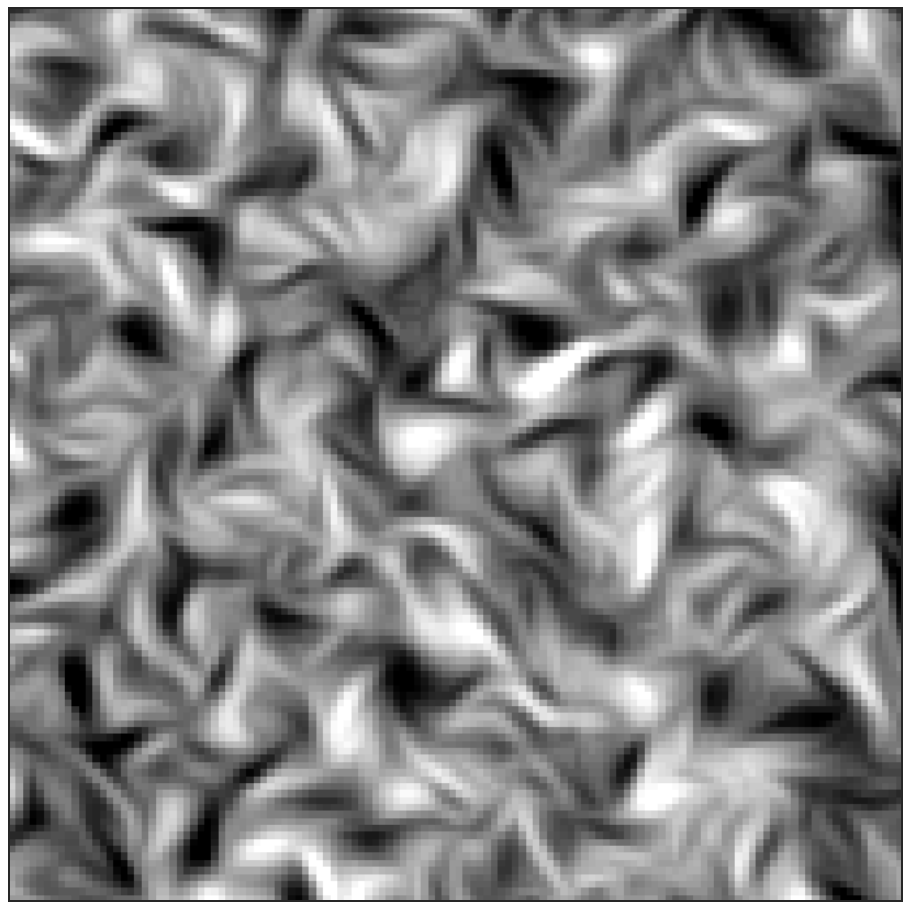}
	\end{subfigure}
	\begin{subfigure}[t]{4cm}
%		\caption{}
		\includegraphics[height=4cm]{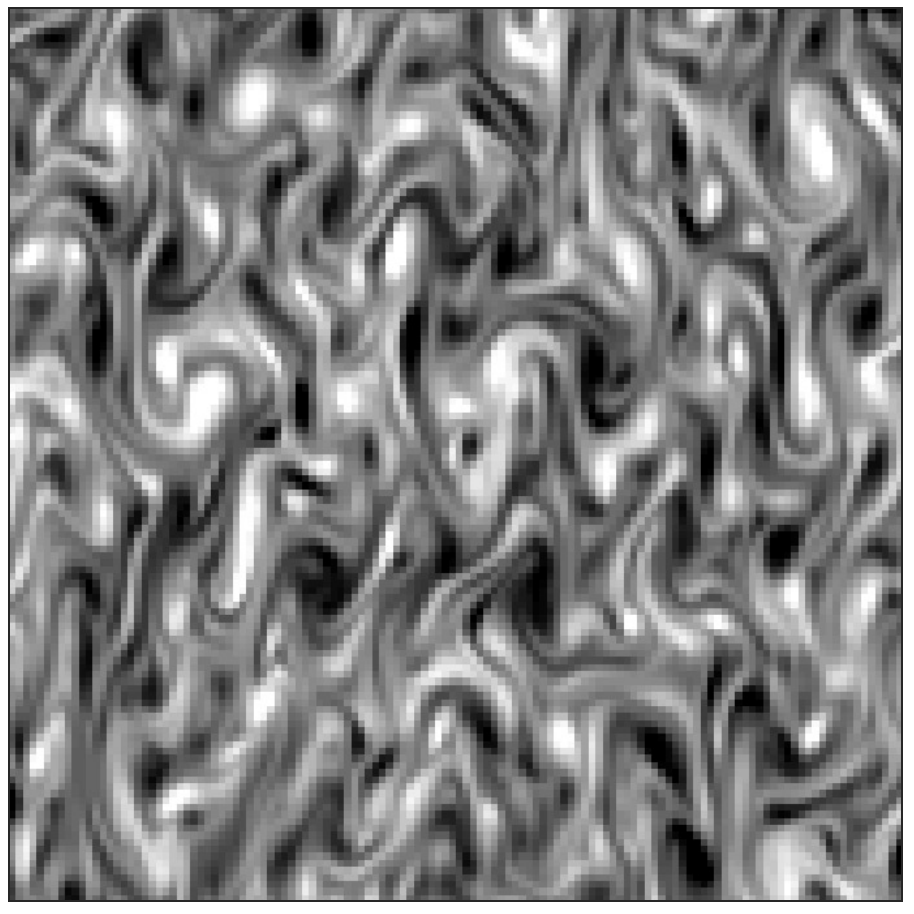}
		\includegraphics[height=4cm]{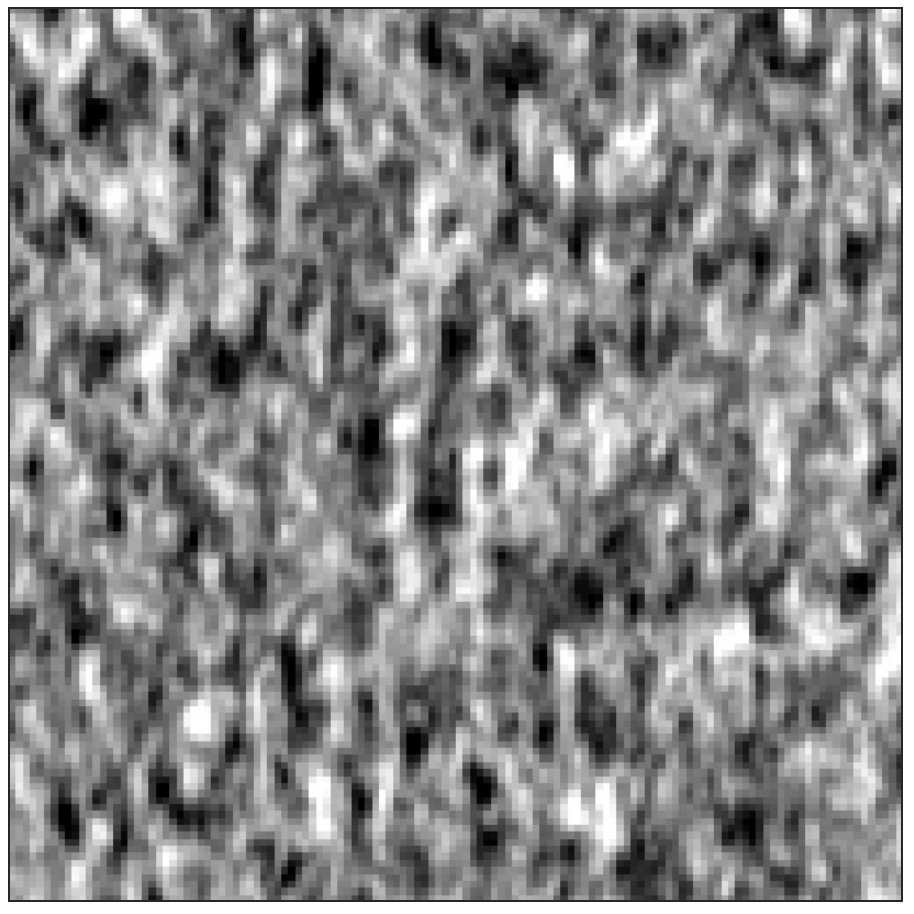}
		\includegraphics[height=4cm]{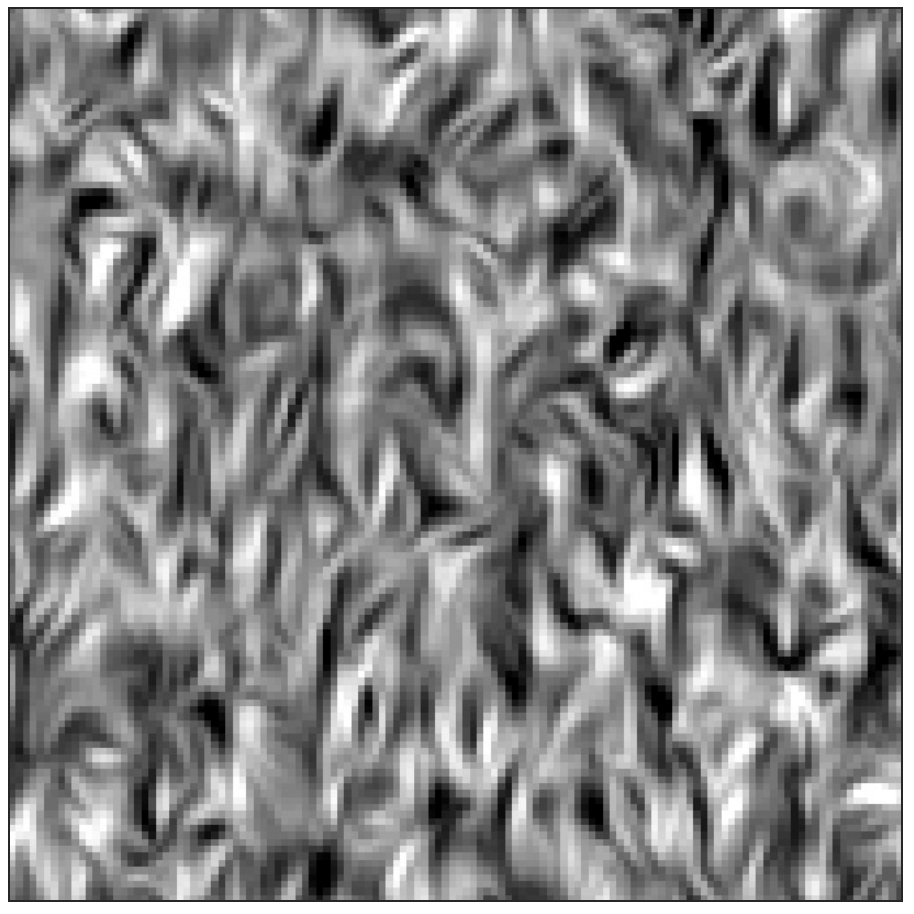}
		\includegraphics[height=4cm]{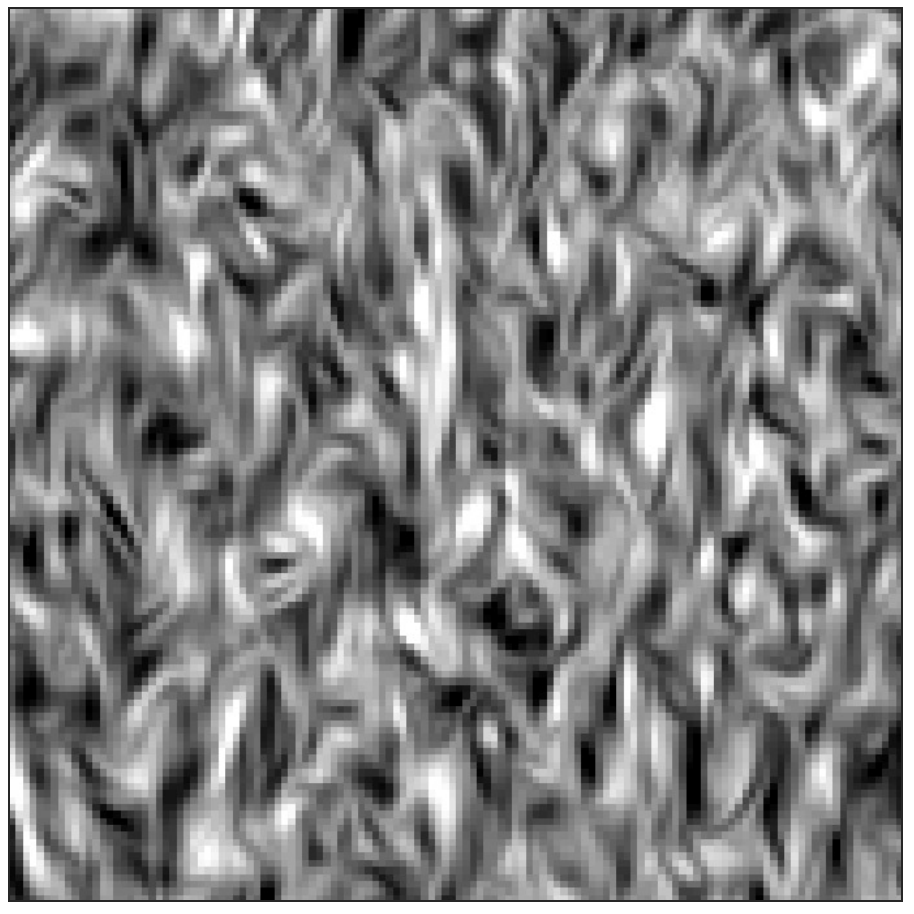}
		\includegraphics[height=4cm]{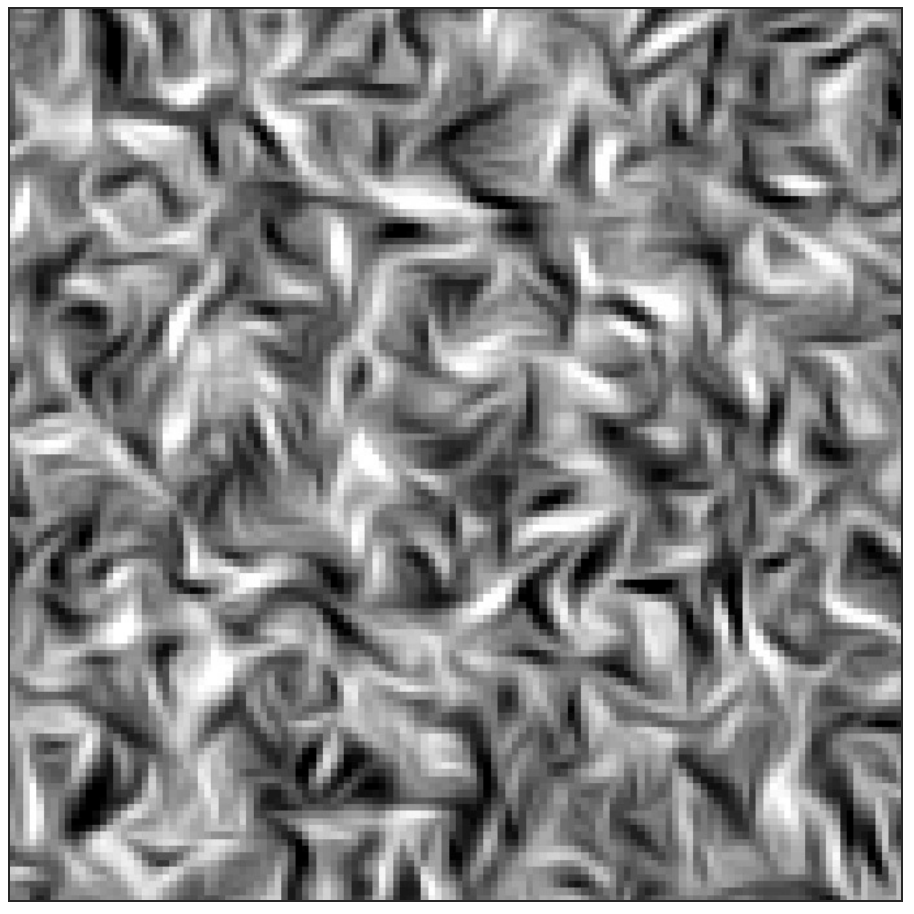}
	\end{subfigure}
	\begin{subfigure}[t]{4cm}
%		\caption{}
		\includegraphics[height=4cm]{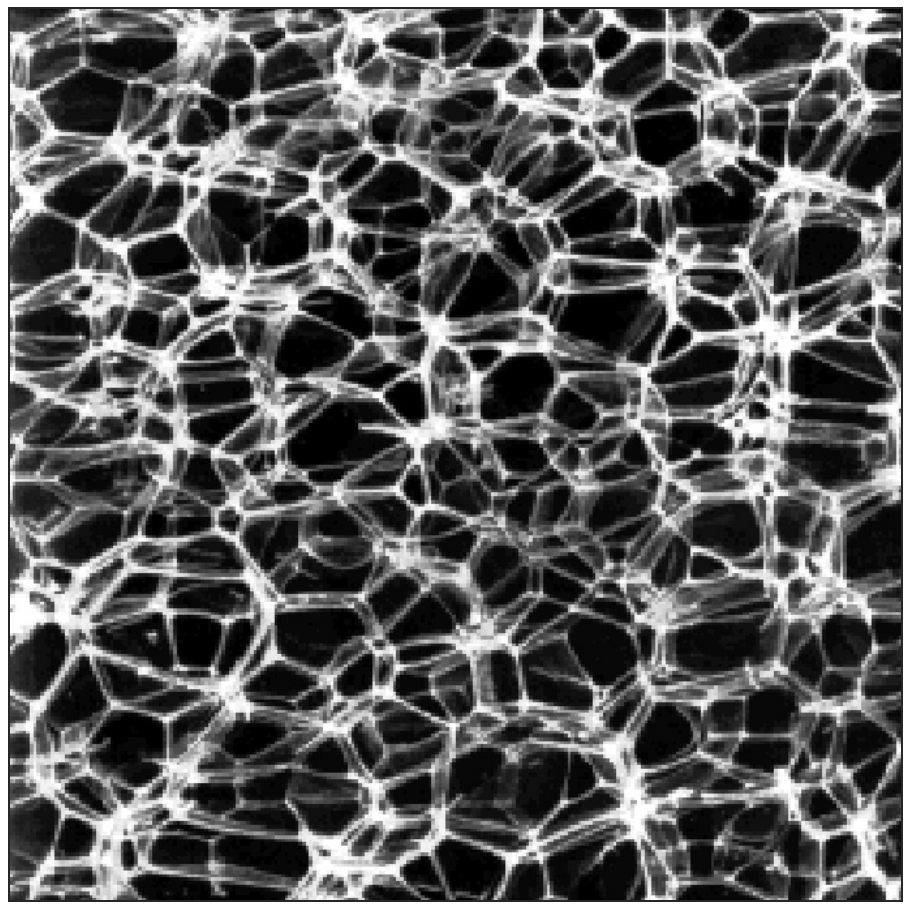}
		\includegraphics[height=4cm]{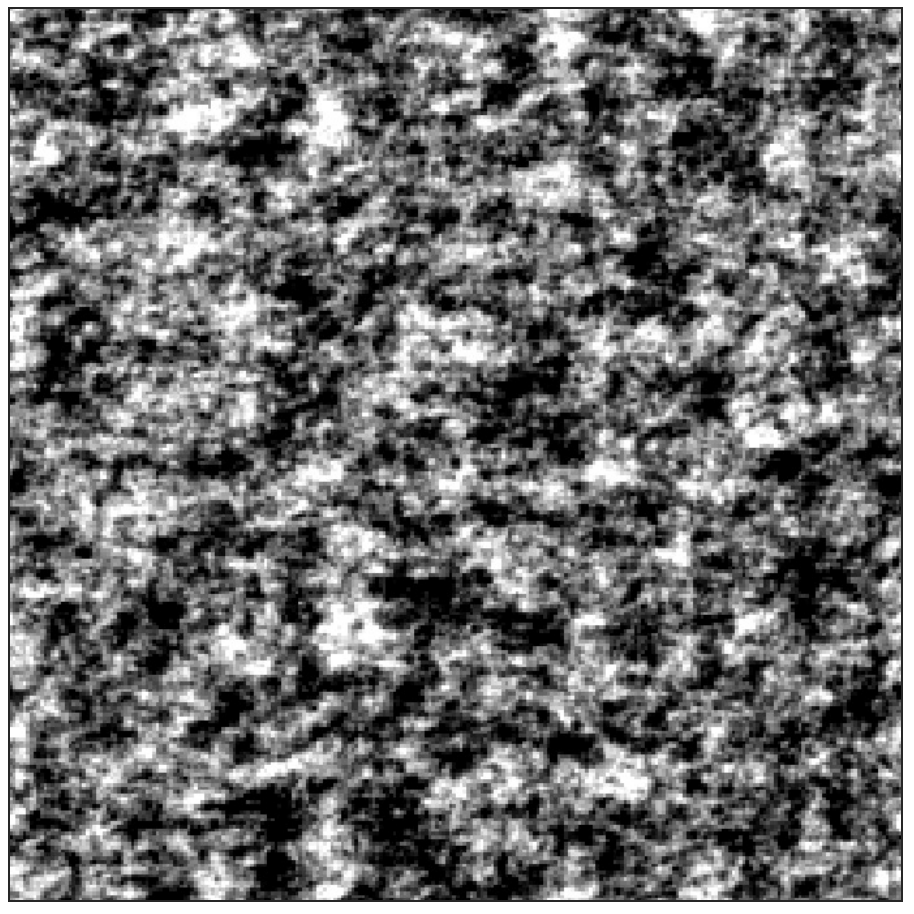}
		\includegraphics[height=4cm]{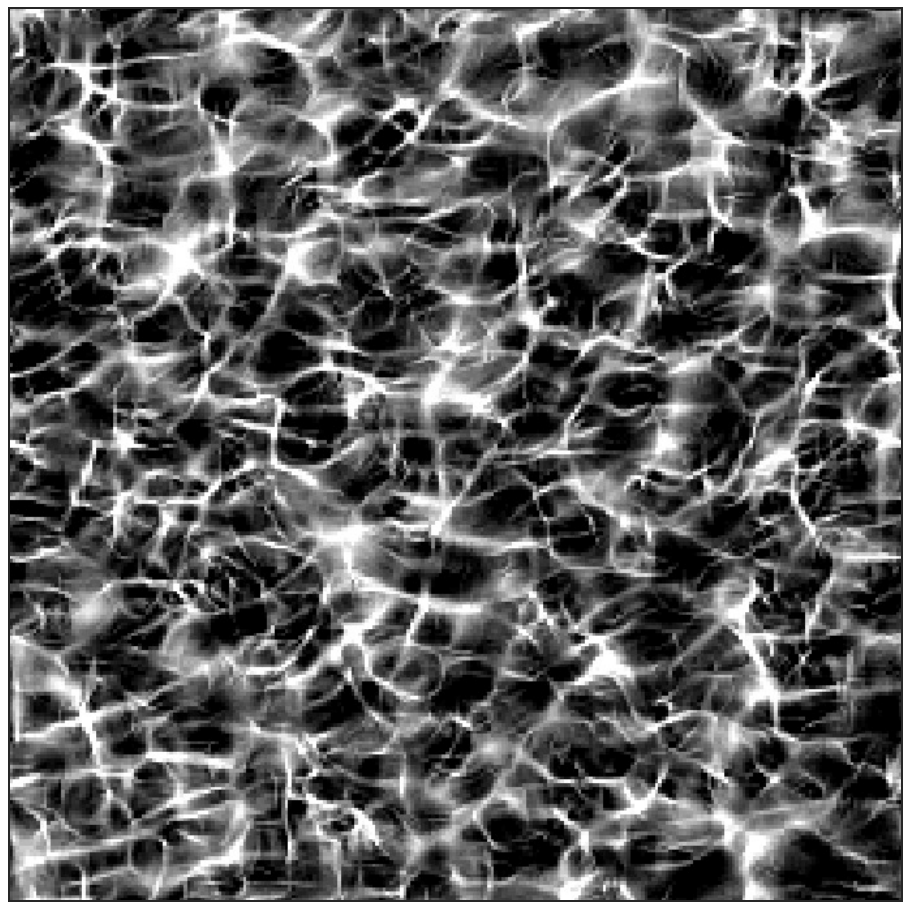}
		\includegraphics[height=4cm]{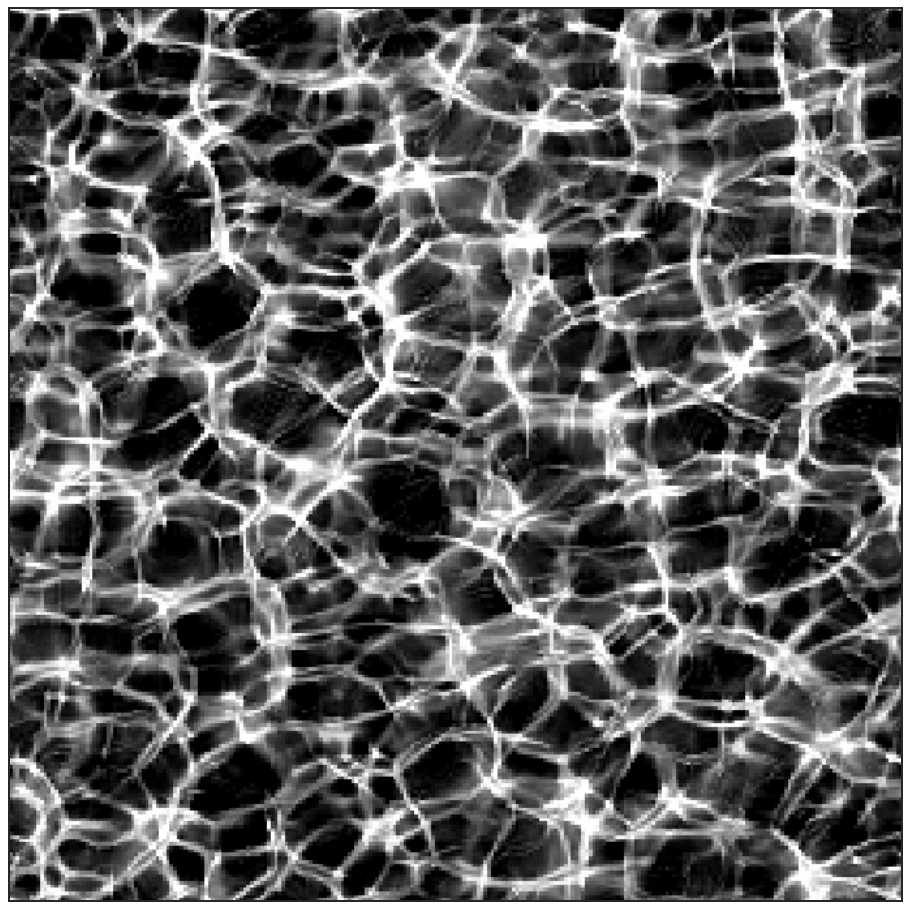}
		\includegraphics[height=4cm]{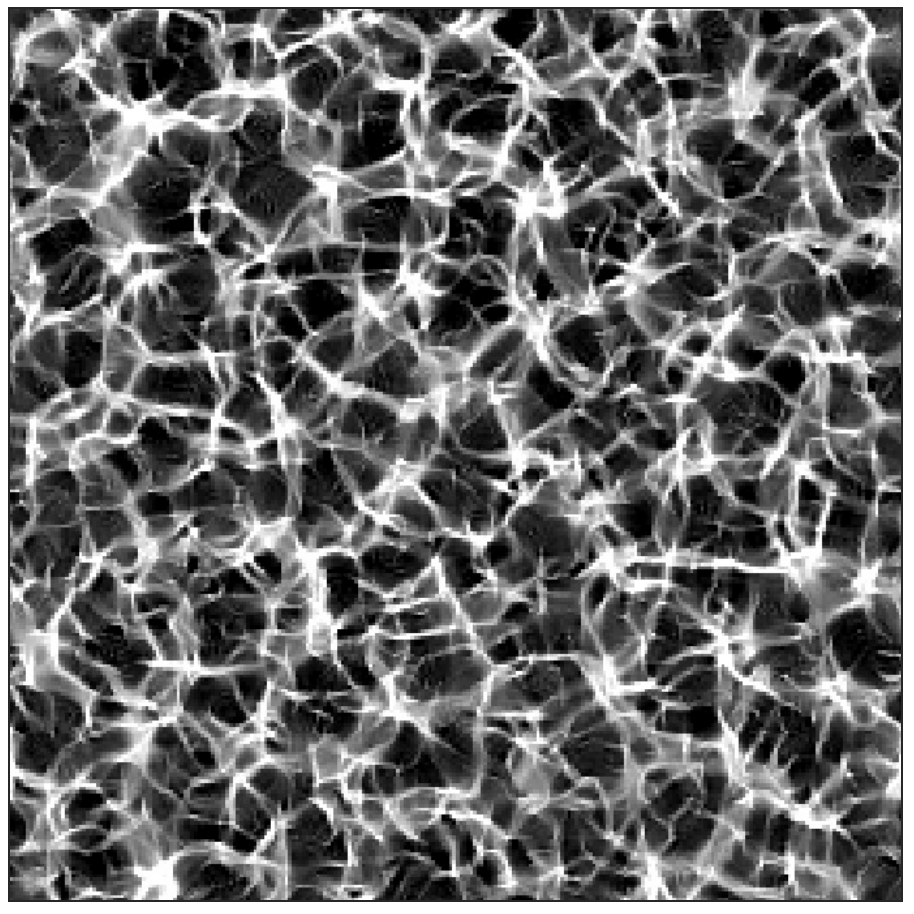}
	\end{subfigure}
	\begin{subfigure}[t]{4cm}
%		\caption{}
		\includegraphics[height=4cm]{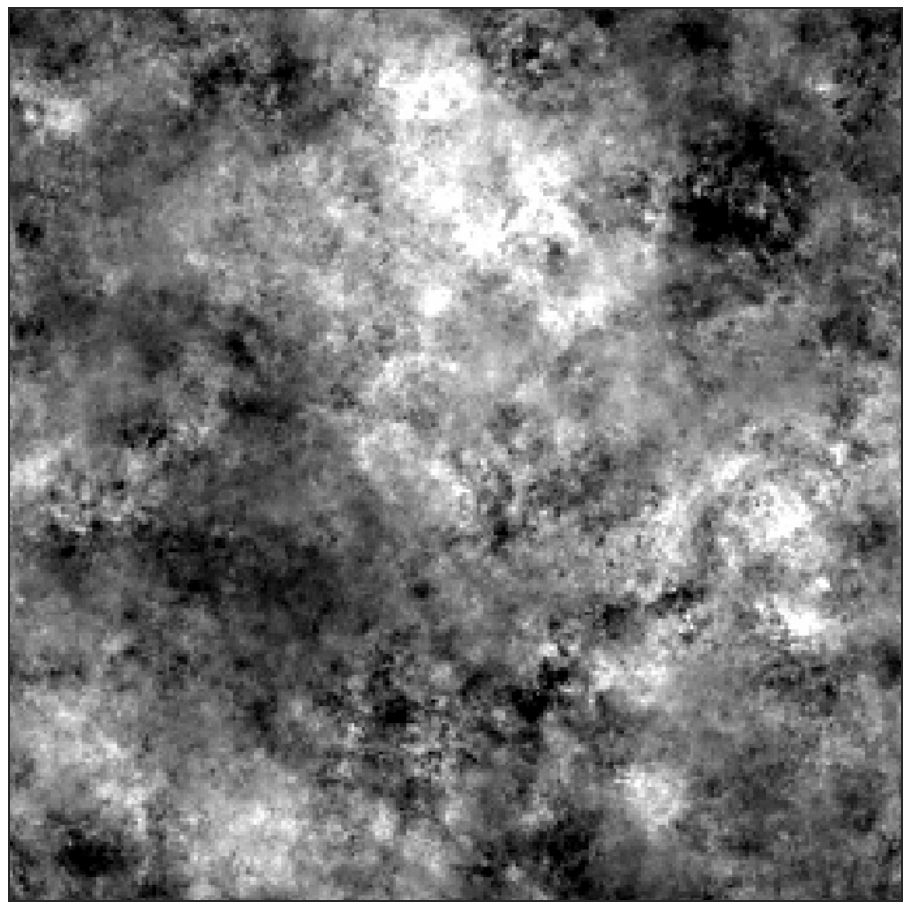}
		\includegraphics[height=4cm]{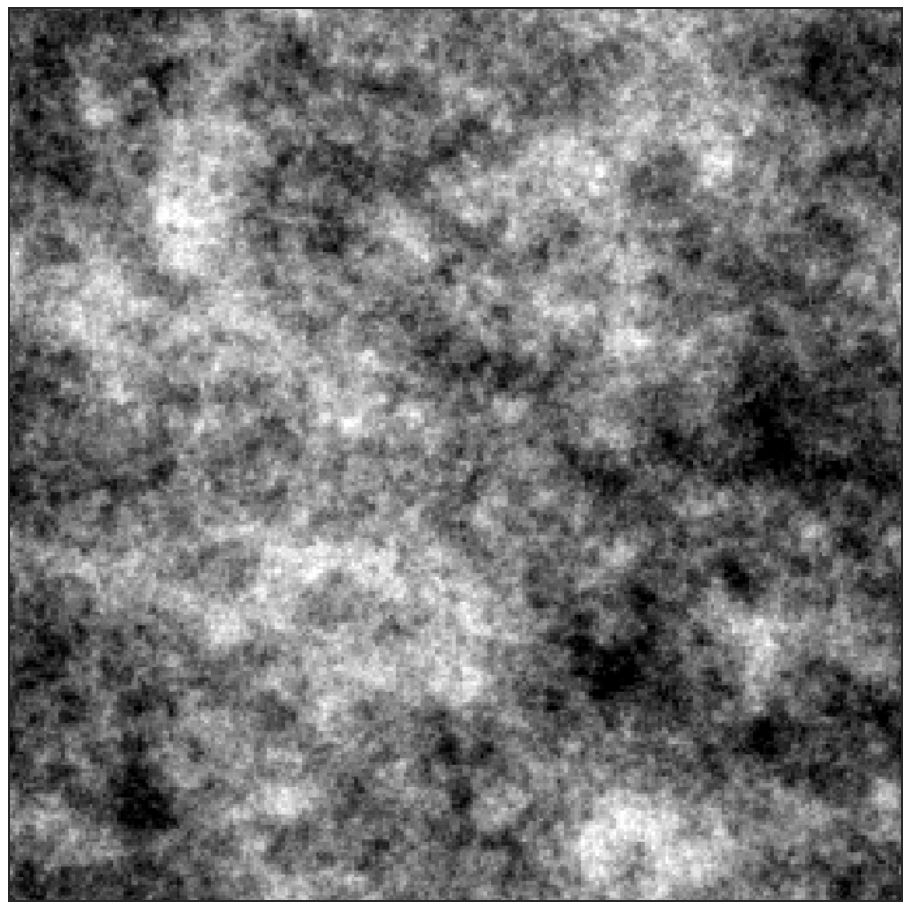}
		\includegraphics[height=4cm]{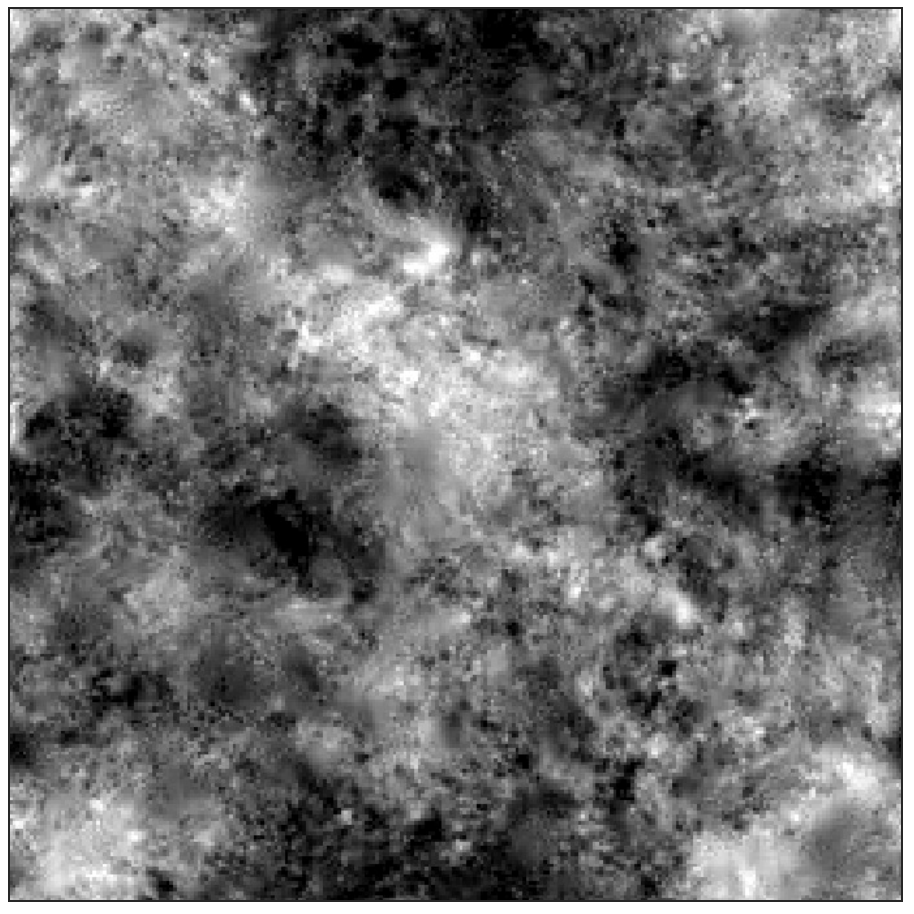}
		\includegraphics[height=4cm]{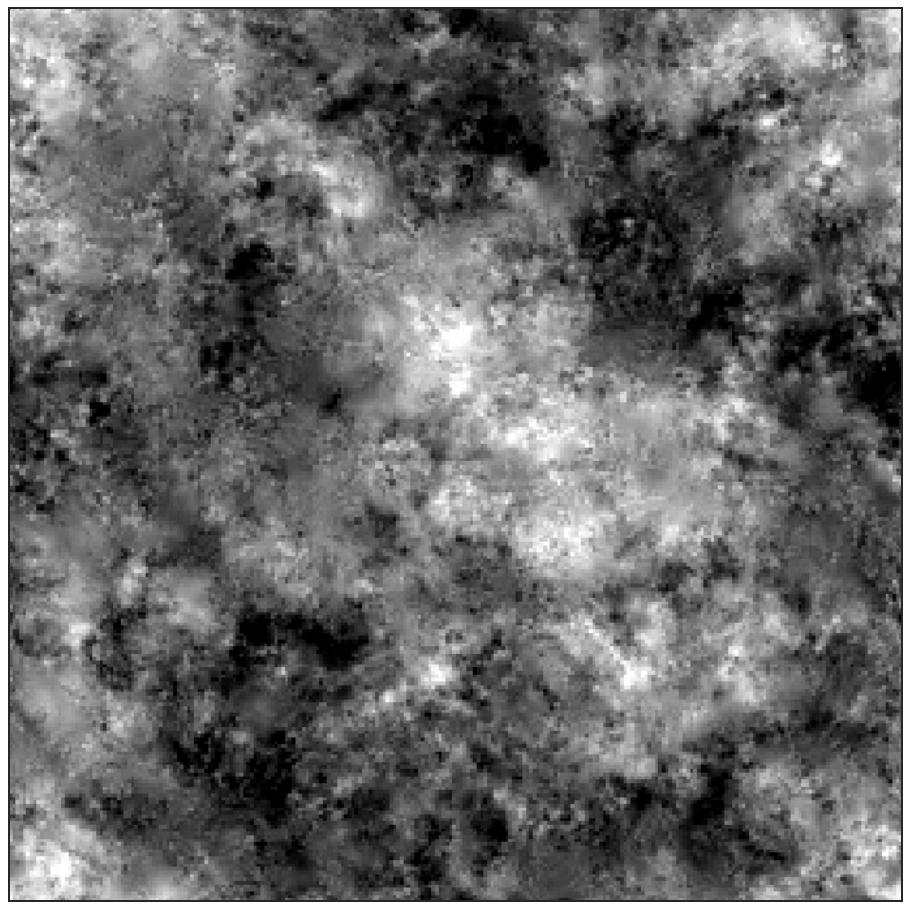}
		\includegraphics[height=4cm]{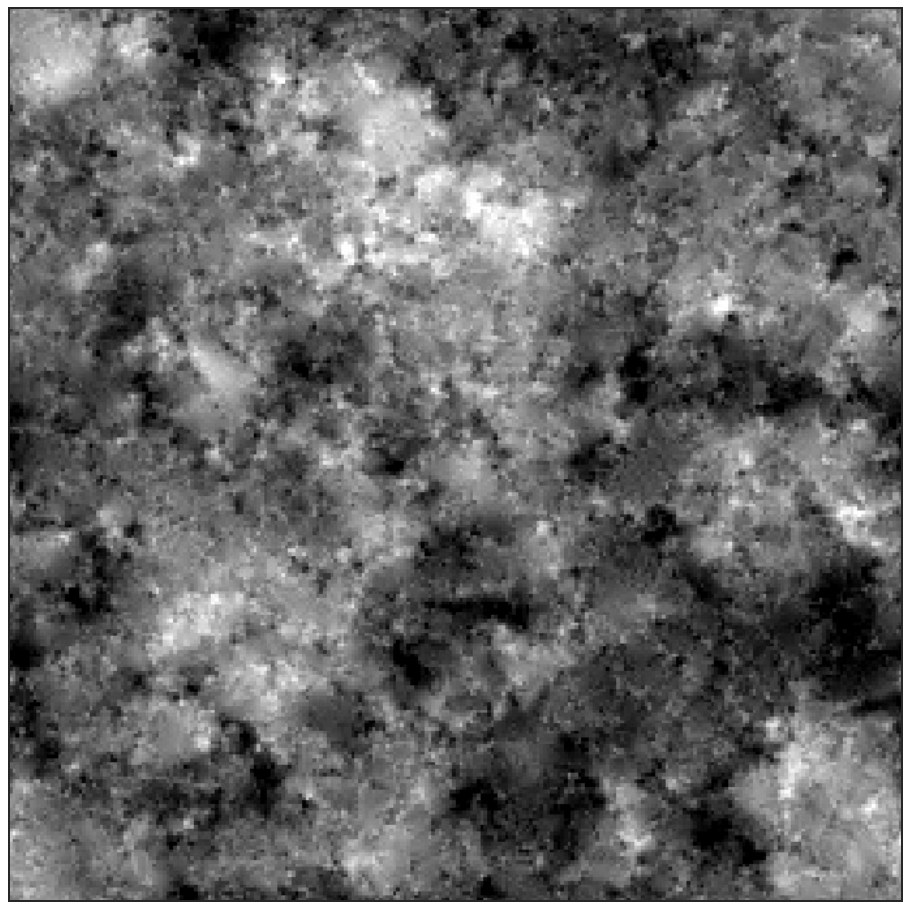}
	\end{subfigure}
}
\caption{The first row shows a realization $\bar x$ of $X$. 
The second to fifth row give realizations of microcanonical models
A, B, C, D computed from $\bar x$. Each column corresponds to a different $X$. First column: Isotropic turbulent vorticity field. 
Second column: Non-isotropic turbulent vorticity field. 
Third column: Bubbles. 
Fourth column: Multifractal Random Walk.
}
\label{fig:synthesis2}
\end{figure}

\paragraph{Visual evaluation}
The quality of different microcanonical models is first
evaluated visually. Among 10 synthesis of each model, we retain the one which yields the smallest loss in (\ref{eq:microobj}).
The top row of Figure \ref{fig:synthesis2} shows a realization $\bar x$
of different stationary processes $X$.
The first and second columns display 
isotropic and non-istotropic vorticity fields
of two-dimensional turbulent flows, with a factor 2 zoom
on the central part of each image.
The third and fourth columns show an
image of bubbles and  a realization
of Multifractal Random Walk \cite{bacry2001multifractal}.
An MRW is a self-similar random process with long range dependencies. 
We model the increments of MRW, which is a stationary process and limit
its maximum correlation scale to $2^5$.
The next rows give realizations of the microcanonical
models $A$, $B$, $C$ and $D$ computed from the same realization
$\bar{x}$ shown at the top. 

As in Figure \ref{Gausssynthesisfig},
the foveal Gaussian model $A$ looses most geometric structures.
On the contrary 
the models $B, C, D$ recover most of this
geometry. The model $B$ captures the correlations of modulus
coefficients across angles
whereas the model $C$ also impose covariance conditions
on phases. There are little visual differences on 
turbulent flows but 
it is more apparent on bubble images.
Model $B$ does not reproduce closed bubbles whereas 
model $C$ recovers bubbles having a better geometry.
The model $D$ is an isotropic version of model $C$.
Its realizations are therefore isotropic.
When $X$ is isotropic then it is visually as 
precise as $C$ but the variance reduction is not visible.
If $X$ is not isotropic, as in 
the turbulence of the second column, then model $D$ does not reproduce
the angular anisotropy.

\subsection{Evaluations of Foveal Covariance Models}
\label{evalucov}

A microcanonical model $\widetilde X$ is conditioned by values
of the covariance of $\UPhi_v (X)$ on a graph whose edges $E$ relate
neighbor vertices $(v,v')$. We differentiate two types of model errors. 
Type I errors result from the use of graph
neighborhoods which are too small.
In this case, the covariance of
$\UPhi(X)$ and $\UPhi(\widetilde X)$ may be different
for $(v,v') \in V^2 - E$.
Type II errors are due to the choice of the representation $\UPhi$.
Such errors are evaluated by comparing
high order moments of $X$ and $\widetilde X$, which are not calculated
by $\UPhi$. These comparisons are performed for $\UPhi = \what \HH \W$.

\paragraph{Wavelet harmonic covariance error}
Comparisons are performed over correlation matrices,
which normalizes covariance values. Type I errors are calculated
by comparing the correlation values of $X$ and $\widetilde X$
over a neighborhood which is much larger
than the ones used by the different models. 
This paragraph considers the case where this neighborhood 
includes all scales and angles, but over a limited spatial range.

We compute correlation
coefficients (\ref{cofrrelasts}) by normalizing
$K_{\UPhi}$ with its diagonal:
\begin{equation}
\label{corrmat}
C_{\UPhi} = D_{\UPhi}^{-1/2} \,
K_{\UPhi}\, D_{\UPhi}^{-1/2} \,~~\mbox{where}~~
D_{\UPhi} = {\rm diag}(K_{\UPhi}).
\end{equation}
We estimate $K_{\UPhi}$ with an empirical average over
$100$ realizations of $X$.
This gives an accurate estimation of both $D_{\UPhi}$ and $C_{\UPhi}$. 
An empirical estimator $\wtilde C_{\UPhi \bar x}$ of
$C_{\UPhi}$ is computed from a single realization $\bar x$ of $X$,
with the same normalization 
\[
\wtilde C_{\UPhi \bar x} =  D_{\UPhi}^{-1/2} \,
\wtilde K_{\UPhi \bar x}\,  D_{\UPhi}^{-1/2}.
\] 
It is calculated on $\Ga^2_0$ where 
$\Ga_0 \subset \Ga$ is a foveal subset of all vertices. It incorporates
correlations between all scales and angles, across a limited spatial
range.

Let $\| C_{\UPhi } \|_{op,\Ga^2_0}$ be the operator norm and hence the largest eigenvalue of the symmetric matrix 
$C_{\UPhi }$ restricted to $\Ga^2_0$. 
The following empirical error measures the estimator error of
$ C_{\UPhi}$ from one realization $\bar x$
\begin{equation}
\label{error20}
\epsilon_{emp} = \frac {  \|  C_{\UPhi} - \wtilde C_{\UPhi  \bar x} \|_{op,\Ga_0} }
{\| C_{\UPhi } \|_{op,\Ga_0}} .
\end{equation}
It is a variance term
due to variabilities of realizations $\bar x$ of $X$. 

\begin{table}
\begin{center}
\begin{tabular}{ |c|c|c| }
\hline
		& Isotropic  &  Non Isotropic    \\
\hline
$\epsilon_{emp}$ &  0.58 (0.05) & 0.68 (0.09)    \\
\hline
$\epsilon_{A}$ & 0.82 (0.01)    &  0.80 (0.01)   \\
\hline
$\epsilon_{B}$ & 0.29 (0.02)     &  0.30 (0.02)   \\
\hline
$\epsilon_{C}$ & 0.24 (0.02)  &  0.25 (0.03)    \\
\hline
$\epsilon_{D}$ & 0.25 (0.02) &  1.6 (0.09)  \\
\hline
\end{tabular}
\end{center}
\caption{Covariance errors (\ref{error2}) 
models A,B,C,D compared to the empirical error 
(\ref{error20}), for the isotropic and non-isotropic 
vorticity fields.
}
\label{tbl:fovelcov}
\end{table}

This empirical error is compared to the estimation error of
$C_{\UPhi}$ computed from the different microcanonical models.
Let $K^{model}_{\UPhi\bar x}$ be 
the covariance matrix of wavelet
harmonic coefficients of a microcanonical model $\tilde X$ instead of $X$. This matrix is estimated from $10$ realizations of $\tilde X$.
We compare $K^{model}_{\UPhi\bar x}$ with $\wtilde K_{\UPhi \bar x}$ by using
the same normalization. We define
$C^{model}_{\UPhi\bar x}$ by normalizing the $K^{model}_{\UPhi\bar x}$
with $D_{\UPhi}$ as in (\ref{corrmat}).
The relative error of the model is defined by
\begin{equation}
\label{error2}
\epsilon_{model} = \frac {  \| 
C_{\UPhi } - C^{model}_{\UPhi \bar x} \|_{op,\Ga_0}}
{\| C_{\UPhi} \|_{op,\Ga_0}} .
\end{equation}
This error has a variance term because the microcanonical model
depends upon a particular realization $\bar x$ of $X$. It 
has also a bias term because the model is calculated from 
covariances over a limited set $E_G \subset \Ga_0^2$.
Optimizing $E_G$ is a trade-off between the variance and bias terms.
If the sufficient statistic set $\LaG \subset \Ga_0^2$ is too small and
is unable to reproduce the correlations 
$C_{\UPhi } $ in $ \Ga_0^2 - \LaG $ then 
$\epsilon_{model}$ is larger than $\epsilon_{emp}$. 
If $ \LaG = \Ga_0^2$ then the variance term dominates
and $\epsilon_{model} = \epsilon_{emp}$.

We define $\Ga_0$ with
$k_{min}=0$, $k_{max}=4$, a maximum
range of scales $\Delta_j = J$ and a maximum range of angles 
$\Delta_\ell = Q/2$, but a small 
translation range $\Delta_n = 2$. 
Its size is $|\Ga_0| = 8025$, so errors are evaluated over
$8025 \times 8025$ correlation values.
Table \ref{tbl:fovelcov} compares the empirical
error $\epsilon_{emp}$ in (\ref{error20}) with the
model error $\epsilon_{model}$ in (\ref{error2}), for models
$A, B, C, D$. We report the
mean of these estimated errors 
by averaging the operator norms
over $10$ realizations $\bar x$ of $X$.
The standard deviation is given in brackets.
These error values are 
consistent with visual evaluations.

Table \ref{tbl:fovelcov} show 
that the Gaussian model A gives a larger error than $\epsilon_{emp}$, 
because
it captures no dependence across scales and angles. It introduces
a large model bias.
For models $B$ and $C$, $\epsilon_{model}<\epsilon_{emp}$. 
The increase of $E_G$ dramatically reduces the model bias, which
clearly appears visually.
These models are only conditioned
by wavelet harmonic covariances over neighbor scales, 
but the microcanonical 
model propagates these constraints across all
scales. It provides a good approximation of covariance values between
far away scales, evaluated in $\Ga_0^2$.
The error of model $C$ is smaller than model $B$, 
suggesting that scale correlations through phase also
play an important role. This also appears in the visual quality of
synthesized bubble images.
The model $D$ has a similar error as model $C$ for the isotropic 
turbulence although it has
a lower variance because estimators are averaged across $Q$
angular directions. This 
indicates that the error is dominated by the model bias term and hence
that a better model would be obtained by further increasing $E_G$. 
This is verified on model $D$ 
by increasing $\Delta_j=J$ and $k_{max}=4$, which
yields a smaller covariance error 0.20 (0.02) 
on the isotropic turbulence.
On the contrary, the model $D$ has an error which is much larger
than model $C$ on the non-isotropic turbulence, which also appears visually.

\paragraph{Long-range spatial correlations}
Our microcanonical models are computed with foveal sufficient statistics sets, which are conditioned over relatively small
spatial neighborhood at each scale $2^j$. Such models can therefore
introduce a large error if there exists long range spatial
correlations between wavelet harmonic coefficients. This paragraph
evaluates such errors.

We compare 
$C_{\UPhi }(v,v')$ and $C^{model }_{\UPhi \bar{x}}  (v,v')$ for 
$v = (\la,k,u)$ and $v' = (\la',k',u')$
with $\la = \la'$, $k = k'$ and for large $|u-u'|$. 
We estimate an average value  
$C^{model}_{\UPhi}$ of $C^{model}_{\UPhi \bar{x}}$ over
$10$ realizations $\bar{x}$. 
The difference between $C^{model}_{\UPhi}$ and $C_{\UPhi }$ 
corresponds to a bias error term.
For $\la = 2^{-j} r_{-\ell} \xi$ we define
$\overline C_{\UPhi} (k,j,a)$ and
$\overline C^{model}_{\UPhi} (k,j,a)$ 
as the maximum values of
$C_{\UPhi }(v,v')$ and $C^{model }_{\UPhi }  (v,v')$ 
over all $\ell$ and all $(u,u')$ at a distance $|u-u'| = 2^j a$,
for $k$ and $j$ fixed. 

\begin{figure}[h]
\centering
	\begin{subfigure}[t]{7cm}
		\caption{}
		\includegraphics[width=3cm]{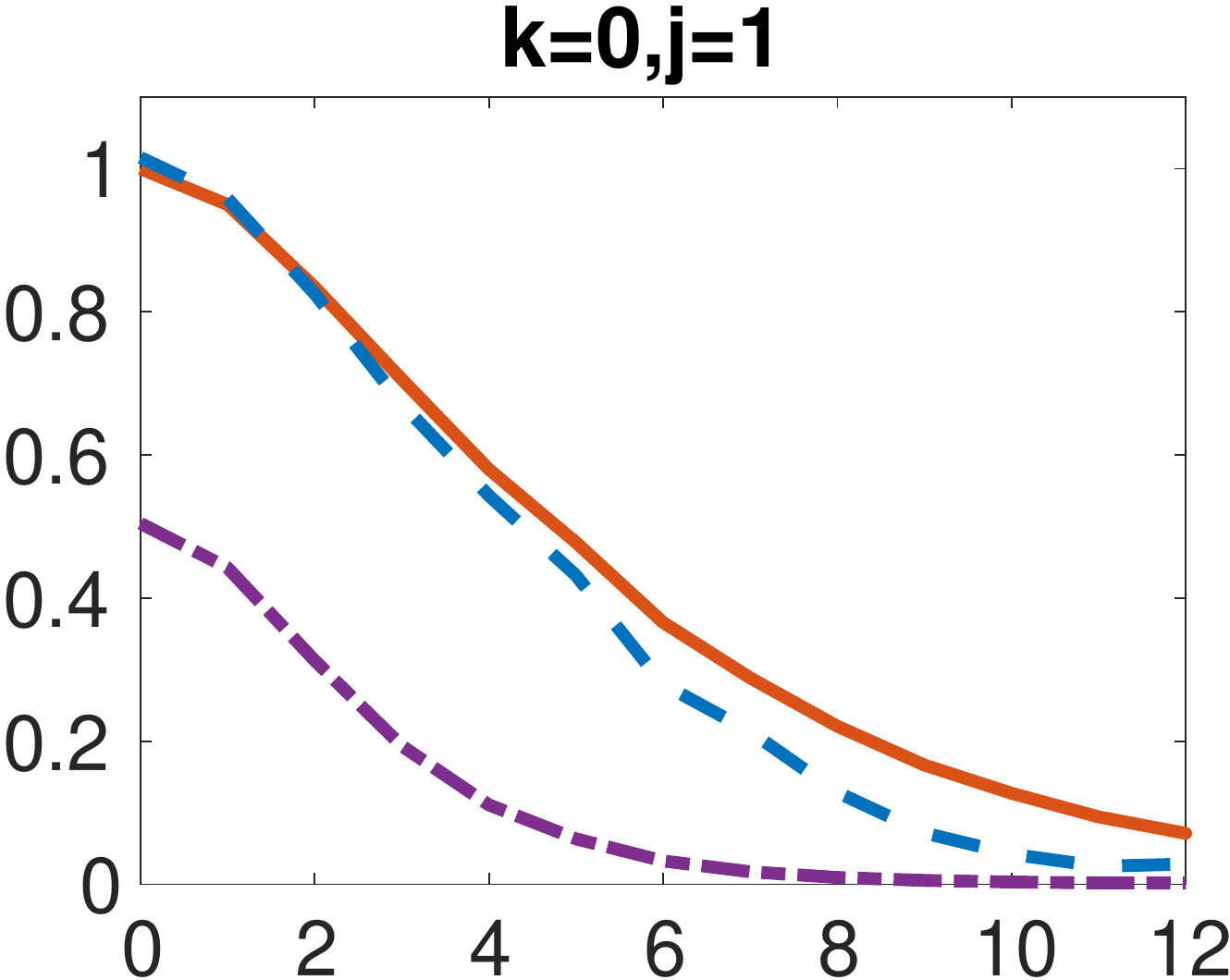}
		\includegraphics[width=3cm]{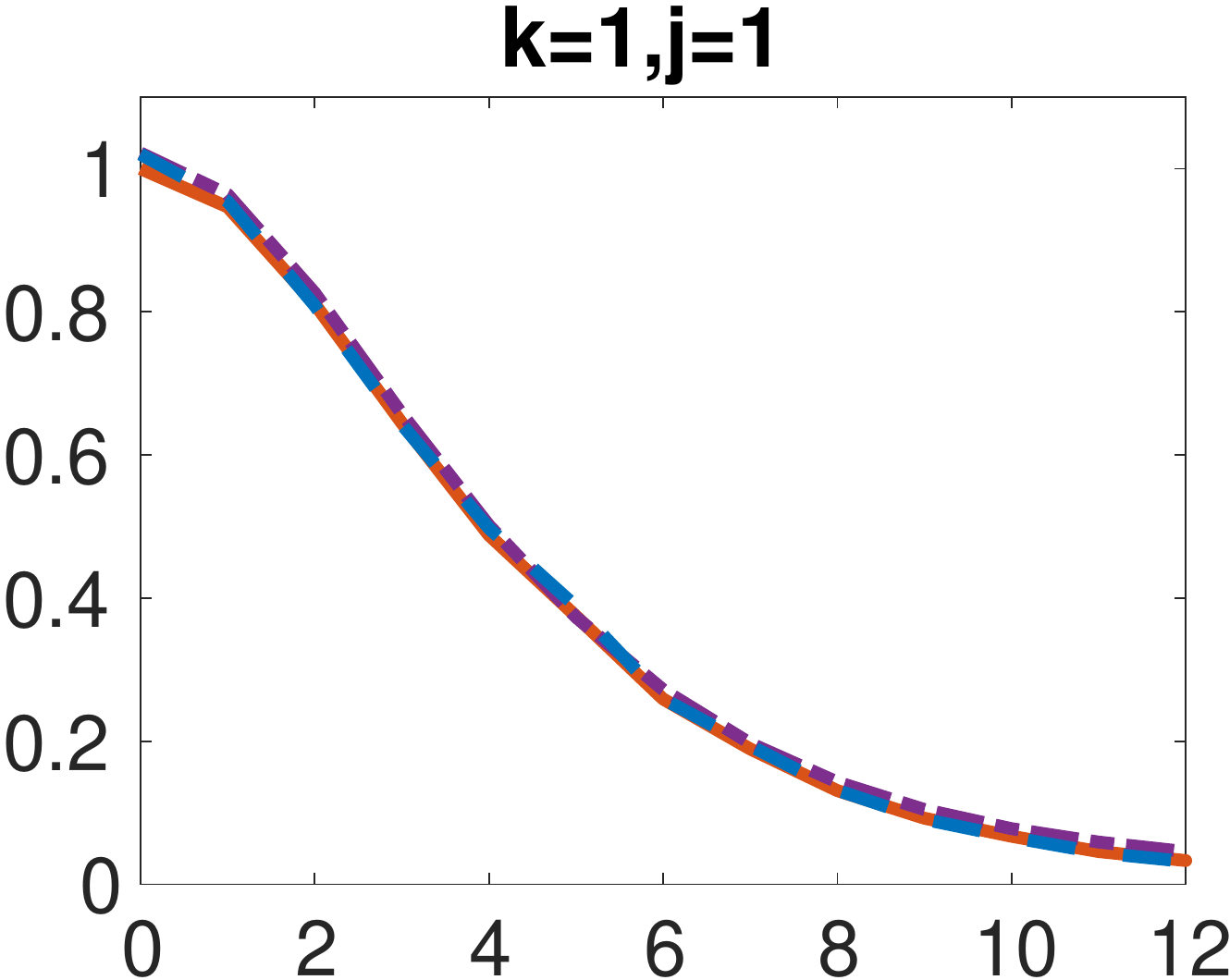} \\
		\includegraphics[width=3cm]{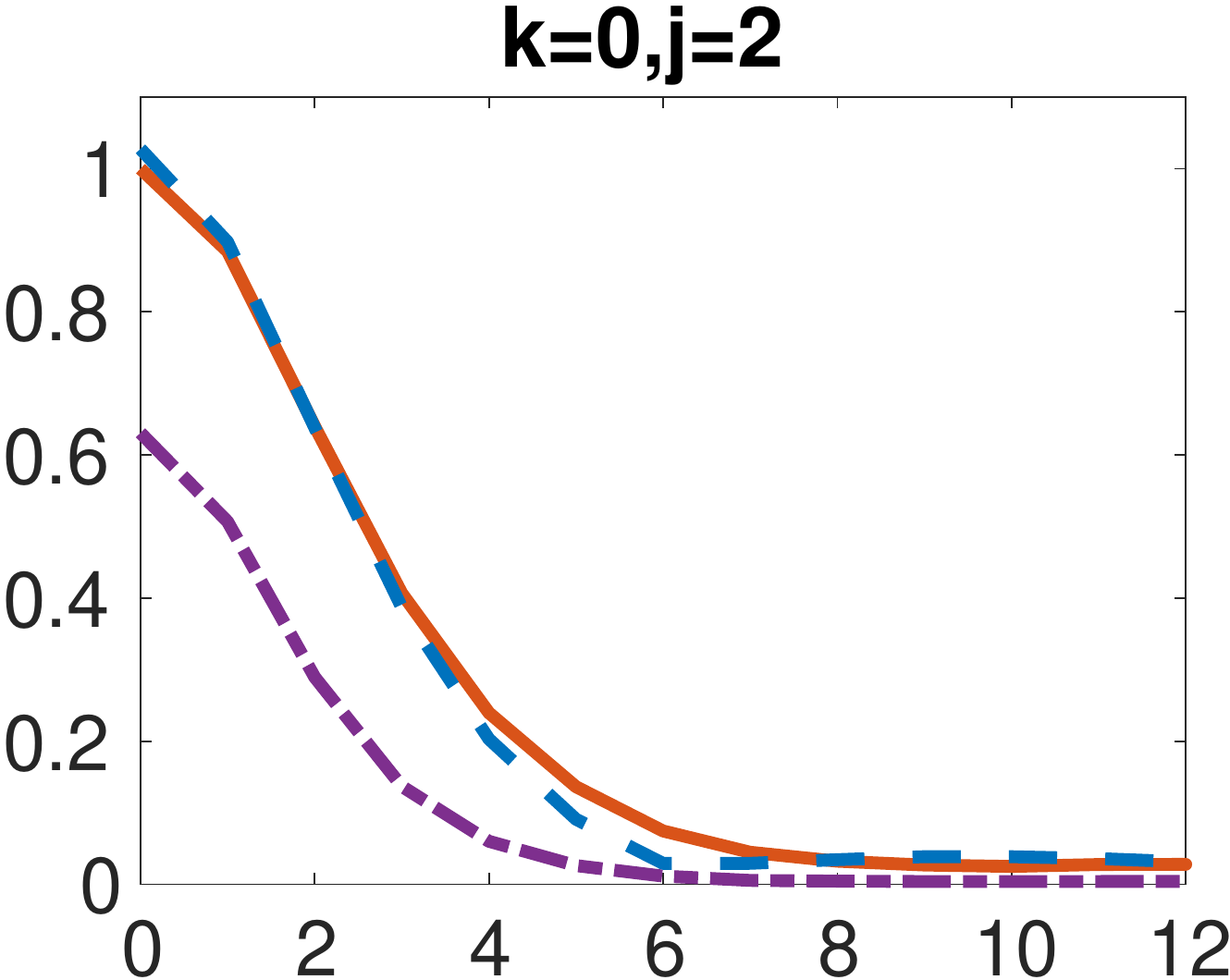}
		\includegraphics[width=3cm]{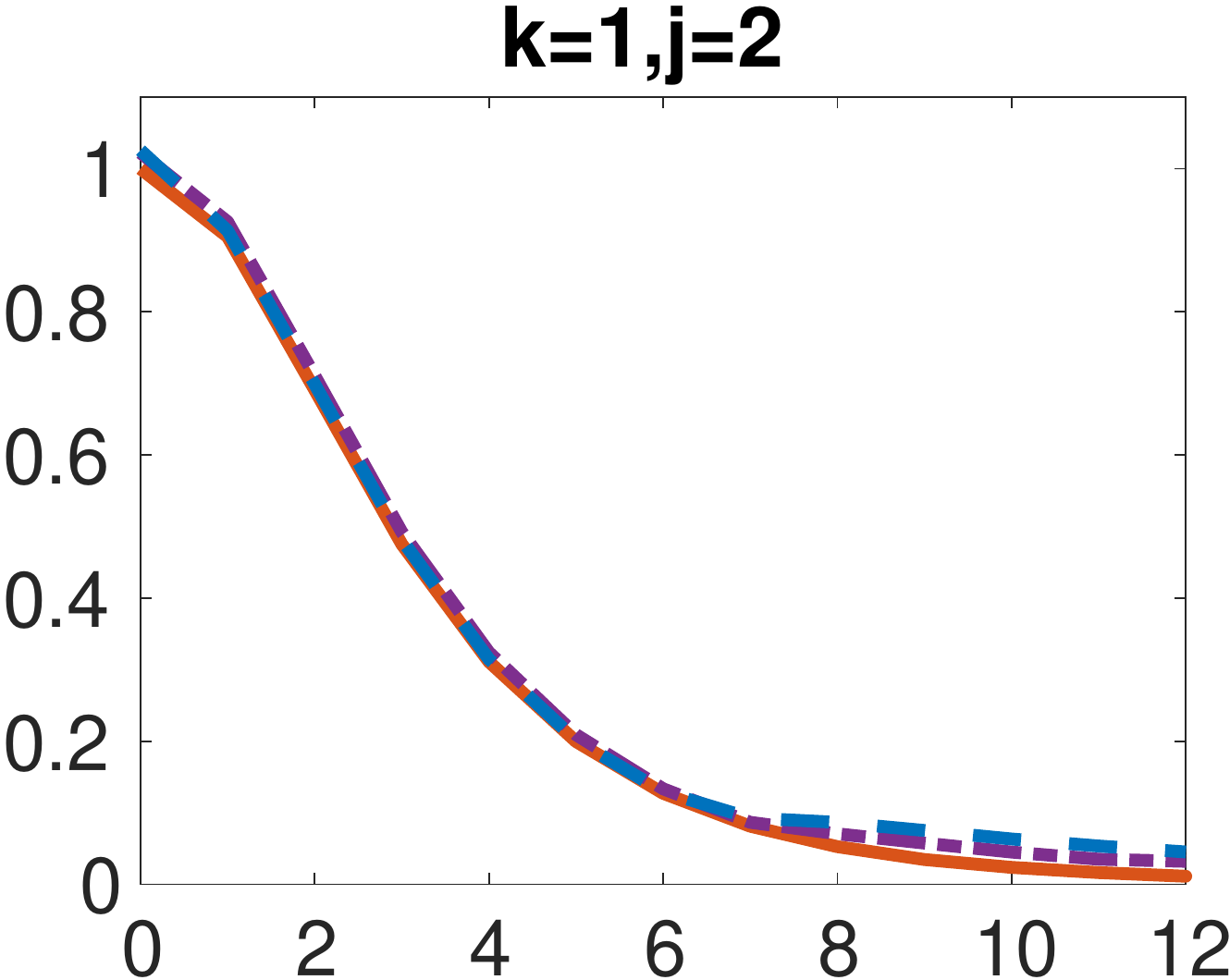} \\
		\includegraphics[width=3cm]{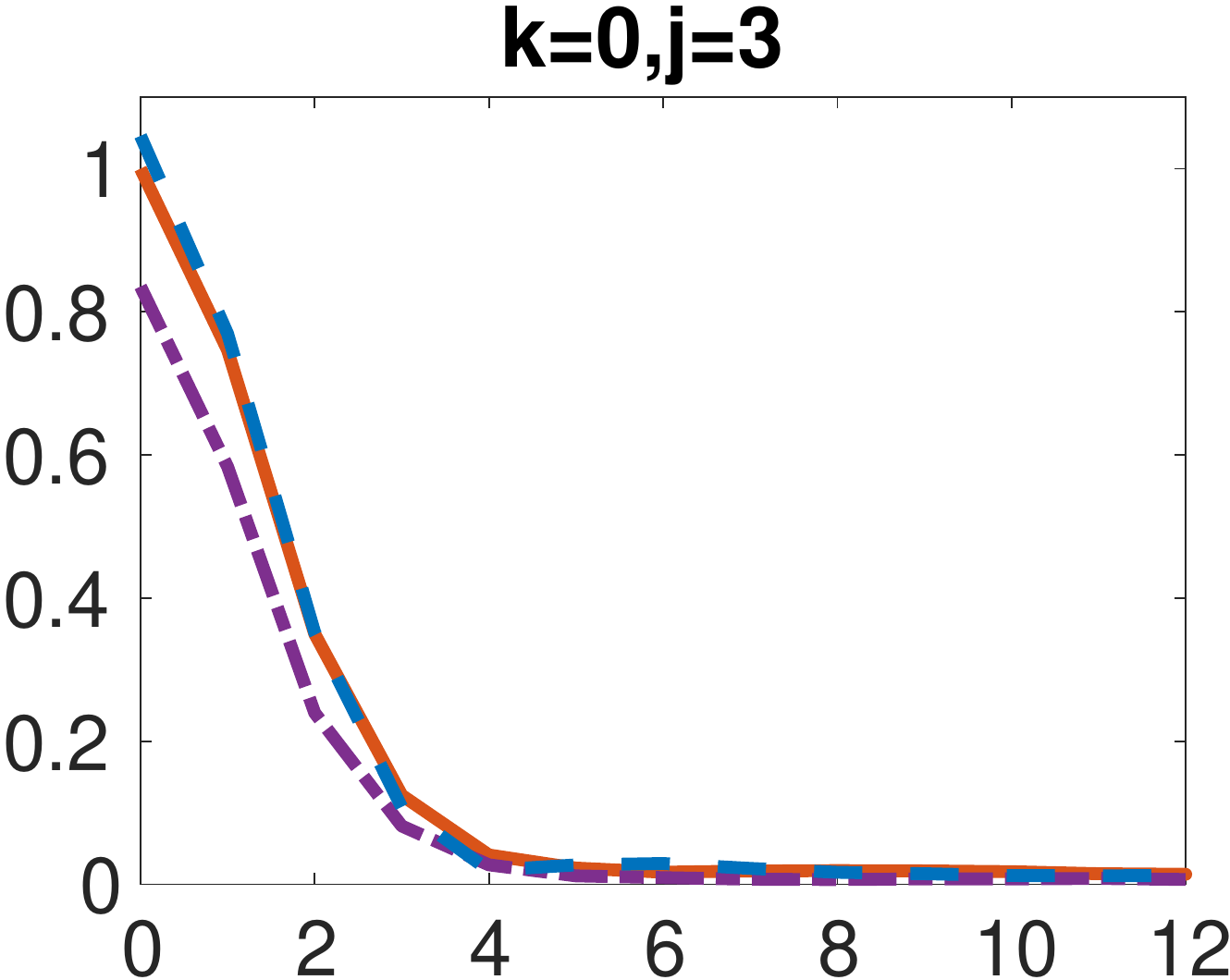}
		\includegraphics[width=3cm]{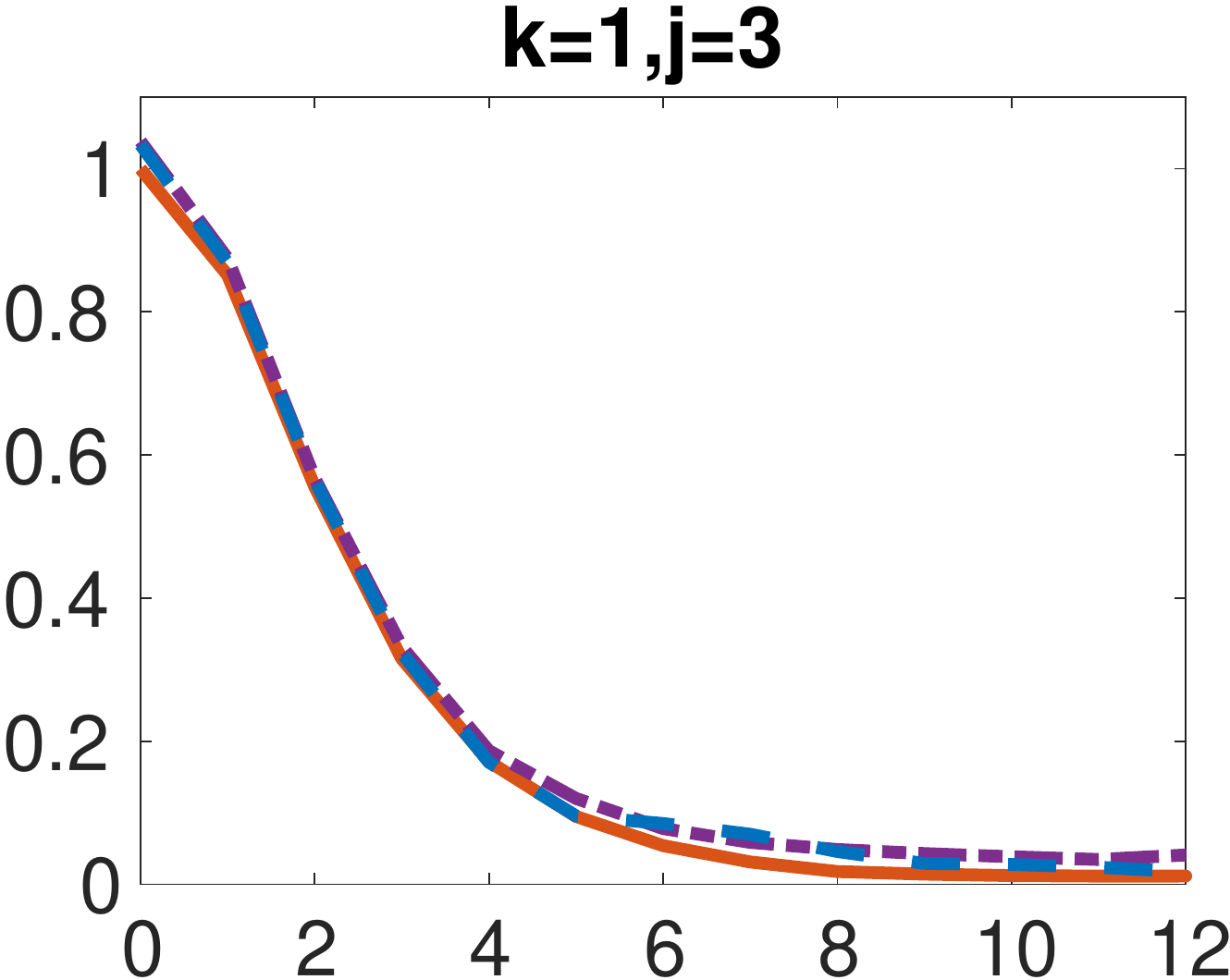}
	\end{subfigure}
	\begin{subfigure}[t]{7cm}
		\caption{}
		\includegraphics[width=3cm]{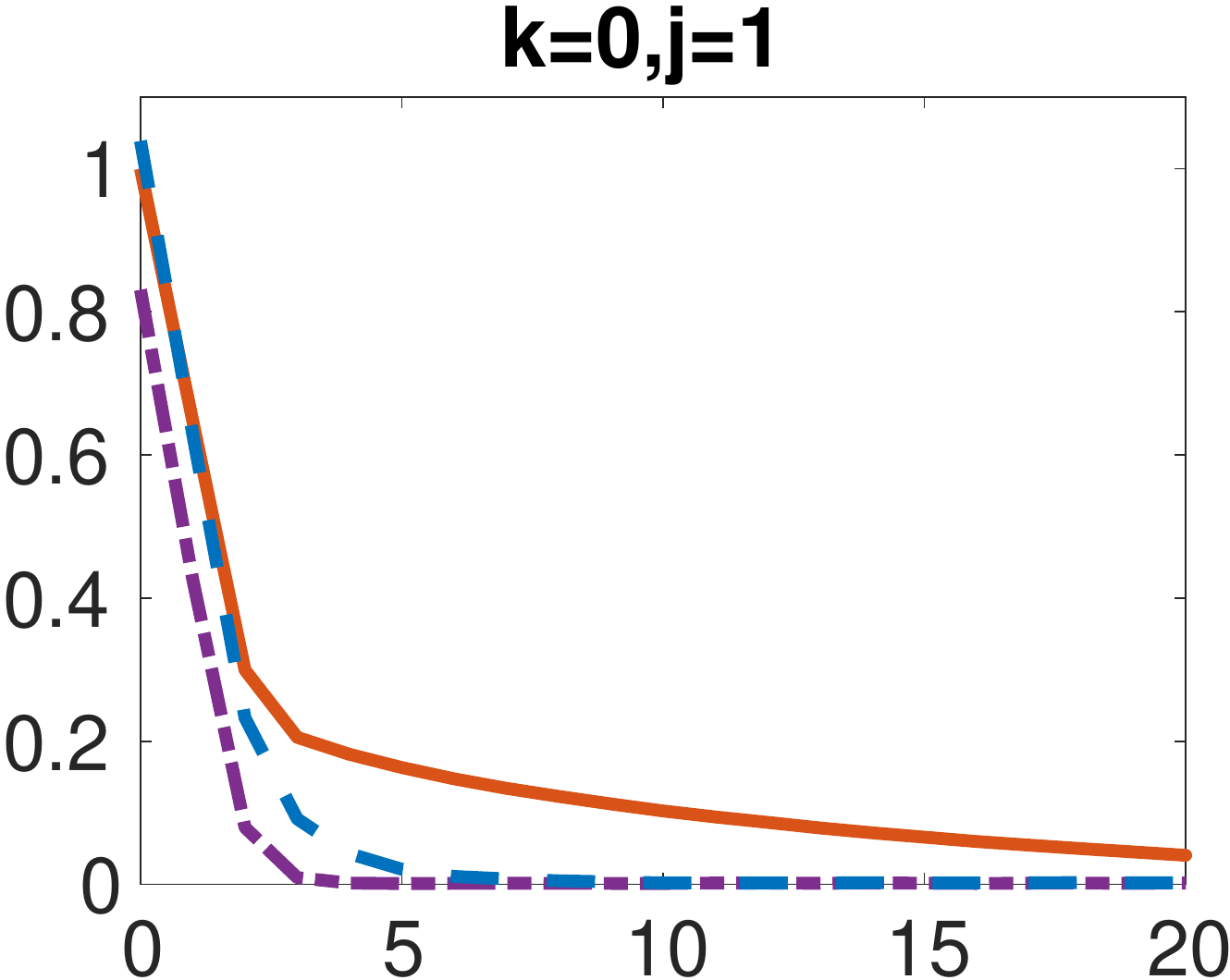}
		\includegraphics[width=3cm]{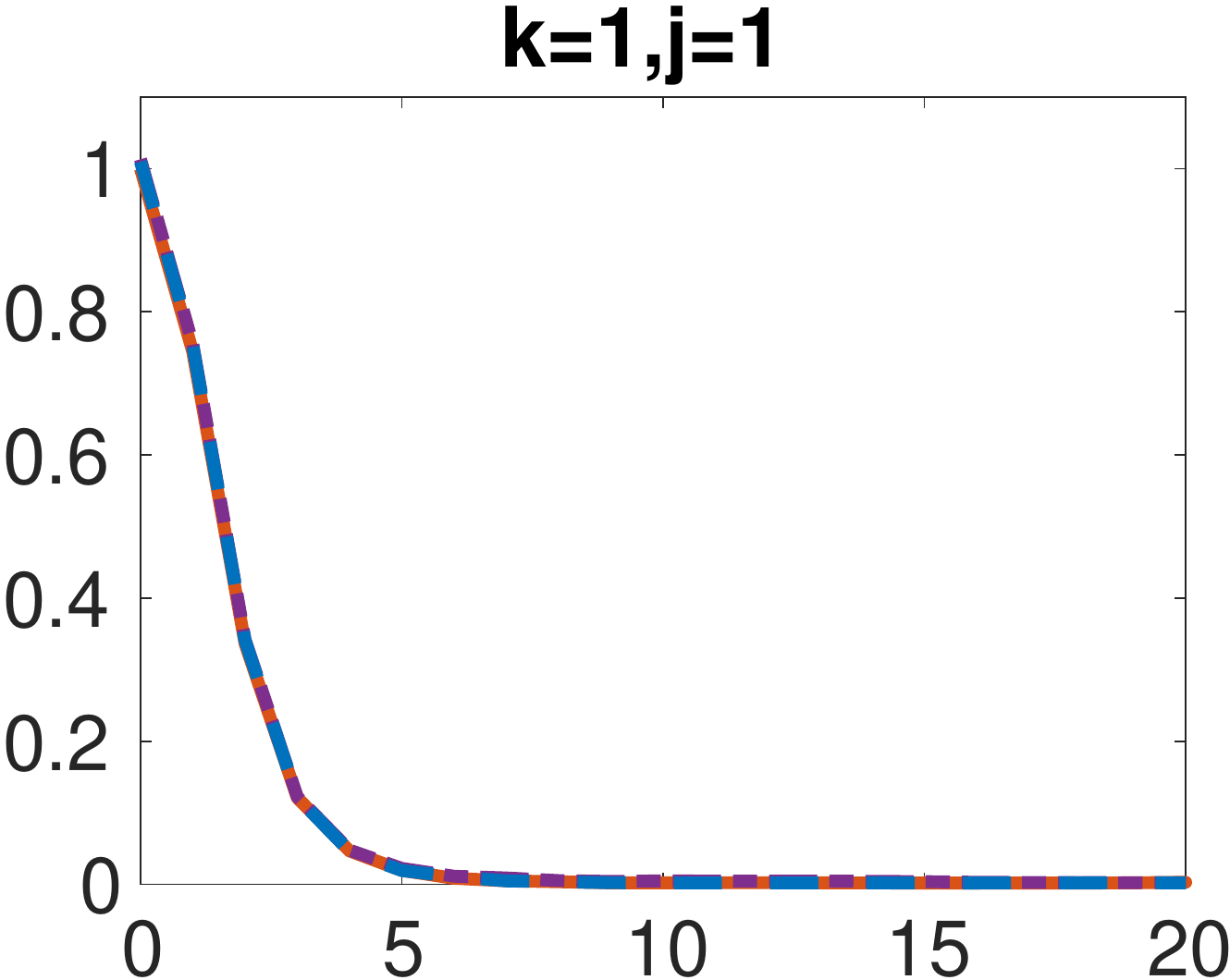} \\
		\includegraphics[width=3cm]{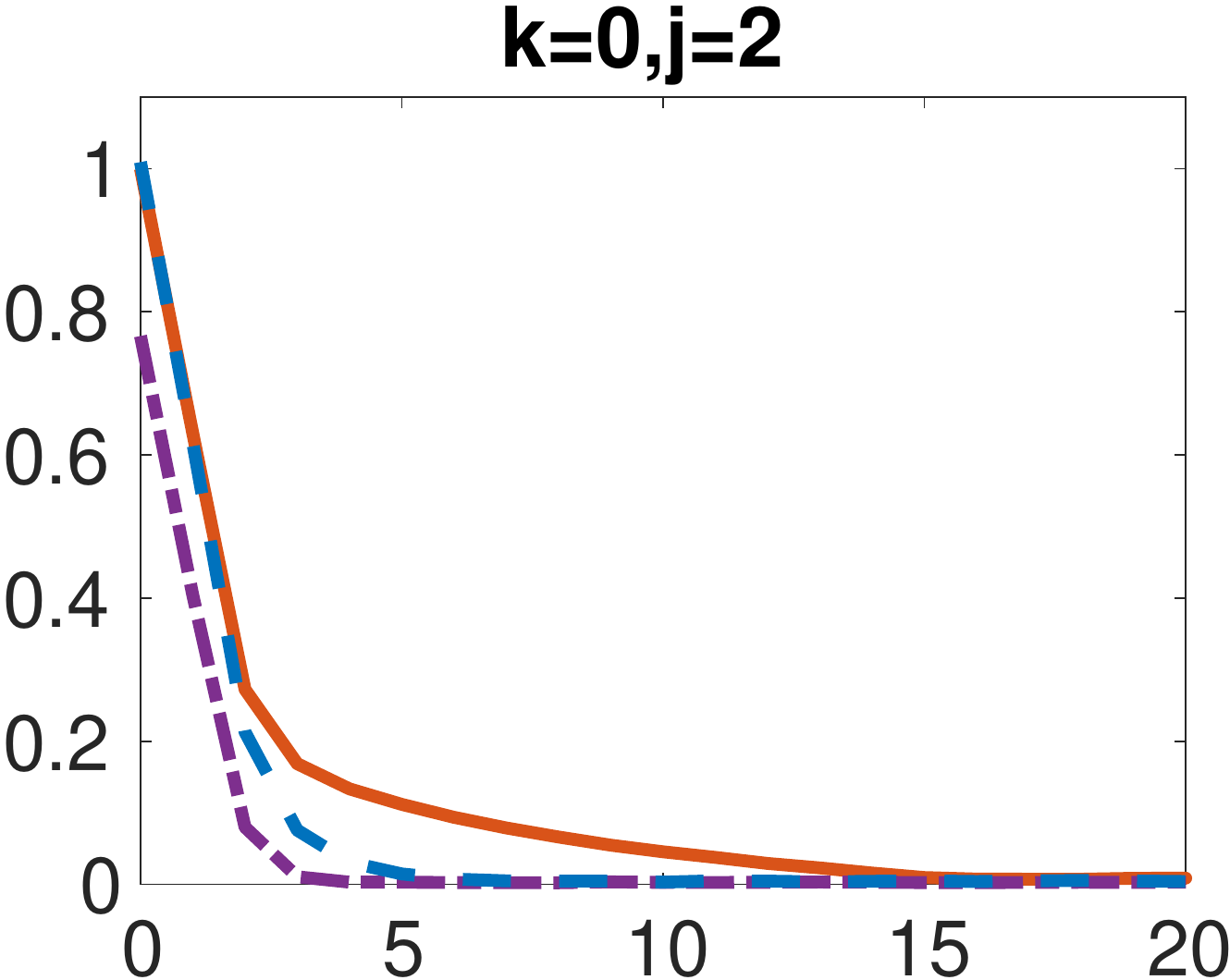}
		\includegraphics[width=3cm]{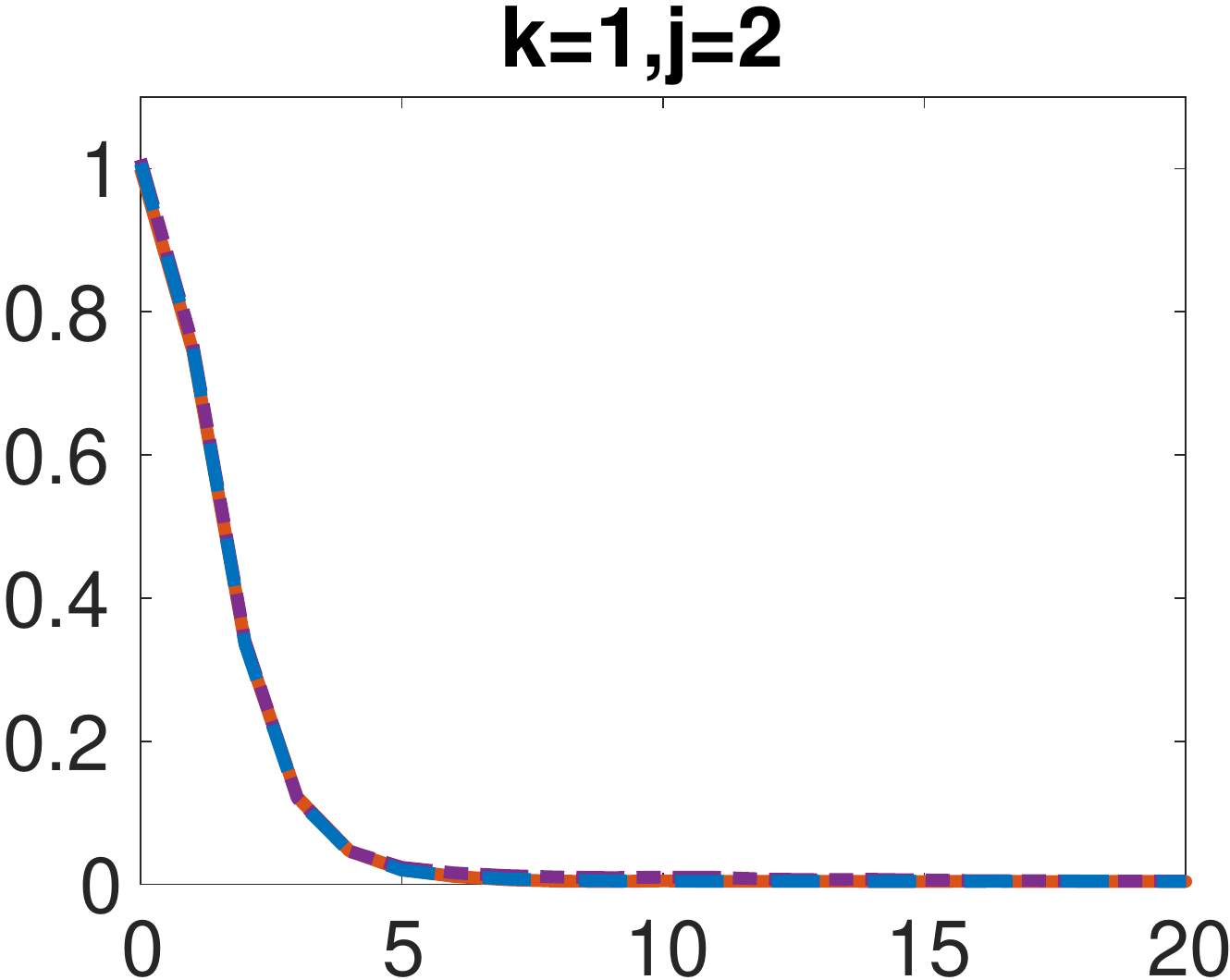} \\
		\includegraphics[width=3cm]{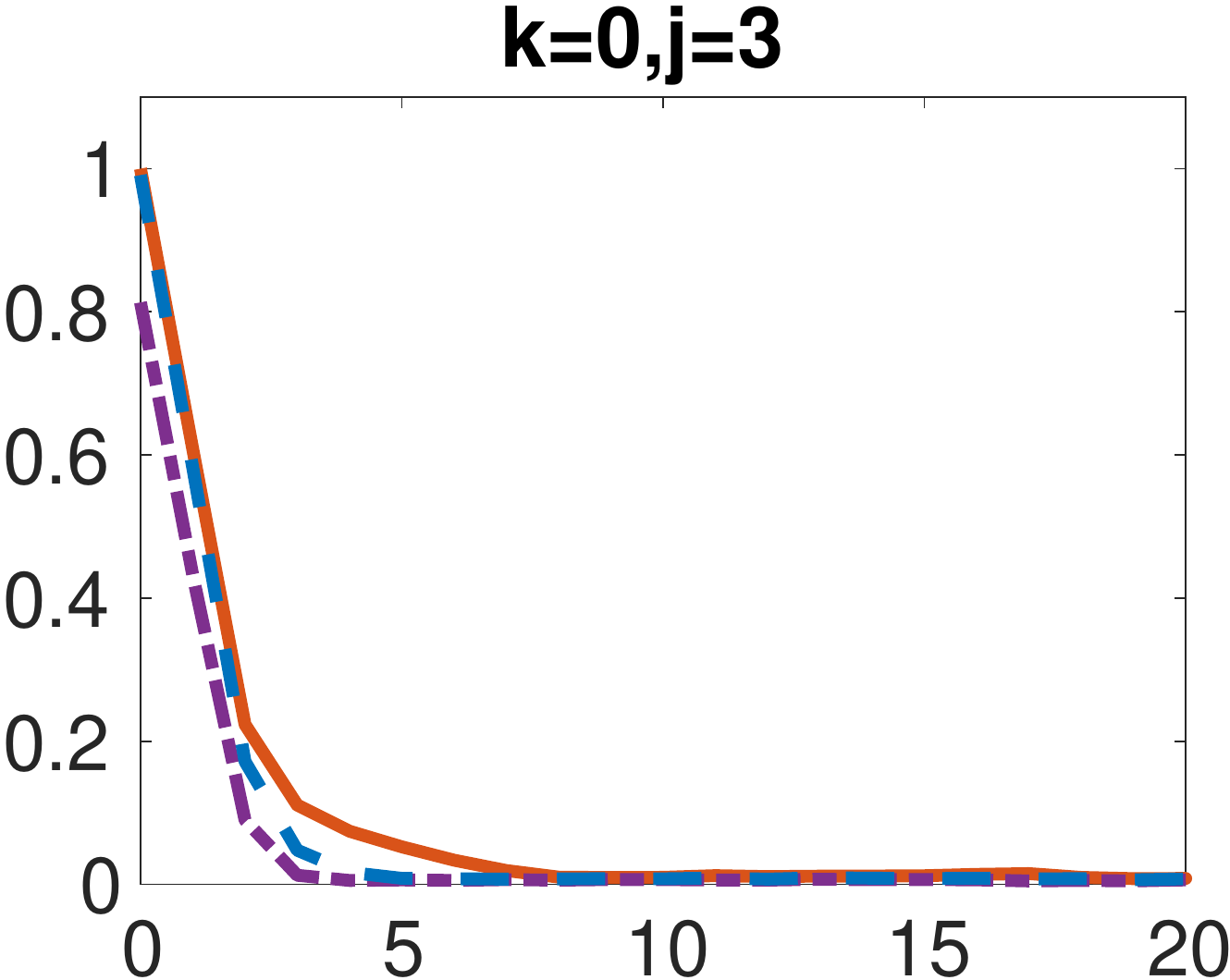}
		\includegraphics[width=3cm]{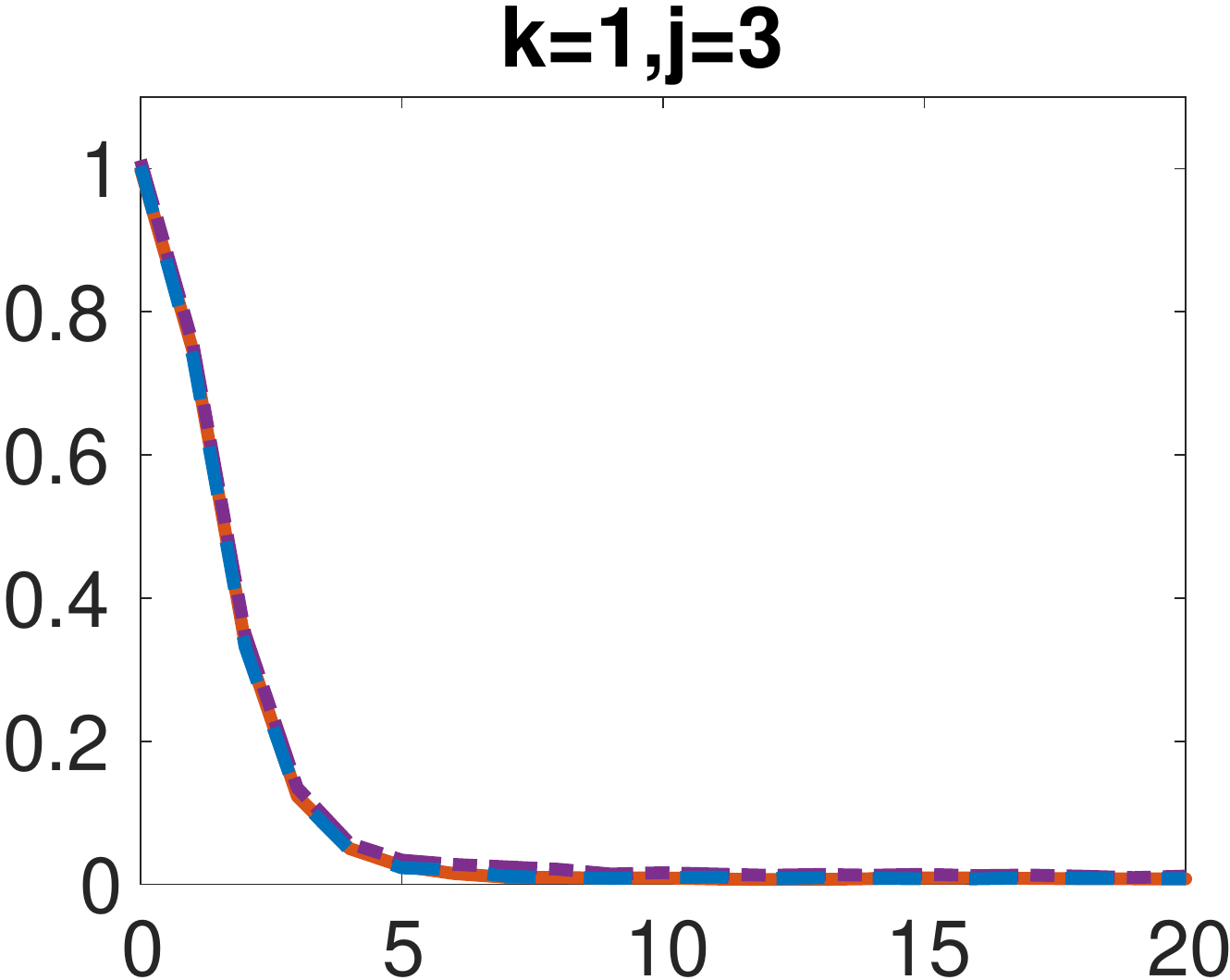}
	\end{subfigure}
\caption{Each graph shows correlation values 
$\overline C_{\UPhi} (k,j,a)$ in full line, as a function of the normalized
distance $a$ for a fixed scale $2^j$ and $k$. The values of
$\overline C_{\UPhi}^{model} (k,j,a) $ are shown
with dash-dot lines for model $A$ and dashed lines for model $D$. 
The value of $k$ and $j$ is given at the top of each graph.
The graphs in (a) are computed for isotropic turbulences and in
(b) for increments of a Multifractal Random Walk.}
\label{fig:corr2}
\end{figure}

Figure \ref{fig:corr2} 
compares the correlation values $\overline C_{\UPhi} (k,j,a)$ and
$\overline C^{model}_{\UPhi} (k,j,a)$ as a function of the normalized
distance $a$, for $k = 0,1$ and $j=1,2,3$.
The notion of short and long range correlations
is defined relatively to each scale $2^j$ through the parameter $a$.
For $k = 1$, it corresponds to correlations of wavelet coefficients,
which have a fast decay when the distance $a$ increases. 
At all scales $2^j$ the long range
correlations of wavelet coefficients for
models $A$ and $D$ are close to the correlations obtained with the
original turbulence and MRW processes.

For $k = 0$, Figure \ref{fig:corr2} 
gives the long range 
correlation of the modulus of wavelet coefficients.
It still has a fast decay for the turbulence field 
but a slow decay for the
MRW. The model $D$ gives a much better approximation 
of long range spatial correlations of turbulence
than model $A$. However there is a residual error at the finest
scale $j=1$ because these foveal models do not capture long range
correlations at fine scales. 
The error is more dramatic for the MRW process 
where the modulus of wavelet coefficients are correlated over much
longer ranges at fine scales.  
As expected, this evaluation shows that long range correlations at 
different scales are not fully
captured by foveal models which are constrained on a correlation range
of the order of $2^j \Delta_n$ at each scale $2^j$. 
One could incorporate 
these long range spatial covariances by increasing $\Delta_n$
but it would increase considerably the model size which is proportional
to $(2 \Delta_n+1)^2$.

\paragraph{High order moments}
Moments of multiple order $q$ can be measured 
with structure functions at different scale $2^j$ \cite{van1996structure,farge1999turbulence}. 
We compute the maximum moment of order $q$ of increments between points $(u,u')$ such that $|u-u'| = 2^{j-1}$
\[
S(j,q) = \max_{|u-u'| = 2^{j-1}} \E(|X(u) - X(u')|^q) .
\]
It is compared with the
structure function $S^{model}_{\UPhi \bar{x}} (j,q)$ calculated for each
microcanonical model $\tilde X$, with an error
\[
\epsilon_{st} (j,q) = \frac{ |   S (j,q) - S^{model}_{\UPhi \bar{x}} (j,q)   |} { |  S (j,q)  |} .
\]
Table \ref{tbl:sferr} reports the mean and standard deviation (in brackets) of $\epsilon_{st} (j,q)$ for the model $A$ and model $D$,
at two different scales $2^j$, for the isotropic turbulence. 
They are estimated over $10$ independent realizations $\bar{x}$.
As expected, the errors increase with the order $q$ because 
foveal covariance model only imposes moments of order $q=1$ and $q=2$. The model $D$ creates no significant error on these high-order moments, since the estimated average error is comparable to the estimator
standard deviation. It indicates that in this case the covariance
terms across scales are sufficient to capture high order
moments. On the contrary,
the model $A$ has significant errors for $q \neq 2$ because
it includes no covariance terms across scales and angles.  
Table \ref{tbl:sferr2} reports the value of these 
errors for the MRW. The
errors of model $D$ remain below the errors of model $A$ but are significant
in this case. This is probably due to the fact that the model $D$ does
not capture well the long range spatial correlations of the MRW,
as shown in Figure \ref{fig:corr2}(b). It 
introduces a large model error which appears in higher order moments.

\begin{table}[h]
	\begin{center}
		%\scalebox{0.8}{
		\begin{tabular}{ |c|c|c|c|c|c| }
			%\multicolumn{3}{c}{(a)} \\
			\hline
			q &1 &2 &3 &4 &5    \\
			\hline			
			$\epsilon_{st}^{A}(1,q)$ & 0.04 (0.01)& 0.03 (0.02)& 0.05 (0.04)& 0.14 (0.05)& 0.24 (0.05)
			\\
			\hline
			$\epsilon_{st}^{D}(1,q)$ & 0.01 (0.01)& 0.02 (0.03)& 0.04 (0.05)& 0.06 (0.07)& 0.10 (0.10) \\
			\hline
			\hline
			$\epsilon_{st}^{A}(2,q)$ & 0.03 (0.02)& 0.03 (0.02)& 0.05 (0.03)& 0.08 (0.05)& 0.14 (0.07)			
			\\
			\hline
			$\epsilon_{st}^{D}(2,q)$ & 0.01 (0.01)& 0.03 (0.03)& 0.04 (0.05)& 0.06 (0.07)& 0.08 (0.09) \\
			\hline
		\end{tabular}
	\end{center}
	\caption{
		The structure function error $\epsilon_{st}(j,q)$ of model A,D for 
		the isotropic turbulent vorticity field as a function of $q$, for $j=1,2$.
	}
	\label{tbl:sferr}
\end{table}

\begin{table}[h]
	\begin{center}
		\begin{tabular}{ |c|c|c|c|c|c| }
			\hline
			q &1 &2 &3 &4 &5    \\
			\hline
			$\epsilon_{st}^{A}(1,q)$ & 0.06 (0.02)& 0.04 (0.03)& 0.16 (0.06)& 0.38 (0.06)& 0.60 (0.04) \\
			\hline
			$\epsilon_{st}^{D}(1,q)$ & 0.02 (0.01)& 0.04 (0.03)& 0.05 (0.05)& 0.13 (0.08)& 0.28 (0.09) \\
			\hline
		\end{tabular}
	\end{center}
	\caption{
		The structure function error $\epsilon_{st}(j,q)$ of model A,D for 
		the increment of MRW process as a function of $q$, for $j=1$.
	}
	\label{tbl:sferr2}
\end{table}

This analysis shows that microcanonical models computed
from phase harmonic covariances in foveal neighborhoods capture important
non-Gaussian properties. It gives
accurate models of large classes of processes such as
the turbulence examples. 
The evaluation methodology
with moments is also able to detect model errors. 
It shows that 
long range correlations of wavelet harmonic coefficients are not
well captured by foveal models. 
To circumvent this issue without increasing too much the model size, one
can capture these long range correlations 
with another wavelet transform at each scale, which defines
a scattering transform \cite{leonarduzzi2019maximum}. It amounts to cascade
several wavelet transforms and non-linearities to define $\UPhi$ as in
a multilayer convolutional neural network. Very good visual texture syntheses
have been obtained with covariances of 
multi-layer convolutional neural network
coefficients \cite{gatys2015texture}, but their 
mathematical properties are not understood.

\appendix

\section{Maximum Entropy Wavelet Covariance model}
\label{appendixmaxent}
We explain how to compute the Lagrangian multiples of a 
maximum entropy stationary Gaussian random vector specified by wavelet covariance values, which is used for the model $A$.

To solve the maximum entropy problem,
we follow \cite{Cover2006} and derive a dual problem to minimize the entropy 
with respect to the Lagrangian multiples.
To simplify the deviation, we assume that $X$ is a zero-mean stationary process, and we consider only the translation group $G$. The wavelet 
transform is indexed by $v = (\la,u)$.
The maximum-entropy wavelet covariance model can be written
\[
	\tilde{p} (x) = \frac{1}{Z}
	 \exp\Big(- \frac 1 2
	 \sum_{(\ga  ,\ga'  ) \in \La}
	 \beta_{\ga ,\ga' } \sum_{g \in G} x \star \psi_{\la} (u-g) \,
	 x \star \psi_{\la'}^* (u'-g  ) \Big).
\]
The sum over $\La$ is real-valued because 
$\beta_{\ga, \ga' } =  \beta_{\ga', \ga }^*$. 

With the Parseval formula, 
the sum over translations in $G$ can be written in the Fourier domain as a sum over frequencies 
 \[
 \tilde{p} (x) = \frac{1}{Z}
 \exp\Big(- \frac 1 {2d}
 \sum_{(\ga,\ga') \in \La} 
 \beta_{\ga,\ga'} \sum_{\om} |\what{x}(\om)|^2 \what{\psi}_{\la} (\om) 
 \what{\psi}_{\la'}^* (\om) \,e^{-i \om.(u-u')} \,  \Big).
 \]

The power spectrum $\wtilde P(\om)$ of $\tilde p$ is
\begin{equation*}
\widetilde P(\om) =  \frac{1}{d} \int | \hat{x} (\om) |^2 \tilde{p} (x) dx = 
\Big( \sum_{(\ga,\ga') \in \La}  \beta_{\ga,\ga'} \, \what \psi_{\la}(\om)\,
\what \psi^*_{\la'}(\om) e^{-i \om.(u-u')} 
\Big)^{-1} .
\end{equation*}

To find the optimal Lagrange multipliers, 
we minimize the entropy
of $\tilde{p}$ with respect to all the dual variables $ \beta_{\ga ,\ga'}$. 
The entropy has a closed form,
\begin{equation}
\label{entroinsdfs}
	H(\tilde{p}) = \frac{1}{2} 
	\sum_{(\ga,\ga') \in \La} 
 	\beta_{\ga,\ga'} \sum_\om 
	\widetilde P(\om) \what{\psi}_{\la} (\om) 
	 \what{\psi}_{\la'} (\om)^*  e^{-i \om.(u-u')} + \log( Z ) . 
\end{equation}
The partition function $ Z$ also has a closed form,
\[
	\log( Z) =  \frac{1}{2} \sum_{\om} \log  \widetilde P (\om) + \frac{d}{2} \log(2\pi).
\]

Note that the $ \beta_{\ga,\ga'}$ should be constrained 
so that $\widetilde P(\om) >0, \forall \om $. 
This constrained optimization problem is addressed 
by setting the entropy loss to infinity if any condition is violated.
The derivative of $H(\tilde{p})$ with respect to $ \beta_{\ga ,\ga'}$ 
is calculated from (\ref{entroinsdfs}). We can thus use an unconstrained optimization solver L-BFGS to solve this problem from the real and imaginary
parts of $ \beta_{\ga ,\ga'}$ and $ \beta_{\ga',\ga}$.
The optimal solution gives the power spectrum $\widetilde P(\om)$ of the Gaussian model $A$.
Samples of model A are obtained 
by sampling a stationary Gaussian random vector 
whose power spectrum is $\widetilde P(\om)$.  

\section{Proof of Theorem \ref{Reluthm}}
\label{Relu}

We show that for the rectifier $\rho(a) =  \max(a,0)$, 
\begin{equation}
\label{nsdfsdfa8sdfs}
%\frac 1 {2 \sqrt{2}} |z - z'| \leq
\| \what \HH (z) - \what \HH (z') \|^2
\leq \frac 1 4 \,|z - z'|^2 , \quad \forall (z,z') \in \C^2.
\end{equation}

By definition, $\HH(z)= \{\rho( \Real (  e^{ i \alpha} z ))\}_{\alpha \in [0,2\pi]}$. 
Since $ \| \what \HH (z) - \what \HH (z') \|^2 =\| \HH( z ) - \HH (z') \|^2$,
we do the proof on $\| \HH( z ) - \HH (z') \|^2$.
We prove the theorem by first showing that for a rectifier $\rho(a) = \max(a,0)$ and any $(a,a') \in \R^2$
\[
|\rho(a) - \rho(a')| + |\rho(-a) - \rho(-a')| = |a-a'| .
\]
Indeed $\rho(a) = a$ and $\rho(-a) = 0$ or $\rho(a) = 0$ and
$\rho(-a) = -a$ . The equality is verified by considering separately the cases where $a$ and $a'$ have same or different signs. For $a' = 0$ we get

\begin{equation}
\label{noi8sdf9s98s2}
|\rho(a)|^2 + |\rho(-a)|^2 = |a|^2 
\end{equation}
and for any $a' \in \R$
\begin{equation}
\label{noi8sdf9s98s}
\frac 1 2 |a-a'|^2 \leq |\rho(a) - \rho(a')|^2 + |\rho(-a) - \rho(-a')|^2 \leq 
|a-a'|^2 .
\end{equation}
For any $z \in \C$, 
$\Real(e^{i (\alpha+\pi)} z) = \Real(-e^{i \alpha} z)$ and
$\Real(e^{i (\alpha+\pi/2)} z) = {\rm Imag}(e^{i \alpha} z)$, so
(\ref{noi8sdf9s98s2}) and 
(\ref{noi8sdf9s98s}) imply that 
\[
\sum_{n=0}^3
|\rho (\Real(e^{i(\alpha + n \pi/2)} z) )|^2 =  |z|^2 .
\]
and
\[
\frac 1 2 |z-z'|^2 \leq \sum_{n=0}^3
|\rho (\Real(e^{i(\alpha + n \pi/2)} z)) - \rho (\Real(e^{i(\alpha + n \pi/2)} z'))|^2 \leq
 |z- z'|^2 .
\]
Integrating both of these double inequalities over $\alpha \in [0,\pi/2]$ gives
\[
 \frac 1 {2 \pi} \int_0^{2 \pi}
|\rho ( \Real(e^{i\alpha} z) ) |^2  d \alpha = \frac 1 4 |z|^2 ,
\]
and
\[
\frac 1 {8} |z - z'|^2 \leq \frac 1 {2 \pi} \int_0^{2 \pi}
|\rho (\Real(e^{i\alpha} z)) - \rho (\Real(e^{i\alpha} z') ) |^2 d \alpha  \leq \frac 1 4 |z-z'|^2 ,
\]
which proves (\ref{nsdfsdfa8sdfs}).

\section{Proof of Theorem \ref{SparseSpecMat}}
\label{covarianceFourier}

Property (\ref{densfi8sdfs}) is proved similarly to (\ref{high-ordercov})
by translating $X$ by 
$\tau \in \Lambda_d$ and by using its stationarity to verify that
$\Cov([\what X(\om)]^k, [\what X(\om')]^{k'})
 = e^{i  (k \om - k'\om'). \tau}\, 
\Cov([\what X(\om)]^k, [\what X(\om')]^{k'})$ so both terms vanish
if $k \om \neq k'\om'$.

Property (\ref{densfi8sdfs8}) is verified for $\om \neq 0$ by 
decomposing nominator and denominator
with $\Cov(A,B) = \E(A B^*) - \E(A)\,\E(B^*)$.
Property (\ref{densfi8sdfs7}) uses the same decomposition and
the fact that $\E(\what X(\om)) = 0$ and 
\[
\E([\what X (\om)]^{k}) = 0 ~~\mbox{if}~~k \om \neq 0~.
\]
This is proved again by translating $X$ by 
$\tau \in \Lambda_d$, which transforms $\what X(\om)$ into
$e^{-i \tau.\om} \what X(\om)$. Since $X$ is stationary, it
does not modify this expected value and hence
$\E([\what X (\om)]^{k}) = e^{-i k \tau.\om} \E([\what X (\om)]^{k})$.
Since this is valid for any $\tau \in \Lambda_d$, 
$\E([\what X (\om)]^{k}) = 0$
if $k  \om \neq 0$ modulo $2 \pi$ in dimension $r$. 

Property (\ref{densfi8sdfs7}) for $\om = 0$ and 
property (\ref{densfi8sdfs000}) are also proved
by decomposing $\Cov(A,B) = \E(A B^*) - \E(A)\,\E(B^*)$ and
showing that 
\begin{equation}
\label{zerofreqnsd}
\E([\what X(0)]^{2k}) = \E(|\what X(0)|)~~,~~
\E([\what X(0)]^{2k+1}) = \E(\what X(0)) = d\, \E(X(u))~~.
\end{equation}
This results
from the fact that $\what X(0) = \sum_{u \in \Lambda_d} X(u)$ is real
so $[\what X(0)]^{2k} = |\what X(0)|$ 
and $[\what X(0)]^{2k+1} = \what X(0)$.

If $X$ is Gaussian then $\what X(\om)$ and $\what X(\om')$ are independent
if $\om \neq \om'$ 
because they are Gaussian random variables with a zero covariance.
It results that $[\what X(\om)]^k$ and $[\what X(\om')]^{k'}$ are also
independent for any $(k,k') \in \Z^2$ 
and hence have a zero covariance if $\om \neq \om'$.
If $\om = \om'$ and $k \neq k'$ then (\ref{densfi8sdfs}) 
proves that their covariance remains zero.
It results that $\Cov([\what X(\om)]^k,[\what X(\om')]^{k'}) \neq 0$ 
only if $(\om,k) = (\om',k')$ and hence that the covariance is diagonalized.
If $\om \neq 0$ then
$\what X(\om)$ is a zero-mean 
complex Gaussian random variable whose real and imaginary parts are
not correlated. As a result
${\E(|\what X(\om)|)^2}/{\E(|\what X(\om)|^2)} = \pi/4$.

\section{Bump Steerable wavelet}
\label{appendixcomplexsteer} 
We review the bump steerable wavelet introduced in \cite{mzr-18}. It is illustrated in Figure \ref{waveletfig} and provides a sparse representation of images with oriented structures. 
A general way of constructing steerable wavelets 
is to use the polar coordinate in the Fourier frequency domain \cite{unseral_steerable13}. 
 
We specify the bump steerable wavelet along the radius 
$| \om |$ and angle $arg(\om)$ for $ \om = | \om | e^{i \cdot arg(\om)}$. 
We assume that the number of angles $Q$ is even and its central frequency is $\la = (\xi_0,0)$. 
Its formula is
\begin{equation*}
\label{bump}
\widehat \psi (\om) =  c \cdot \exp \Big (  \frac {- (|\om | - \xi_0 )^2} { \xi_0 ^2 - (|\om |- \xi_0 )^2} \Big)\, 1_{[0,2  \xi_0  ]}  (| \om |)  \cdot  \cos^{Q/2-1} ( arg(\om) ) 1_{ arg(\om) < \frac{\pi}{2} },
\end{equation*}
where $c$ is a normalization constant. The radial function along $| \om |$ is a bump function which is a compact-support approximation of a Gaussian window. The same angular window function along $ arg(\om)$ is used in  \cite{steerableSimoncelli}. 

As in \cite{mzr-18}, the father wavelet $\phi$ is an isotropic Gaussian window function,
\begin{equation*}
\widehat \phi (\om)  =  \exp \Big (  - \frac { |\om |^2} {2 \sigma ^2} \Big) .
\end{equation*}
We choose $\xi_0 = 1.7 \pi$, $\sigma = 0.702 \times \sqrt{2} \times 2^{-0.55}  \xi_0$ and $c=1.29^{-1} 2^{\frac Q 2-1} \frac{ (\frac Q 2-1)! } { \sqrt{ (\frac Q 2) ( Q-2)! } }$. These hyper-parameters are also used in \cite{mzr-18}.
For the wavelet transform $\W x$ with oversampling, we obtain 
numerically the frame constants $A_{\W} =2.0$ and $B_{\W} =  4.6$ for $d=128^2$, $J=5$ and $Q=16$.
It is therefore a complete and stable representation.
These mother and father wavelets also 
satisfy $\psi(u_1,-u_2) = \psi(u_1,u_2)$ 
and $\phi(u_1,-u_2) = \phi(u_1,u_2)$.

\section{Proof of Theorem \ref{thm:cgsym}}
\label{algosym}

Observe first that $\mu_0$ is invariant to the action of
any $g \in G$. Indeed,
the probability measure $\mu_0$ of a Gaussian white noise is uniform on
an ball of $\R^d$ centered in $0$. As a result $\mu_0$ is invariant to
the action of any linear unitary operator and hence invariant to the
action of any $g \in G$ which is linear and unitary. 
Since $\mu_0$ is
invariant to the action of $g$, if the gradient descent is covariant
to the action of $g$ then the derivations of
Theorem 3.4 in \cite{bruna2018multiscale} prove that
the probability measure
$\mu_t$ of $x_t$ is invariant to $g$ for $t \geq 0$.

We thus need to show that 
the L-BFGS gradient-descent algorithm is covariant to the action 
$g \in G$ 
at each iteration (before stopped by line-search conditions). 
Let $\tilde{x}_t$ and $x_t$ 
be two sequences generated by the gradient descent 
from the initial conditions $\tilde x_0$ and $x_0$.
The covariance to the action of $g$ means that if
$\tilde{x}_0 = g. x_0$ then 
$\tilde{x}_t = g. x_t$ for any $t \geq 0$. 
Although invariant properties of a gradient-based algorithm are well studied in the optimization literature \cite{OptimBook,FletcherOptim}, we give below a prove for completeness.

L-BFGS  is a variant of the quasi-Newton method BFGS. It estimates the inverse Hessian matrix of the objective function $f$ based on 
\[
s_{t} = x_{t+1} - x_{t}~~\mbox{and}~~y_t =  \nabla f(x_{t+1}) -  \nabla f(x_{t}). 
\]
Assume there is a limited memory size $m$ to 
compute an approximation $H_t^m$ of the Hessian, 
from $ s_{t-a}$ and $y_{t-a}$ for $ 1 \leq a \leq m$. 
At each iteration $t$, the algorithm chooses an appropriate step-size $\alpha_t$ satisfying strong Wolfe conditions \cite{OptimBook} and updates $x_t$ with
\begin{equation*}
x_{t+1} = x_t - \alpha_t H_t^m \nabla f(x_t) .
\end{equation*}

Since $f(x) = \| \widetilde K_{\UPhi x} - \widetilde K_{\UPhi \bar{x}} \|_\LaG^2$ and $\wtilde K_{\wH \W g.x}  = \wtilde K_{\wH \W x}$, 
it results that $f(g . x)=f(x)$. It implies that
\begin{equation}
\nabla f (g . x) = g . \nabla f(x) .
\end{equation}

We prove the covariance property by induction. We suppose
that $\tilde{x}_{t'} = g .  x_{t'}$ for $ t' \leq t$ in order to prove 
that $ \tilde{x}_{t+1} = g . x_{t+1}$.
By definition, 
\begin{equation*}
\tilde{x}_{t+1} = \tilde{x}_t - \tilde{\alpha}_t \tilde{H}_t^m \nabla f(\tilde{x}_t) .
\end{equation*}

The induction assumption implies that $\tilde{s}_{t'} = g .  s_{t'}$ and $\tilde{y}_{t'} = g .y_{t'}$ for $ t' \leq t$. It is
therefore sufficient to verify that
\[
\tilde{H}_t^m = g . H_t^m . g^T, \quad \tilde{\alpha}_t  = \alpha_t, \quad \forall t \geq 0
\]
Each $H_t^m$ is defined recursively by
\[
	H_t^a = V_{t+a-m-1}^T H_t^{a-1} V_{t+a-m-1} + \rho_{t+a-m-1} s_{t+a-m-1} s_{t+a-m-1}^T,
\]
for $ 1 \leq a \leq m$ and for all $t \geq 0$
\[ 
V_{t} = Id - \rho_{t} \,s_{t}\,  y_{t}^T
~~\mbox{and}~~
 \rho_{t} = (s_{t}^T \,y_{t})^{-1}.
\]
We thus have $\tilde{V}_{t} = g .  V_{t} .  g^T$.
By convention, $ H_0^{0} = Id$ and  $ H_t^{0}= (s_{t-1}^T\, y_{t-1})^{-1} Id$ for $t \geq 1$. We thus conclude that $\tilde{H}_t^m = g H_t^m g^T$. 
Note that if $t \leq m$, the index $t+a-m-1$ is negative,
and in this case $V_{t+a-m-1}=Id$ and $\rho_{t+a-m-1}=0$. 

To verify that $\tilde{\alpha}_t  = \alpha_t$, we observe 
that the strong Wolfe-line search conditions are invariant to any 
linear unitary $g \in G$. 
Indeed these conditions are invariant to any affine operator \cite{OptimBook}.

\section{Proof of Proposition \ref{prop:covsign}}
\label{proofcovsign}

If $p$ is invariant to an action of $G$ then
$ M_{\wH \W}$ and $K_{\wH \W}$ are not modified if $X$ is transformed
into $g.X$. Similarly, we verify from 
(\ref{phiFouriensfmean6}) and (\ref{phiFouriensf6}) 
that
$ \widetilde M_{\wH \W}$ and $\widetilde K_{\wH \W}$ 
are not modified if $\bar x$ is transformed
into $g.\bar x$. The proposition results are obtained by
relating 
$[ g.x \star \psi_{\la} ]^k$ with $[x \star \psi_\la (u)]^k$ and applying
this relation to $x = X$ and $x = \bar x$.

(i). If $g. x = -x$ then
\begin{equation}
\label{conoi8usdfase}
[ g.x \star \psi_{\la} (u) ]^k = (-1)^k [x \star \psi_\la (u)]^k.
\end{equation}
For $v = (\la,k,u)$ we thus derive that
$ M_{\wH \W} (v) = (-1)^k M_{\wH \W} (v) $
and $ \wtilde M_{\wH \W} (v) = (-1)^k \wtilde M_{\wH \W} (v) $ and thus
vanish if $k$ is odd. Similarly, for 
$v' = (\la',k',u')$ we verify that
$ K_{\wH \W} (v,v') = (-1)^{k+k'} K_{\wH \W} (v,v')$
and
$ \wtilde K_{\wH \W \bar x} (v,v') = 
(-1)^{k+k'} \wtilde K_{\wH \W \bar x} (v,v')$
which vanishes if $k+k'$ is odd.

(ii) If $g.x (u) = x(-u)$ then
\begin{equation}
\label{conoi8usdfase2}
[ g.x \star \psi_{\la} (u) ]^k = [x \star \psi_\la (u)]^{k*} 
\end{equation}
because $\psi_{\la} (-u)= \psi^*_{\la} (u)$ and $x$ is real.
Since $\what h(k)$ is real, 
$ M_{\wH \W} (v) = \what h(k)\,\E([X \star \psi_\la (u)]^{k})^*  =  M_{\wH \W} (v)^*$ is real. The same property applies to
$ \wtilde M_{\wH \W} (v)$. Applying (\ref{conoi8usdfase2}) to 
$ K_{\wH \W} (v,v')$  and
$\wtilde K_{\wH \W \bar x} (v,v')$  also proves that they are real.

\section{Proof of Theorem \ref{prop:covrot}}
\label{proofcovrot}

Since $\la = 2^{-j} r_{-\ell} \xi$ the 
index of $\UPhi = {\what \HH \W}$ can
be written $v = (j,\ell,k,u)$. 
We compute the covariance $K_{\what \HH \W}(v,v')$
for $v' = (j',\ell',k',u')$ with $u' = u$. Since the
distribution is stationary the covariance value remains
the same if we set $u = u' = 0$.

To prove the theorem we first observe that if
$g.x = r_\eta x$ then
\begin{equation}
\label{angul}
[ g.x \star \psi_{ \la } (0) ]^k = [  x \star \psi_{ r_\eta \la } (0) ]^k .
\end{equation}
If $X$ is isotropic then 
$K_{\wH \W}  (v,v')$ and 
$\wtilde K_{\wH \W \bar x}  (v,v')$ are invariant by rotations.
The covariance property (\ref{angul}) thus implies that they are
a function of the difference of angles $\ell - \ell'$. It is therefore
a periodic convolution kernel along angles. It results that the angular dependence
is diagonalized by a Fourier transform along the angle variable,  
which  proves (\ref{rotation2}) for $\UPhi = \F_\ell\,\what \HH  \W$. 

A central reflection is a rotation by $\pi$. Since the distribution of
$X$ is invariant to all rotations it is invariant to a central reflection.
Property (ii) of Proposition (\ref{prop:covsign}) implies that
$K_{\what \HH  \W}$ is real. 

Let us consider the line reflection $g$ relatively to the 
horizontal axis. 
Since $\psi(u_1,-u_2) = \psi(u_1,u_2)$ 
and $\phi(u_1,-u_2) = \phi(u_1,u_2)$ we verify that
for any $\la = 2^{-j} r_{-\ell} \xi$ and $\la' = 2^{-j} r_{\ell} \xi$ we have
\begin{equation}
\label{psiflipsymm}
		\psi_{\la}(u_1,u_2) = \psi_{\la'} (u_1,-u_2) . 
\end{equation}
As a result, if $\la = 2^{-j} r_{-\ell} \xi$ then
\begin{equation}
\label{refl}
[   (g.x) \star \psi_{\la} (0) ]^k = [ x \star \psi_{\la'} (0) ]^k~~
\mbox{with}~~ \la' = 2^{-j} r_{\ell} \xi.
\end{equation}

If $K_{\what \HH  \W}$ is invariant
to line reflections then (\ref{refl}) implies that
$K_{\what \HH  \W}$ is not modified when $(\ell,\ell')$ is transformed
into $(-\ell,-\ell')$. Since $K_{\what \HH  \W}$ is real and invariant
to a change of sign of $(\ell,\ell')$ it results that the 
Fourier coefficients of $K_{\F_\ell \what \HH  \W}$ are also real.

\bibliographystyle{unsrtModif}

\bibliography{ref}

\end{document}